\documentclass[draft]{agujournal2019}
\usepackage{url} 
\usepackage{lineno}
\usepackage{soul}
\soulregister\ref7
\soulregister\cite7
\soulregister\citeA7
\soulregister\citeNP7

\usepackage{amssymb}

\usepackage{tabulary}

\journalname{Journal of Advances in Modeling Earth Systems (JAMES)}

\begin{document}

\title{Machine learning emulation of precipitation from km-scale UK regional climate simulations using a diffusion model}

\authors{Henry Addison\affil{1}, Elizabeth J. Kendon\affil{2,1}, Suman Ravuri\affil{3}, Laurence Aitchison\affil{1}, Peter A.G. Watson\affil{1}}

\affiliation{1}{University of Bristol, Bristol, UK}
\affiliation{2}{Met Office Hadley Centre, Exeter, UK}
\affiliation{3}{NVIDIA Corporation}

\correspondingauthor{Henry Addison}{henry.addison@bristol.ac.uk}


\begin{keypoints}
\item We present a probabilistic machine learning method (CPMGEM) for emulating computationally expensive km-scale regional climate models
\item Our emulator of a km-scale UK climate model reproduces well the climatology of daily precipitation conditioned on coarse climate variables
\item Our emulator complements the Met Office's high-resolution UKCP Local dataset by allowing realistic and computationally cheaper samples of daily precipitation

\end{keypoints}

\begin{abstract}
High-resolution climate simulations are valuable for understanding climate change impacts. This has motivated use of regional convection-permitting climate models (CPMs), but these are very computationally expensive. We present a convection-permitting model generative emulator (CPMGEM), to skilfully emulate precipitation simulations by a 2.2km-resolution regional CPM at much lower cost. This utilises a generative machine learning approach, a diffusion model. It takes inputs at the 60km resolution of the driving global climate model and downscales these to 8.8km, with daily-mean time resolution, capturing the effect of convective processes represented in the CPM at these scales. The emulator is trained on simulations over England and Wales from the United Kingdom Climate Projections Local product, covering years between 1980 and 2080 following a high emissions scenario. The output precipitation has a similar spatial structure and intensity distribution as in the CPM simulations. The emulator is stochastic, which improves the realism of samples. We include some evidence about the emulator's skill for extreme events with return times up to $\sim$100 years. We demonstrate successful transfer from a ``perfect model'' training setting to application using GCM variable inputs. It captures the main features of the simulated 21st century climate change, but exhibits some error in the magnitude. 
We also show that the method can be useful in situations with limited amounts of high-resolution data. Potential applications include producing high-resolution precipitation predictions for large-ensemble climate simulations and producing output based on different GCMs and climate change scenarios to better sample uncertainty.
\end{abstract}

\section*{Plain Language Summary}


Climate models allow us to explore how the climate may change in the future. High-resolution climate models divide the atmosphere into very small boxes (e.g. 2.2 km across). This allows understanding of the changing climate on a local scale, which is important for effective preparation. However, they are slow and expensive to run. This hinders their use for exploring the spread of possible changes at a local scale, particularly for rare events, like heavy rainstorms. We demonstrate a novel use of a generative AI technique to emulate the daily rainfall output of a high-resolution climate model covering England and Wales. That is, to predict the high-resolution daily rainfall when given output from a cheaper, lower-resolution climate model. Our emulator does this faster and more cheaply than the high-resolution climate model and outputs realistic maps of daily rainfall over England and Wales. It reproduces important statistics of the rainfall distribution from the high-resolution climate model, including examples of the heaviest rainfall events and the main features of the predicted change over the 21st Century. This emulator could be used to produce high-resolution predictions for large sets of coarse climate model output and enable better understanding of potential changes in extreme rainfall.

%
%

%


%
%
%
%

\section{Introduction}

\begin{figure}[htbp]
    \centering
    \includegraphics[width=\linewidth]{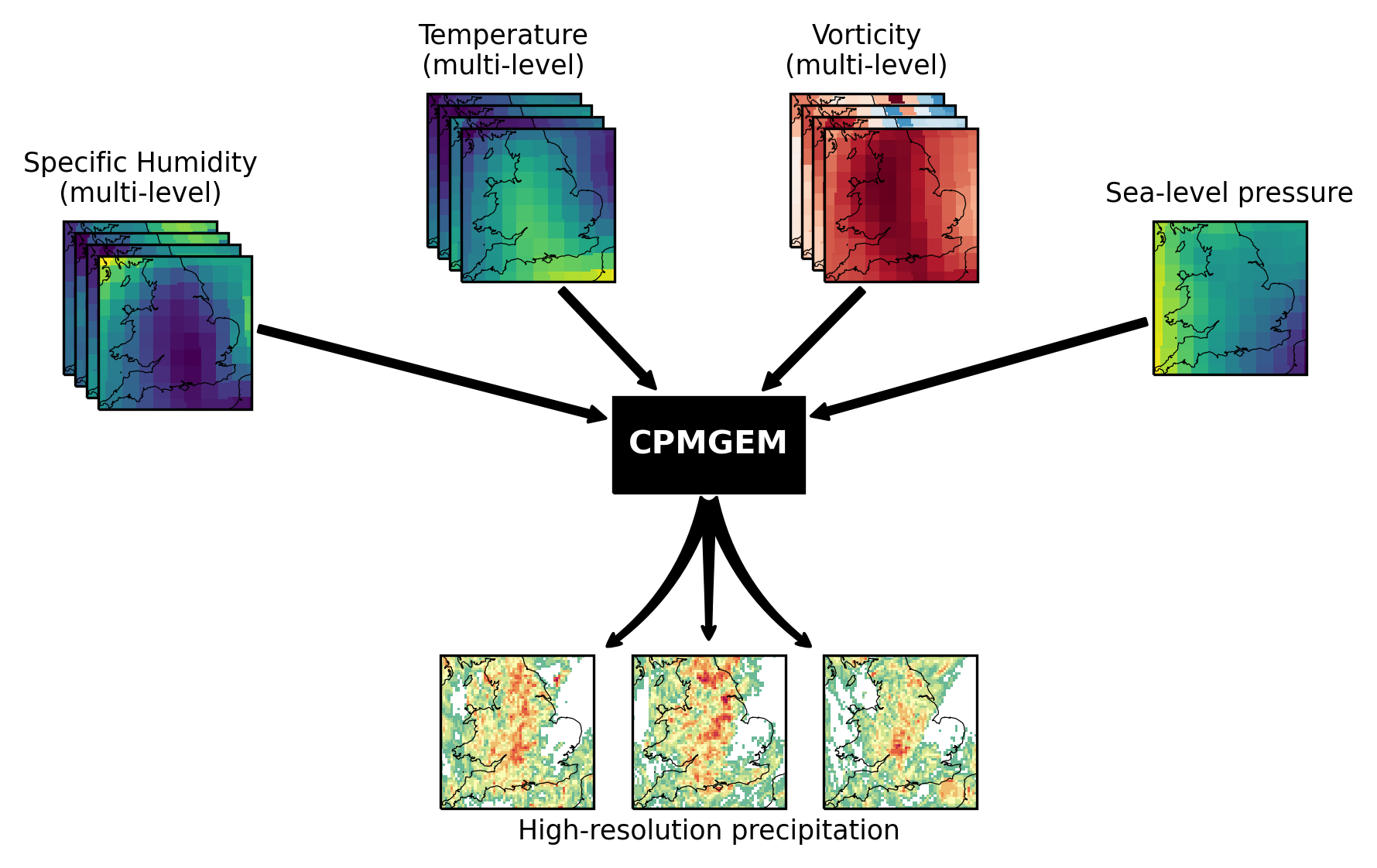}
    \caption{\textbf{Schematic diagram of the inputs and outputs of the emulator.} The emulator is trained to stochastically generate samples of high-resolution, daily mean precipitation over England and Wales (bottom panels). The target is for these samples to have properties matching output from the Met Office UK convection-permitting model. The emulator is stochastic, and can generate any number of samples for a single set of inputs. For input fields, the emulator takes variables at the same 60km grid spacing as the global climate model runs used to drive the CPM. These input fields are pressure at mean sea-level and specific humidity, temperature and vorticity at 250, 500, 700 and 850hPa (all daily means).}
    \label{fig:ai-schematic}
\end{figure}

Understanding precipitation at a local scale is highly important for planning better climate change adaptation measures. Key challenges include representing the fine-scale structure of precipitation and effects such as convection.  Convection is a key driver of many extreme events, but occurs at scales far below the typical resolutions of global climate models (GCMs) \cite{kendon2012vhrrcmprecip}. This has motivated using regional climate model (RCM) simulations at resolutions high enough to capture these processes, with GCMs providing boundary conditions \cite<e.g.,>{kendon2021ukcpscienceupdate}. These are known as regional convection-permitting models (CPMs). However, running a CPM is very computationally expensive, limiting sampling of climate change scenarios and extreme precipitation events. Furthermore, it is highly technically challenging to run a CPM with output from different GCMs \cite{sobolowski2023EUROCORDEXdesign}, limiting exploration of uncertainty in future climate projections at a local scale. Overcoming these challenges would allow for a much more comprehensive understanding of the range of potential severity of future extreme precipitation, for example through studying extreme events with large-ensemble GCM simulations \cite<e.g.,>{maher2021largeensembles,leach2022extremesamplegeneration} and covering the range of climate change projections across different GCMs and emissions scenarios \cite<e.g.,>{eyring2016CMIP6}.

Here, we demonstrate that a machine learning (ML) emulator of a CPM can produce realistic, high-resolution daily-mean precipitation simulations for England and Wales, conditioned on coarse-resolution weather states produced by a GCM. Our emulator is based on a diffusion model, a state-of-the-art generative ML approach \cite{sohldickstein2015difforigin, song2021sbgmsde}. We call it ``CPMGEM" (CPM Generative EMulator). A schematic diagram is shown in Figure~\ref{fig:ai-schematic}. 
We trained the emulator using output from the Met Office United Kingdom (UK) CPM, the first set of national-scale CPM climate simulations, which were produced as part of the UK Climate Projections (UKCP) Local product \cite{kendon2021ukcpscienceupdate, kendon2023uklocaltrends}.
It produces samples at a small fraction of the computational cost of running a CPM at a resolution that is fine enough to produce information about precipitation at small enough spatial scales for applications such as flood modelling \cite<e.g.>{bates2023ukfloodingrisk}.
The emulator output is at 8.8km grid spacing, coarser than the 2.2km grid spacing of the CPM, but this reflects the scale of its better-resolved features, which generally span multiple grid boxes \cite<e.g.,>{klaver2020effectiveresolution}, and coarsened CPM output remains more realistic than the output of an RCM with a similar $\sim10\mathrm{km}$ grid spacing \cite{kendon2017cpmprecipimprovements}.

There exists much previous work on developing statistical methods to predict high-resolution precipitation \cite<e.g.,>{schoof2013downscalingclimatology, maraun2018downscalingbook}. However, these methods have generally struggled to produce predictions with realistic spatial structure \cite{maraun2019sddownscaling, Widmann2019sptialvariability}, which is important for predicting impacts such as flooding \cite<e.g.,>{scahller2020resolutionrole, archer2024CPMflooding}. One particular challenge is representing the stochastic aspect of downscaling, where the coarse-scale weather state does not uniquely define the state at a finer scale \cite<e.g.,>{maraun2017climchangesimbiascorrection, maraun2019sddownscaling}. We refer to the difference between the high-resolution precipitation and the best possible deterministic prediction given the coarse inputs as the stochastic component. This component can have a very complex structure at fine scales, particularly when small-scale convection plays a substantial role. This is very difficult to represent with conventional statistical or deterministic ML approaches.

The limitations of conventional statistical methods have motivated application of ML methods, which have been found to be able to make precipitation predictions with high-quality spatial structures. Some previous work has applied ML to downscale coarse-resolution precipitation predictions to capture high-resolution details~(e.g., \citeNP{Vosper2023cgandownscalingcyclones, Harris2022cgandownscaling, Leinonen2020GANsd, vandal2018mldownscaling, sha2020unetdsprecip, lopezgomez2025rcmdiffusiondownscaling}). However, a CPM has potential to also improve the coarse-scale structure of precipitation events, particularly when convection plays a large role. Some studies have examined ML methods that do not use coarse-resolution precipitation as an input, learning to make predictions based on other fields that specify the weather state, potentially allowing learning of this added value on large-scales. These have primarily focussed on deterministic approaches \cite<e.g.,>{wang2023dlforbiascorrectanddownscaling, doury2024precipemulator}. However, these do not produce output that fully captures fine-scale spatial structures. Other works have found that stochastic ML models of precipitation using coarse precipitation as an input can capture fine detail well, indicating that use of such stochastic methods is promising(e.g., \citeNP{Vosper2023cgandownscalingcyclones, ravuri2021deepgenprecip, Harris2022cgandownscaling, hess2022phiganesmprecip, rampal2025cganadvlosssensitivity}).
Recent work has begun exploring the application of stochastic ML methods to predict precipitation based on non-precipitation predictor variables \cite{addison2022neuripsccai,mardani2023residualdiffkmscaledownscalingpublished,ling2024eastasiadiffusion,rampal2024rcmemulatorsreview,saoulis2025fathomdiffusion}, the approach we use here.

Whilst many statistical downscaling studies have focused on reproducing properties of observations (e.g., \citeNP{gutierrez2019sdcomparison,bettolli2025stationcnnberngamma,hess2025scaleadaptiveconsistencymodels, aich2024diffdownbc}), there is growing interest in learning to emulate high-resolution numerical models \cite<e.g.,>{walton2015hybriddownscaling, boe2022hybriddownscalingRCMemulator, doury2023RCMemulator, doury2024precipemulator,kendon2025MLemulatorsperspective}. The goal is to produce output similar to that of the numerical models at much reduced cost, enabling more complete sampling of climate change uncertainty and internal variability. 
One key advantage of learning to reproduce the output of high-resolution models is the ability to learn the effects of climate change from a physically-based model. For example, in the UK, this includes complex effects such as summer rainfall becoming more concentrated into shorter, more intense downpours with climate change \cite{kendon2014heaviersummerrain, kendon2021ukcpscienceupdate}, which may be difficult to capture using a method trained on observations alone \cite{rampal2024cganfutureextrapolotaion}. Another advantage is having potentially larger datasets for training, which can help with learning a good representation of extreme events.
It remains an on-going challenge to develop emulators with good skill when extrapolating to unseen climates \cite{rampal2024rcmemulatorsreview,sun2024detdldownscalingreview}, such as outputs from different GCMs or from different time periods to those seen in training. Since our training dataset contains simulations for years between 1980 to 2080, in our case we use future simulation data in the training of the emulator and target producing rainfall for climate states within or close to this range, avoiding the need to extrapolate to very different climates. This can then be applied to downscale other similar GCM simulation datasets where downscaling with the CPM is not practical, such as large-ensembles of GCM runs of the 21st century.

A practical CPM precipitation emulator needs to satisfy several key requirements together, including being able to capture the effects of the complex dynamical processes represented in CPMs, realistically representing spatial structures of resulting precipitation at both small and large scales, with sufficiently fine resolution for impacts modelling, and being able to use predictors from GCMs. There are still open questions about the capability of machine learning models to achieve these aims concurrently. Methods to emulate RCMs with parameterized convection (e.g., \citeNP{rampal2025cganadvlosssensitivity, rampal2024cganfutureextrapolotaion, lopezgomez2025rcmdiffusiondownscaling, doury2024precipemulator}) may not necessarily transfer well to emulating the more realistic physics of CPMs. For example, being able to represent the stochastic component of precipitation realistically may be more important and challenging, such as in weather situations when small-scale thunderstorm activity dominates. Previous work emulating CPMs has focussed evaluation on weather prediction skill \cite{mardani2023residualdiffkmscaledownscalingpublished} or has used climate models and evaluations driven by ERA5 \cite{tomasi2025CPMdiff, miralles2022CPMGANwind, annau2023CPMdetGANwind}, leaving open the question of whether ML emulators of climate models at this scale are able to recreate climate statistics using GCM outputs as predictors, as is necessary in practice for downscaling GCM simulations. 
It has also not been shown whether the methods considered in these studies represent the climate change signal well, which is important for the goal of projecting future precipitation changes, or whether the methods work well for extreme weather.

Here, we show that our UK CPM emulator performs well at matching climatological precipitation statistics from the CPM when given coarse weather state information, both when this is coarsened CPM data (Section~\ref{sec:results:cCPM_predictors}) and when this is output from the driving GCM (Section~\ref{sec:gcm-xfer}). We use a relatively large set of CPM simulations for training and testing, enabling relatively precise quantification of biases. We also use our large dataset to provide some evaluation of our emulator's ability to represent extreme weather events with return periods up to $\sim$100 years, event intensities that are typically important in high-stakes applications (e.g., \citeNP{weaver2017climaterisk,watson2022mlextremes}). We furthermore evaluate how well the emulator has learnt the 21st century climate change signal when using GCM predictors, finding that it captures the main features, but that it exhibits some error in the magnitude (Section~\ref{sec:results:ccs}). We also show that the method can perform well when limited amounts of CPM data are available (Section~\ref{sec:results:small_data}). The method is described in detail in Section~\ref{sec:method} and we summarise the results and discuss directions for future work in Section~\ref{sec:conc}.

\section{Materials and methods}\label{sec:method}


\subsection{Data}

CPM data for training and evaluation are taken from the UKCP Local product \cite{kendon2023uklocaltrends, kendon2021ukcpscienceupdate} and the GCM data is from UKCP18 Global simulations that were used to produce boundary conditions for the CPM \cite{murphy2018ukcp18land}.

\subsubsection{Target CPM precipitation}
\label{sec:method:cpmpr}

We aim to emulate the daily mean precipitation output over England and Wales from the Met Office UK CPM.
This dynamical model has a grid spacing of 2.2km covering the UK. The dynamical downscaling is done in two steps.
To provide boundary conditions for the UK CPM, Met Office GCM simulations with 60km grid spacing are first dynamically downscaled using a 12km RCM \cite{kendon2021ukcpscienceupdate}. The domain for the 12km RCM is the EURO-CORDEX grid \cite{eurocordexdomain} which covers Europe and parts of the North Atlantic and North Africa (see \url{https://cordex.org/domains/cordex-region-euro-cordex/} for full definition). The boundary conditions for the UK CPM are then derived from the RCM simulations. The CPM has a skilful representation of convective processes and captures small-scale rainfall systems that are not resolved in the driving GCM \cite{kendon2012vhrrcmprecip, kendon2014heaviersummerrain, kendon2020futureCPMwinterprecip}.
Note that our emulator just uses input variables on the scale of the GCM grid to directly emulate the UK CPM, and output from the 12km RCM is not used.

We use data from 12 ensemble members. Each member uses a GCM/RCM pair with unique parameter settings, but an identical CPM, and we treat their outputs as being exchangeable. 
Note that the ensemble members are not meant to be aligned in time. The driving GCM ensemble members will develop their own circulation over the long duration of their simulation runs.
The simulations follow the Representative Concentration Pathway 8.5 (RCP8.5) climate forcing scenario. We use data from three time periods provided in the initial 2021 release of UKCP Local, which we refer to as ``Historic'' for 1981--2000, ``Present'' for 2021--2040 and ``Future'' for 2061--2080. This gives a total of 720 years of data, which allows development of an emulator that can learn to represent extreme events, and also allows evaluation on such cases.

The simulations use a 360-day year, consisting of twelve 30-day months.
Simulations begin on 1st December of a given year, so we take each numbered year to run between 1st December to 30th November (e.g. 1981 means 1st December 1980 to 30th November 1981).

We use daily-mean data from the CPM. For our target high-resolution precipitation, we coarsen the CPM output to 8.8km grid spacing using conservative interpolation and extract the England and Wales domain. The scale of resolved features in dynamical climate models is generally several times the grid box spacing \cite{klaver2020effectiveresolution}, so this coarsening captures the better resolved scales. This coarsening also allows a good trade-off between resolution, domain size and compute requirements. The emulator can run well on commodity hardware such as a 10GB NVIDIA GeForce RTX 2080 Ti graphics processing unit. This precipitation resolution in space and time is superior to that recently used for state-of-the-art UK flood risk modelling \cite{bates2023ukfloodingrisk}, so it is at a useful scale for climate impacts assessment, while the domain is large enough to visualise features such as fronts and mesoscale convective organisation. There are no clear barriers to increasing the spatial resolution and domain size given sufficient computing resources.

Climate model output is known to have biases compared to observations.
The root mean squared (RMS) relative time-mean bias in precipitation for the Met Office UK CPM over the whole UK land domain was found to be 14.6\% in winter and 14.5\% in summer when driven by ERA-Interim reanalysis \cite{dee2011erainterim} (25.7\% in winter and 15.6\% in summer when driven by the GCM) \cite{kendon2021ukcpscienceupdate}.
This is an improvement on the bias for the coarser driving models.
For our purposes, successful emulation means recreating CPM behaviour, including biases compared to observations. If appropriate, a user can then apply bias-correction using the same methods as would be applied to CPM output.

\subsubsection{Coarse predictors}
\label{sec:methods:coarse}

Ideally we could learn a mapping directly from GCM variables, with 60km grid spacing, to high resolution CPM precipitation. However, synoptic-scale features in a CPM and its driving GCM simulation are not always well-aligned, due to internal variability in the CPM and intermediate RCM (also discussed by \citeA{doury2023RCMemulator,vandermeer2023unetrcmemulator, banomedina2024emulatorxferxai, rampal2024rcmemulatorsreview}). For example, the GCM may simulate precipitation in the west of the UK whilst it is in the east in the CPM on the same day due to a difference in the exact positioning of a weather front. This means that a mapping directly between GCM and CPM variables is very noisy. \citeA{banomedina2024emulatorxferxai} find an emulator trained using coarsened fine-scale data as inputs is better able to transfer to other GCMs. \citeA{vandermeer2023unetrcmemulator}, however, find it better to downscale GCM inputs when the GCM to RCM mapping is learnt directly but their GCM and RCM datasets may show better alignment due to their longer time-scales (monthly rather than daily). Therefore, we follow the same approach as \citeA{doury2023RCMemulator} and train using data from the CPM that has been coarsened to the GCM resolution using conservative interpolation (referred to as ``cCPM''). This ensures that during training, the weather features in the coarse- and high-resolution data are aligned. Once trained, the emulator can use either coarsened CPM variables or GCM variables as inputs. 

We have selected predictors of precipitation based on those that have been found useful in previous statistical downscaling work (e.g., \citeNP{gutierrez2019sdcomparison}) and based on the understanding of the physical drivers of precipitation \cite{chan2018precippredictors}. We use daily-mean predictor variables, as higher time resolution output is not generally as widely available from GCM simulations. Since we train the emulator using coarsened CPM variables as input with the aim to apply it using GCM variables, it is important to select variables that are represented to a similar degree of realism in the CPM and GCM. It is also important to select variables where the direction of causal influence is primarily from the variables to the high-resolution precipitation. We avoid using the coarse-resolution precipitation or variables in the boundary layer (below $\sim$850hPa) as predictors, since these are not represented sufficiently realistically in the GCM. We tested several choices of coarse-resolution predictor variables during the emulator development. In our final design, we use pressure at mean sea level and three variables on levels in the free troposphere: specific humidity, temperature and relative vorticity of the horizontal wind components. For these multi-level variables, we use a range of pressure levels (250, 500, 700 and 850 hPa), similar to previous works (e.g., \citeNP{gutierrez2019sdcomparison, doury2023RCMemulator}). We use the vorticity rather than the full wind field because vorticity has been found to be a good predictor of convection \cite{chan2018precippredictors} and the divergent component of the horizontal wind and the vertical wind component may be more strongly affected by feedbacks from convection. Vorticity is calculated from eastward and northward wind components on the coarse 60km grid. The full set of coarse input variables to the emulator are:
\begin{itemize}
    \item pressure at mean sea level
    \item specific humidity at 850, 700, 500 and 250 hPa
    \item vorticity at 850, 700, 500 and 250 hPa
    \item temperature at 850, 700, 500 and 250 hPa
\end{itemize}

We use the coarse-resolution predictors across the whole England and Wales domain. Note these are different to the inputs used by the CPM, for which atmospheric GCM data is supplied at the boundary of the intermediate 12km RCM. We describe our method as an ``emulator'' in the sense that it is developed to predict high-resolution precipitation corresponding to a given GCM simulation that has the same properties as precipitation from the CPM. This approach has the benefit of using the physics embedded in the GCM to predict reasonably realistic weather states at coarse-resolution.

\subsection{ML Models}

\subsubsection{CPMGEM: Diffusion model-based emulator}
\label{sec:methods:cpmgem}

Figure~\ref{fig:ai-schematic} shows an overview of the emulator's inputs, outputs and domain. Inputs and outputs cover the same spatial domain. The inputs are the coarse variables described above. The emulator is stochastic and can generate an arbitrary number of samples of high-resolution precipitation for given coarse-resolution inputs. This means the emulator can learn to represent the inherent stochastic component of the downscaling task. Ideally, the distribution of these samples matches the distribution of high-resolution precipitation that the CPM would produce over many runs using the same driving GCM data, though we cannot directly test this.

The emulator is a diffusion model based on the work of \citeA{song2021sbgmsde}. The training process of a diffusion model proceeds by first adding noise in many steps to target samples (precipitation fields in our case) so that they resemble pure noise after the final step. Then a neural network is trained to assist computing of a process which reverses this noising (again over many steps) and thus can be used to turn pure noise into samples from the target distribution (see \ref{sec:app:diff} for a fuller description). 

Diffusion models have shown excellent performance in problems in the image domain \cite{dharwial2021diffbeatsgans, song2021sbgmsde}, including work demonstrating their effectiveness in stochastic super-resolution \cite{saharia2023sr3,li2022srdiff}.
Diffusion models offer a good trade-off over other deep generative models in terms of sample sharpness (variational autoencoders and normalising flows tend to produce blurry samples), sample diversity (generative adversial networks, GANs, can suffer from mode collapse mode collapse \cite{grover2018flowgan}) and sampling cost (more efficient than AR models and NFs with growing image/grid size).
GANs are prone to issues with training instability \cite{creswell2018ganreview} making them harder to use.

The hyperparameters, network architecture and training procedure we use are the same as those of \citeA{song2021sbgmsde}.
For our emulator we use \citeA{song2021sbgmsde}'s sub-Variance-Preserving (``sub-VP") formulation of the stochastic differential equation (SDE) that defines the noise-adding process in their framework.
\citeA{song2021sbgmsde} note that the sub-VP SDE ``can perturb any data distribution to standard Gaussian under suitable conditions''. This is important for an distribution like precipitation.
In practice the diffusion process is split into discrete steps. We used 1000 steps, which was sufficient for both training and sampling, matching the number used by \citeA{song2021sbgmsde} for their 64 x 64 pixel image generator.
We use their NCSN++ configuration (which is an improved version of their original Noise Conditional Score Network based on the U-Net architecture \cite{Ronneberger2015unet} used by \citeA{ho2020ddpm}) as the backbone neural network. We have adapted it to use conditioning information by creating an input to the network based on the coarsened variables and the target field that is to be denoised. We do this by regridding the coarsened variables to the target grid using a nearest neighbour approach and then stacking these upsampled fields with the target field. We added a final single-channel convolutional layer so that the output (the reduced noised version of our noisy target field in the inputs) matches the size of our target.
This final network has \(\sim\)63M parameters, which requires approximately 7GB of GPU memory to perform the forward/backward passes needed in training.

We train for 20 epochs on the training dataset (batch size  of 16, to keep within a 10GB GPU memory limit) using the Adam optimizer with learning rate 0.0002. Further training did not show improvements to the loss computed on the validation set nor to our evaluation metrics computed on samples based on the validation set (see Section~\ref{sec:methods:splits} for how data was split into training, validation and test sets).
We keep the learning rate, gradient clipping, and learning rate warm-up schedules used by \citeA{song2021sbgmsde} and also their use of exponential moving average to help stabilize training as well as potentially resulting in a more generalizable model \cite{moralesbrotons2024ema}.
We use the Euler-Maruyama method, a simple first-order method \cite{bayram2018sdemethods}, to solve the reverse SDE for generating samples with the fitted network.

The diffusion model produces samples of high-resolution precipitation fields directly. Unlike latent diffusion \cite{rombach2022ldm}, it does not operate on a compressed latent space and unlike work by \citeA{mardani2023residualdiffkmscaledownscalingpublished} it does not predict a stochastic residual to be combined with another prediction.

Training  and sampling are feasible using a single commodity GPU. Training for 20 epochs took approximately 41 hours using a 10GB NVIDIA GeForce RTX 2080 Ti.
It took approximately 5 hours on the same hardware to generate a set of samples using the predictors from a single ensemble member of the test dataset (9 years; see Section~\ref{sec:methods:splits} for how data is split into training, validation and test sets). A 20 year block of emulated projections would thus take around 11-12 hours.
By comparison, each 20 year simulation of the CPM requires 6 months using several hundred CPUs.

\subsubsection{Comparison methods}
\label{sec:methods:comparison}

We compare results against a deterministic U-Net \cite{Ronneberger2015unet}. The U-Net architecture has been shown to perform well in similar climate downscaling settings (e.g., \citeNP{doury2023RCMemulator, vandermeer2023unetrcmemulator}). 
We use a U-Net based on the original architecture by \citeA{Ronneberger2015unet} with adjustments to match the resolution and number of input and output channels of our dataset (64x64 resolution, 13 input channels, 1 output channel). As with the diffusion emulator, the coarse input variables are regridded to the target grid using a nearest neighbour approach so the input and output resolutions are the same.
The network has \(\sim\)17M parameters. We train for 100 epochs using a mean squared error (MSE) loss function (again when the validation set loss stopped improving; this is a different loss function to that used with the diffusion model, so the number of epochs is not expected to be the same). Whilst performance may be improved by using an alternative loss function \cite<e.g.,>{doury2024precipemulator}, the advantages and disadvantages of particular loss functions are not yet fully clear, and an approach has not yet been found that results in a U-Net producing predictions with realistic small-scale structure, one of our primary aims. So we have chosen to focus on the commonly applied MSE loss here.
We also considered the larger, enhanced U-Net network architecture within by the diffusion model, but we found no improvement when trained with the deterministic MSE loss.

We also compare emulator predictions with output from a statistical method, Bias Correction and Spatial Disaggregation (BCSD) method \cite{woodHydrologicImplicationsDynamical2004, maurerUtilityDailyLargescale2010}. 
BCSD first applies a bias correction step which maps the distribution of the coarse input predictor variable to the distribution of the coarsened target variable. Then a spatial disaggregation step applies a multiplicative day-of-year mean anomaly between the coarsened and full-resolution target variable to the bias-corrected output of the first step.
Here, we used the implementation of \citeA{lopezgomez2025rcmdiffusiondownscaling} applied separately to each time period (to better match the assumption of a stationary climate).
In our implementation, the steps are fitted using day-of-year means and standard deviations computed from the training data set (combining data from all 12 ensemble members and using a window of 3 days either side; see Section~\ref{sec:methods:splits} for how data is split).
Our BCSD implementation was fitted using the cCPM precipitation as the predictor and the high resolution precipitation as the target, which we refer to as ``BCSD\_cCPM''. In testing, it was applied to cCPM precipitation from the test dataset. Note this is a deterministic method that uses coarse precipitation as a predictor,  unlike our emulator, so it receives more direct information about the target precipitation when tested using cCPM inputs.

\subsection{Training}

\subsubsection{Training, validation and test datasets}
\label{sec:methods:splits}

The training dataset was used to fit the model parameter values (including for transforming the variables, described below). Results for different emulator designs (e.g. choices of predictor variables and transformations) were checked on the validation dataset and used to select the final version. All results shown in this manuscript are based on the test dataset, giving an unbiased estimate of the emulator's quality.

For each 20 year time period, the training dataset includes 14 years in total (70\%) and the validation and test datasets include 3 years each (15\%). Across all three time periods and 12 ensemble members, this sums to 504 simulated years of training data and 108 simulated years in each of the validation and test datasets. These datasets were constructed by selecting whole seasons (spring, summer, autumn and winter) at random, with an equal number of each season in each dataset. For each selected season we take data from all 12 CPM ensemble members, so that each subset has equal proportions across the ensemble members. Note the GCM ensemble members are independent of each other except for using the same climate forcings, so in a given season the ensemble samples a broad range of weather states. For example, the test dataset is formed of three randomly chosen springs, summers, autumns and winters from each time period (combining data from all 12 ensemble members for the chosen days). Selecting data by season means that we have minimal data leakage between the training, validation and test datasets due to auto-correlation between days, but also means the data in each subset cover a wide range of years and sample a representative set of climatic conditions, following \citeA{schulz2021dlvsnumweather}.
To ensure the ML models are not benefitting from auto-correlation in the data providing information about the test days, we repeated our evaluation excluding the first and last 15 days of each season from the test set. This ensures at least a 15-day gap between any example seen in training and examples used for evaluation, a duration at which auto-correlation in precipitation becomes very small. We found this made no material difference to the results (not shown). As the aim of this work is to emulate the Met Office's UK CPM so that it may be applied to other GCM simulations similar to those performed in UKCP18, we do not hold out any ensemble members or time periods from the training dataset, enabling learning from all of them.

\subsubsection{Variable transformations}
\label{sec:methods:varxfms}

We apply transformations to the data to improve the emulator's performance. To produce target data for training the emulator's neural network, we first take the square root of the precipitation to produce a less skewed distribution than the raw precipitation, which has many zero values and a long tail. (An alternative approach is a log-based transformation (e.g., \citeNP{Harris2022cgandownscaling}), but this led to unrealistic samples being output by our diffusion model.)
Then we linearly map the square-rooted values so that those in the training dataset lie in the interval [-1, 1]. This is inverted upon sampling, and any negative value increased to 0 before squaring (this clipping affects about 15\% of values, but in practice all are very small: if they were positive values, they would all correspond to precipitation amounts less than 0.15 mm/day, with 90\% of cases under 0.003 mm/day).

Input variables are standardised to have a mean of zero and standard deviation of one (when pooled for all grid boxes, ensemble members and times) in the coarsened CPM training subset.

\subsubsection{Adjustment of GCM inputs}
\label{sec:methods:gcm-adj}

When using predictor variables from the GCM, it was found to be beneficial to adjust their mean and standard deviation to match those of the coarsened CPM variables, based on the training dataset. This is to account for systematic differences between the GCM and CPM variables at coarse resolution.
This was done for each coarse grid box separately. For \(x_{g}\), the value of a variable at grid box \(g\), the adjusted value of the variable, \(\tilde{x}_{g}\), is defined as:

\begin{equation}
    \tilde{x}_{g} = \left( \frac{x_{g} - \bar{x}^{GCM}_{g}}{s^{GCM}_{g}} \right) s^{CPM}_{g} + \bar{x}^{CPM}_{g}
\end{equation}
where \(\bar{x}^{GCM}_{g}\) and \(s^{GCM}_{g}\) are the mean and standard deviation of the values of \(x\) taken from GCM at grid box \(g\) from the training dataset over all three time periods, and \(\bar{x}^{CPM}_{g}\) and \(s^{CPM}_{g}\) are the equivalent for CPM-sourced values of the variable.
This is only applied to emulator inputs and so not applied to precipitation.

The adjusted GCM inputs are then standardised as in Section~\ref{sec:methods:varxfms}, based on the mean and standard deviation of the pooled adjusted GCM training dataset. \citeA{kendon2025MLemulatorsperspective} discuss further the effects of this adjustment.

\subsection{Evaluation diagnostics}

We test two inference modes of the CPMGEM emulator:
\begin{itemize}
\item CPMGEM\_cCPM, which uses coarsened CPM variables as inputs, as used in training.
\item CPMGEM\_GCM, which uses GCM variables as inputs, corresponding to how CPMGEM would be used in practice.
\end{itemize}
In the coarsened cCPM setting, we make comparisons with the deterministic U-Net emulator, denoted by ``U-Net\_cCPM'', and with bilinear interpolation of the coarsened precipitation (denoted ``cCPM Bilinear'').
For the stochastic CPMGEM emulator, we generated 6 samples for each day in both inference modes.
To avoid confusion with the ensemble of CPM simulation runs, we will use the terms ``samples'' or ``sample set'' to refer to the range of outputs from the emulator, and ``ensemble members'' to refer to different simulations of the CPM.

Below, we describe evaluation diagnostics that require more detailed explanation than can be given in the results section.

\subsubsection{Radially Averaged Power Spectral Density}
\label{sec:methods:rapsd}

We use RAPSD to compare the complexity of structures (the variability over different spatial scales) produced by different approaches \cite{harris2001rapsd, sinclair2005rapsd}.
To compute the power spectrum of a precipitation field, we compute the 2D Fourier transform of the field and convert to power by multiplying by complex conjugate. 
We then radially average over nested annuli centred on the origin in the 2D spatial frequency space to create a 1D vector of power at different frequency bands. Finally to compare difference approaches, we compute the mean for each frequency band over all samples.

\subsubsection{Spread-Error plot}
\label{sec:methods:spread-error}


To evaluate the size of the stochastic component of the diffusion model emulator, we follow the approach of \citeA{leutbecher2008ensembleforecasting} and produce a spread-error plot, also discussed by \citeA{haynes2023uncertainty}. The idea is that if the CPM simulation output is statistically similar to samples from the emulator for a given GCM state, as in the ideal case, then the relationship with the mean of the samples should be identical for CPM and emulator outputs. Then the root mean square difference between the emulator ensemble mean and the CPM output (the root mean square error, RMSE, of the ensemble mean) and the root mean square difference between the emulator ensemble mean and the individual emulator samples (the root mean square spread, RMSS) should be equal. To account for the finite ensemble size of the emulator output, RMSS has a correction factor of \(\frac{n+1}{n-1}\) applied, where \(n\) is the number of emulator samples generated for each simulator example (in our case, \(n=6\)).

This is also the case if samples are binned into intervals of the RMSS. The spread-error plot compares the RMSE and RMSS within these bins, with values calculated for every grid box and example in the test dataset. We compute the squared spread of emulator samples and squared error for the mean of the samples for each grid box in every example of the test subset. We group these pairs of spread and error into into bins containing an equal number of values according to this spread and plot the RMSS and the corresponding RMSE for the predictions in each bin.

\subsubsection{Bootstrapping domain mean change confidence intervals}
\label{sec:methods:bootstrapping}

We use bootstrapping \cite{efron1982bootstrap} to estimate the sampling variability of the change in seasonal domain mean precipitation going from Historic to Future time periods for our emulator in Section~\ref{sec:results:ccs}. In particular, to estimate 95\% confidence intervals of the values. For a given season and time period, we sample with replacement from the set of 36 seasonal domain means across the ensemble members and time period (12 ensemble member by 3 years) to produce paired estimates of both the CPM domain mean and emulator domain mean (so sampling the same set of GCM states for each). By repeating this resampling 100,000 times for both time periods we estimate the 0.025 and 0.975 quantiles of the change in the mean from Historic to Future time periods for the CPM and emulator.

\section{Results}\label{sec:results}

Below, we present an evaluation of our emulator's performance at reproducing the UK CPM's properties.
We first test the emulator using CPM data coarsened to the 60km GCM grid as input (CPMGEM\_cCPM), to correspond to the inputs used in training.
Our next set of tests use predictors from the GCM (CPMGEM\_GCM) to evaluate how well the emulator transfers to the challenge of predicting high-resolution precipitation when CPM data is not available, including capturing the climate change response. 
Finally, we evaluate the performance when there is a limited amount of training data.

Some further results are also presented by \citeA{kendon2025MLemulatorsperspective}, relating to their discussion of methods for emulating regional climate models more generally.

\subsection{Evaluation using coarsened CPM predictors}
\label{sec:results:cCPM_predictors}

Here, we evaluate our emulator using coarsened CPM data as inputs and compare its predictions with those from the deterministic U-Net using the same inputs (``U-Net\_cCPM'') and the statistical method BCSD applied to coarsened CPM precipitation (``BCSD\_cCPM''). See Section~\ref{sec:methods:comparison}.

\subsubsection{Evaluation on full domain}

\begin{figure}[htbp]
    \centering
    \includegraphics[scale=1.0]{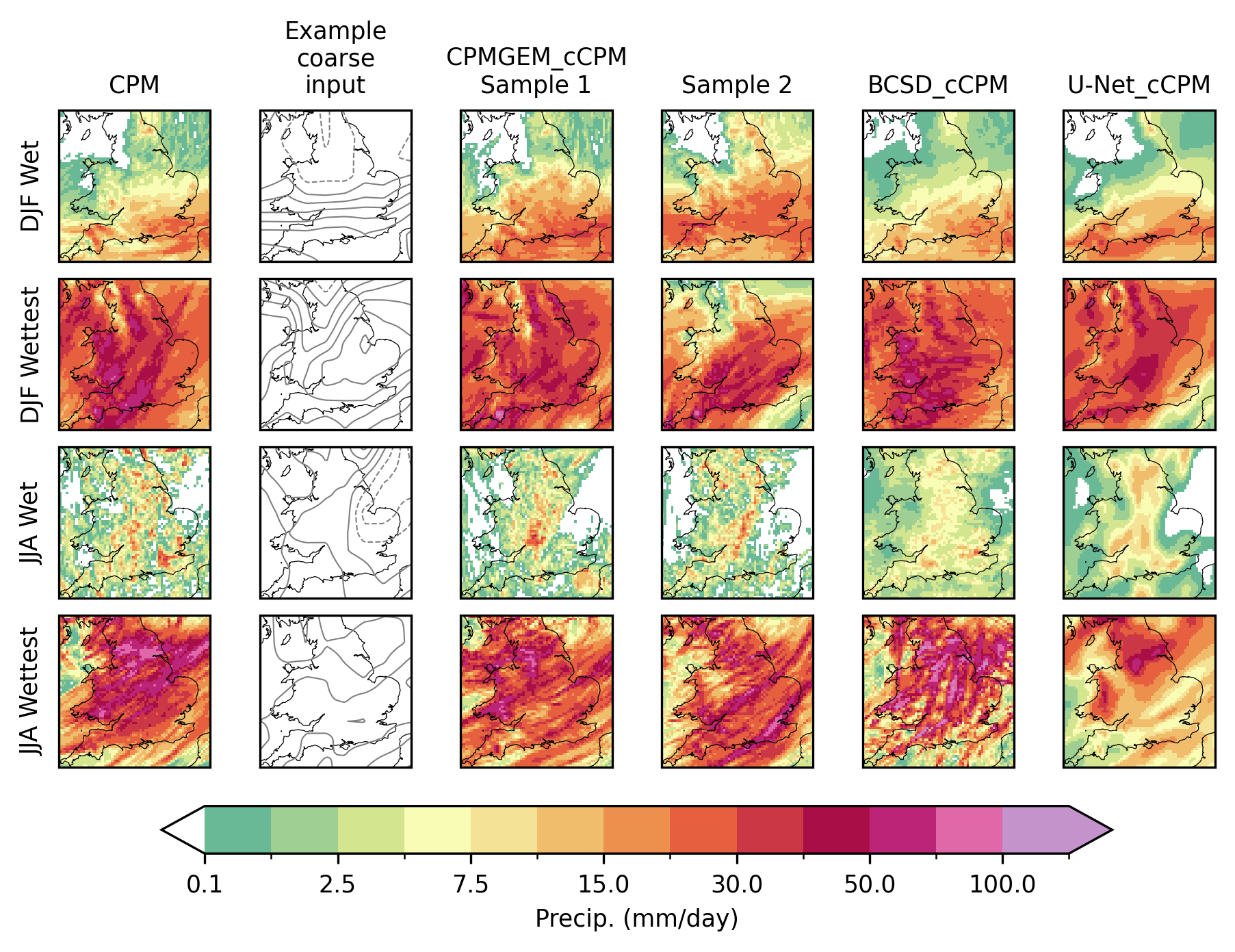}
    \caption{\textbf{Examples of predictions of daily-mean precipitation.} The first row shows results for a wet day in winter (December--February, DJF; the 80th percentile of the domain-mean). The second row shows the wettest winter day in the 108 year test dataset. The third and fourth rows are similar but for summer (June--August, JJA). The first column is the precipitation from the convection-permitting model (CPM). The second column is an example coarse resolution input field, the 850hPa vorticity. Contours, in grey, are drawn in steps of \(2\times10^{-5}\textrm{s}^{-1}\) between \(-10^{-4}\textrm{s}^{-1}\) and \(+10^{-4}\textrm{s}^{-1}\) with dashed lines for negative values and solid lines for positive. Columns 3 and 4 are samples chosen at random from the emulator using coarsened CPM atmospheric variables as predictors (CPMGEM\_cCPM). Column 5 is the output of applying BCSD to the coarsened CPM precipitation. Column 6 is the prediction by U-Net. Note that the highly stochastic nature of precipitation downscaling means samples from the diffusion model are not expected to match the CPM precipitation in full detail, but to represent the distribution of plausible precipitation fields for the given low resolution predictors, where the CPM simulation output is a single example.}
    \label{fig:CPM-samples}
\end{figure}

We show example predictions of daily mean precipitation in Figure~\ref{fig:CPM-samples}. This shows examples on days with substantial precipitation, selected according to the domain-mean value in the CPM. The first row is a wet day in winter, December--February (``DJF Wet'', the 80th percentile across domain means for all winter examples in the test subset). The second row is the wettest winter day (``DJF Wettest'', the winter example from the test set with the maximum domain-mean). The domain-mean precipitation in this example is only expected about once every 108 years, making this a test of the models' skill for extreme events of intensities relevant to planning climate resilience \cite{weaver2017climaterisk,watson2022mlextremes}. The third and fourth rows are similar but for summer, June--August (``JJA Wet'' and ``JJA Wettest''). Here, ``JJA Wet'' is an example chosen to have precipitation close to the 80th percentile and also a large region of convective showers, to illustrate the performance for this weather type. The other three examples show precipitation patterns characteristic of fronts.

The first column is the target CPM precipitation. The second column is vorticity at 850hPa derived from coarsened wind fields from the CPM. It is an example of the coarse inputs used by the emulator and provides a visualization of one of the features it sees.

The third and fourth columns show two samples from the diffusion model (randomly chosen from the six). These demonstrate the emulator's ability to produce daily-mean precipitation samples with realistic fine detail. For DJF Wet it recreates the high intensity structures in the southern half of the domain whilst also having realistic lower intensity showers further north. For JJA Wet, it predicts small clusters of high intensity, similar to those seen in the CPM output. 
The predictions for the extreme examples in the second and fourth rows also have similarly realistic spatial structure to the CPM.
Note, there is a substantial stochastic component of precipitation that is unpredictable given just the coarse-scale state and selecting the most extreme day in CPM output will tend to select a day when this component is positive and relatively large.
Therefore it would not necessarily be expected that typical samples from the emulator for the same day would reach as intense values as in the CPM. (Evaluation of the overall intensity distribution across all days, including whether the emulator accurately reproduces the frequency of extreme values, is presented later in this section.)

The fifth column shows the results of applying the BCSD method to the coarsened precipitation. They show evidence of the fine resolution structures that BCSD is able to add. However, the structures can be different to those of the CPM and CPMGEM.
For example, in the DJF Wettest and JJA Wettest cases (rows 2 and 4) it is missing the structures oriented southwest-northeast.
In the JJA Wet example (row 3), BCSD does not re-create the small high-intensity convective events. Further examples of days with high summertime convective activity, illustrating the more realistic samples produced by CPMGEM in these situations compared to BCSD, can be found in Figure~S8 in Supporting Information.

The rightmost column shows predictions from the deterministic U-Net\_cCPM. While it also mostly correctly represents the large-scale patterns (like the fronts in rows 2 and 4), it does not recreate the fine level detail of the CPM simulations. This is similar to other deterministic downscaling models \cite<e.g.,>{doury2024precipemulator, sha2020unetdsprecip}. This is very noticeable in the JJA Wet case, for which it predicts a much smoother rainfall field, with lower peak intensity than seen in either the CPM or CPMGEM\_cCPM.

The differences between the two CPMGEM\_cCPM samples indicate the size of the stochastic component as learnt by the emulator. Whilst the large-scale features are fairly similar, as expected since the samples are conditioned on the same coarse-scale input, the locations of the heaviest precipitation amounts generally differ, which may result in very different local effects simulated by an impacts model.
This indicates that the stochastic component is generally large and it is important that it is represented. The CPMGEM\_cCPM samples also illustrate how the emulator could generate different realisations of a given day's precipitation, perhaps to explore the impact if the heaviest downpours occurred in different locations, for example in areas with critical infrastructure. (This could be an interesting application of a similar emulator trained on observations and applied to historical weather events, although we do not explore this here.)

\begin{figure}[htbp]
    \centering
    \includegraphics[scale=1.0]{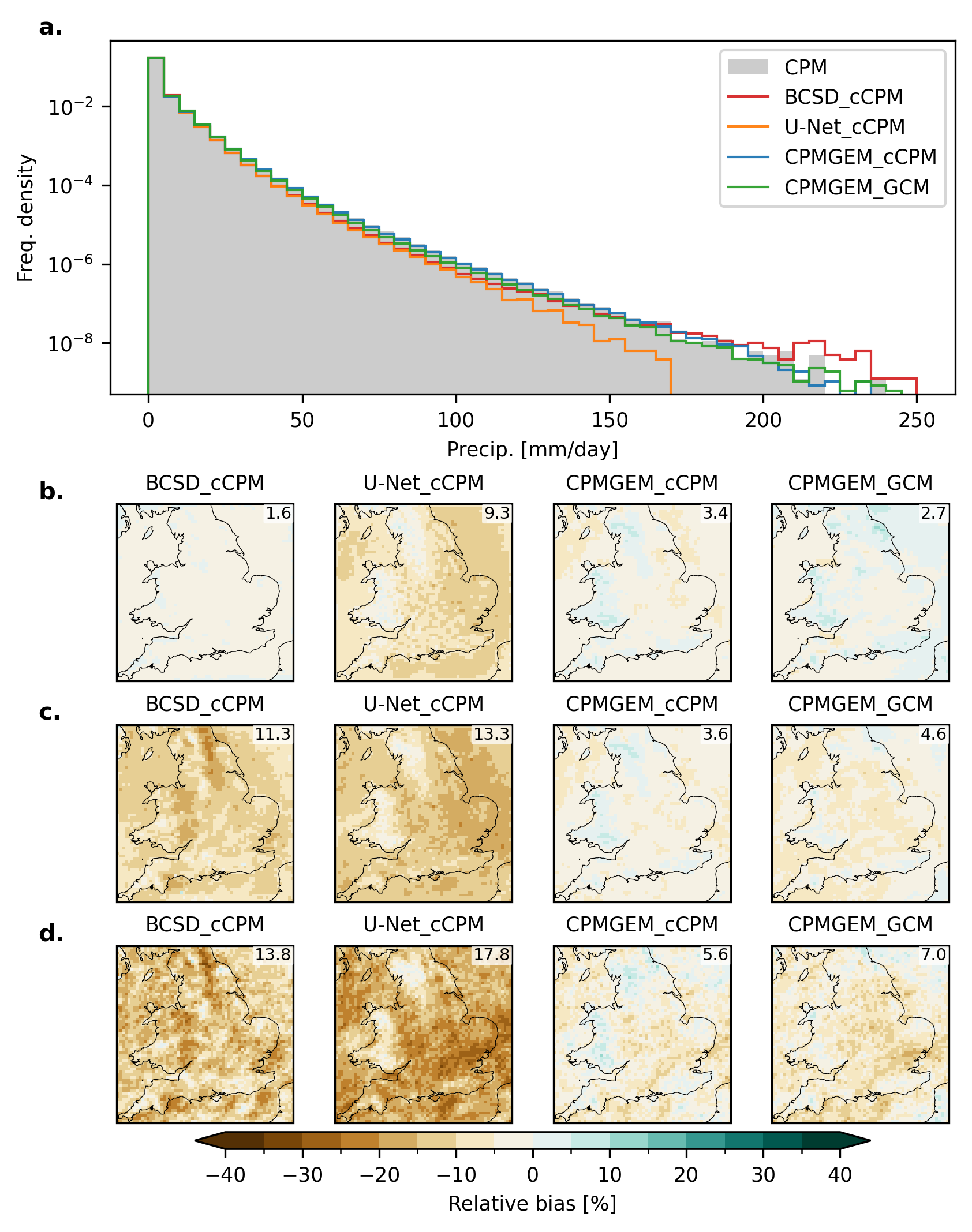}
    \caption{\textbf{Statistical properties of predictions.} (a) Histograms of precipitation values on the 8.8km grid. The grey shaded area is the frequency density of the target CPM precipitation. The lines show frequency densities from the diffusion model emulator acting on coarsened CPM and GCM inputs respectively (``CPMGEM\_cCPM'' (blue) and ``CPMGEM\_GCM'' (green)), BCSD\_cCPM (red) and U-Net\_cCPM (orange). Note the vertical axis is logarithmic. (b) Relative mean bias as a percentage of the CPM mean for each model. Root mean squared bias (\%) over the domain is shown in top right of each spatial plot. (c) Same as (b) but for standard deviation bias. (d) Same as (b) but for bias of 99.9th percentile (mean of the 99.9th percentile over six sample runs for CPMGEM). All plots combine data from all three time periods.}
    \label{fig:CPM-GCM-dist}
\end{figure}

As well as producing realistic samples, it is important that the frequency distribution of precipitation produced by the emulator matches that of the CPM. Figure~\ref{fig:CPM-GCM-dist}(a) shows the frequency distribution of precipitation across all 8.8km grid boxes for the CPM, the downscaling models and the results of applying BCSD to cCPM precipitation. The distribution of CPMGEM\_cCPM matches that of the CPM distribution very closely up to intensities of over 200 mm/day, the extreme tail of the test set (there are only two CPM simulated values beyond the maximum shown). Note the log scale of the density axis.

The distribution of BCSD\_cCPM also matches that of the CPM distribution up to high intensities, showing that BCSD reproduces the distribution of the CPM given coarse precipitation, including being able map the far lower extremes of coarsened precipitation to intensities found in the extremes of the CPM precipitation distribution. This is an expected strength of a method using quantile mapping. However, there is evidence that it is under-estimating the frequency of events in the 50 to 100 mm/day range (where it is closer to the U-Net\_cCPM histogram).

U-Net\_cCPM has too low frequencies of precipitation more intense than $\sim$30 mm/day.
Potentially this could be improved by using a loss function designed to increase prediction of extreme values \cite<e.g.,>{doury2024precipemulator}. However, it would also be expected that the stochastic aspect of CPMGEM\_cCPM contributes to generating a more accurate frequency of extreme values.

Importantly, the CPMGEM\_cCPM emulator also predicts a similar frequency of wet days as the CPM. The frequency of days with more than 0.1mm of precipitation in individual 8.8km grid boxes is 52.6\% for CPMGEM\_cCPM versus 53.0\% for the CPM (66.4\% versus 67.4\% in winter and 36.9\% for both cases in summer).

The precipitation distribution varies across the UK, due to effects such as elevation. Figure~\ref{fig:CPM-GCM-dist}(b), (c) and (d) show spatial maps of the biases in the mean, standard deviation, and 99.9th percentile.
Spatial maps of bias in RX1Day are included in Supporting Information (Figure~S7) and are similar to plots for 99.9th percentile.
The the root mean squares of these biases over space are shown in the top-left of each panel in order to quantify their typical sizes.
CPMGEM\_cCPM displays small biases over the whole domain.
The mean biases of CPMGEM\_cCPM relative to the CPM are much smaller than the mean bias of the CPM relative to observations (see Section~\ref{sec:method:cpmpr}).
The emulator has learnt the differences in precipitation properties between nearby locations, below the scale of the coarse-resolution inputs.

BCSD\_cCPM has very small biases in the mean, as expected given it directly attempts to correct the mean in both its bias correction and spatial disaggregation steps. However, it has consistent underestimation of the standard deviation and 99.9th percentile over the whole domain.
U-Net\_cCPM has a consistent dry mean bias everywhere and a too low standard deviation (similar to findings of \citeA{doury2024precipemulator} when using the same MSE loss), again indicating the likely role of the stochastic part of the precipitation variability in the diffusion model emulator.
U-Net\_cCPM also consistently underestimates the 99.9th percentile, indicating that it underestimates the more extreme intensities.

\begin{figure}[htbp]
    \centering
    \includegraphics[scale=1.0]{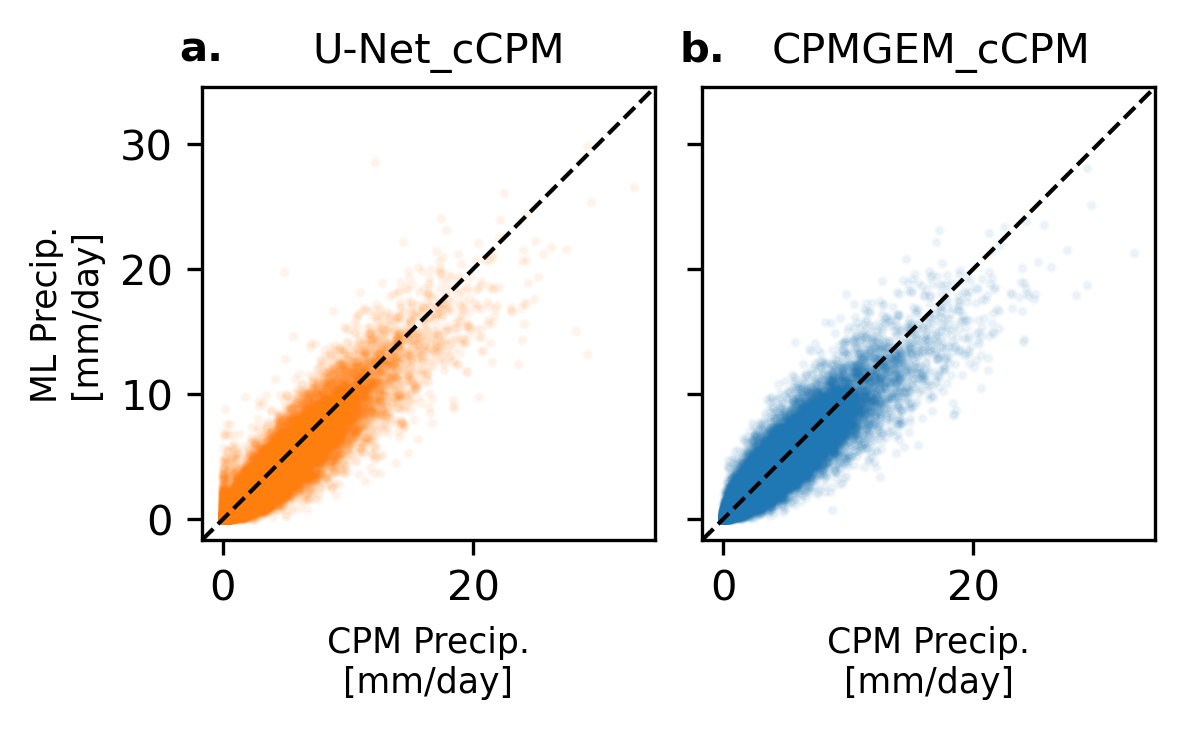}
    \caption{\textbf{Scatter plot of daily domain-mean precipitation for predictions versus target CPM values}. Predictions from a) U-Net\_cCPM and b) CPMGEM\_cCPM. All plots combine data from all three time periods.}
    \label{fig:CPM-scatter}
\end{figure}

As well as matching the climatological distribution of CPM precipitation, the emulator must also learn the dependence of the precipitation distribution on the coarse input variables. We examine this using a scatter plot of domain-mean precipitation from CPMGEM\_cCPM against that from the CPM target (Figure~\ref{fig:CPM-scatter}(b)). We use the domain-mean because this reduces the variability from the stochastic component sufficiently to be able to clearly see that the predicted and target precipitation are well-correlated for CPMGEM\_cCPM (0.94 Pearson correlation coefficient). Note that since precipitation likely has a substantial component that is effectively stochastic and unpredictable given only coarse-scale variables, a perfect correlation could not be expected. 
Also note that if the emulator had not learnt a skilful relationship between the coarse inputs and precipitation and was acting like an unconditional generative model, the correlation with CPM output would not be expected to be large.
The degree of correlation is similar for U-Net\_cCPM (Figure~\ref{fig:CPM-scatter}(a); 0.94 Pearson correlation coefficient). 
This indicates that the deterministic component of CPMGEM\_cCPM (identified conceptually with the mean over many samples) is similarly skilful to predictions from U-Net. Both appear to learn a skilful relationship between the inputs and total precipitation.
The stochasticity in the CPMGEM\_cCPM output likely contributes to its predictions having slightly more spread.
(BCSD\_cCPM has an almost perfect correlation with the CPM for domain means, which is unsurprising as this method is given coarsened precipitation as an input, and  domain-mean precipitation is just the sum of this input over the grid.)

Figure~S1 also shows a scatter plot of predicted versus CPM domain maxima to explore the correlation for heavier precipitation intensities. In this case, the relationship between CPM and CPMGEM\_cCPM appears centred about the ideal $y=x$ line, with correlation 0.81, similar to the case for domain-mean precipitation. But for U-Net\_cCPM there is a clearer bias towards predictions being below the $y=x$ line, indicating frequent underestimation of the maximum by the U-Net emulator.
This corroborates findings of underestimation by U-Net\_cCPM of the highest precipitation intensities shown above.

To indicate relationships on a finer spatial scale, Figures~S3-S6 in Supporting Information show scatter plots for two individual grid boxes, at the locations of London and Lancaster respectively, which are cities in southeast and northwest England. Results are shown for winter and summer to also indicate differences between seasons. 
As with Figure~\ref{fig:CPM-scatter} for domain means, points for CPMGEM\_cCPM appear centred around the ideal $y=x$ line but with a spread, as would be expected for a stochastic model. At both locations, plots show more spread in predictions in summer than winter, likely due to the increased proportion of less predictable convective rainfall in this season. However, the emulator appears to have learnt the relationship between coarse atmospheric conditions and precipitation at both individual grid points with comparable skill. BCSD has a closer correlation between predictions and CPM with less spread than CPMGEM, which may be expected as it is given coarse-scale precipitation as a predictor and is a deterministic method, but the considerable scatter in the relationships is suggestive of there being a substantial stochastic component of variability in the CPM precipitation.

\begin{figure}[htbp]
    \centering
    \includegraphics[scale=1.0]{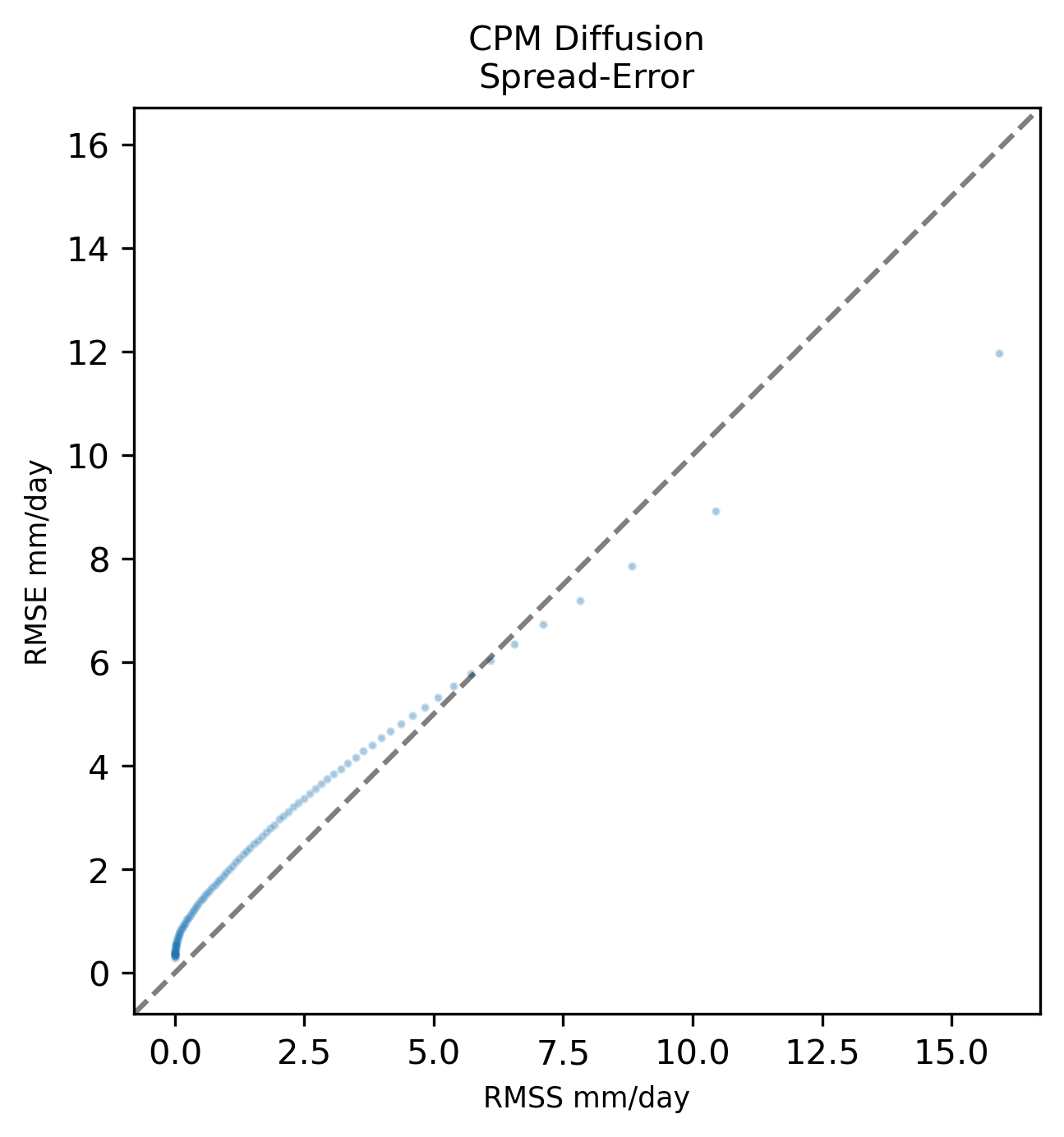}
    \caption{\textbf{Spread-error plot of CPMGEM\_cCPM}. To indicate the calibration of the stochastic component based on spread and error at individual grid boxes (see Section~\ref{sec:methods:spread-error} for details). This is calculated from data from all three time periods. The dashed line shows the ideal case where spread is equal to error. Points above the line suggest under-dispersion in the emulator samples and below over-dispersion.}
    \label{fig:CPM-spread}
\end{figure}

It is also important to evaluate whether the size of the stochastic component of the diffusion model emulator is appropriate. Following the approach of \citeA{leutbecher2008ensembleforecasting}, we produce a spread-error plot (see Section~\ref{sec:methods:spread-error} for more details). Figure~\ref{fig:CPM-spread} shows the root mean square error (RMSE) versus the root mean square spread (RMSS) for individual grid boxes from samples binned according to the spread of the prediction samples for CPMGEM\_cCPM. Ideally, these should be equal for all days with prediction spread falling within a given bin, taking into account adjustments for having a finite ensemble size.
The figure shows evidence that this relationship is being approximately followed, suggesting that CPMGEM\_cCPM is reasonably well-calibrated and can differentiate between situations that are relatively predictable (low spread and error) and less predictable (high spread and error). The spread-error ratio over the whole test set is 0.99.
For the cases up to a spread of 5 mm/day, the emulator is slightly under-dispersive (spread-error ratio of 0.76), as can be seen for the points with lower spread sitting just above the dashed line. There is evidence of larger over-dispersion for the less predictable cases (spread-error ratio of 1.14 for cases with spread over 5 mm/day).
We estimated sampling variability in both the spread and the error by recomputing them from bootstrapped resamplings of the test dataset and found that this is very small (not shown).

\begin{figure}[htbp]
    \centering
    \includegraphics[width=\textwidth]{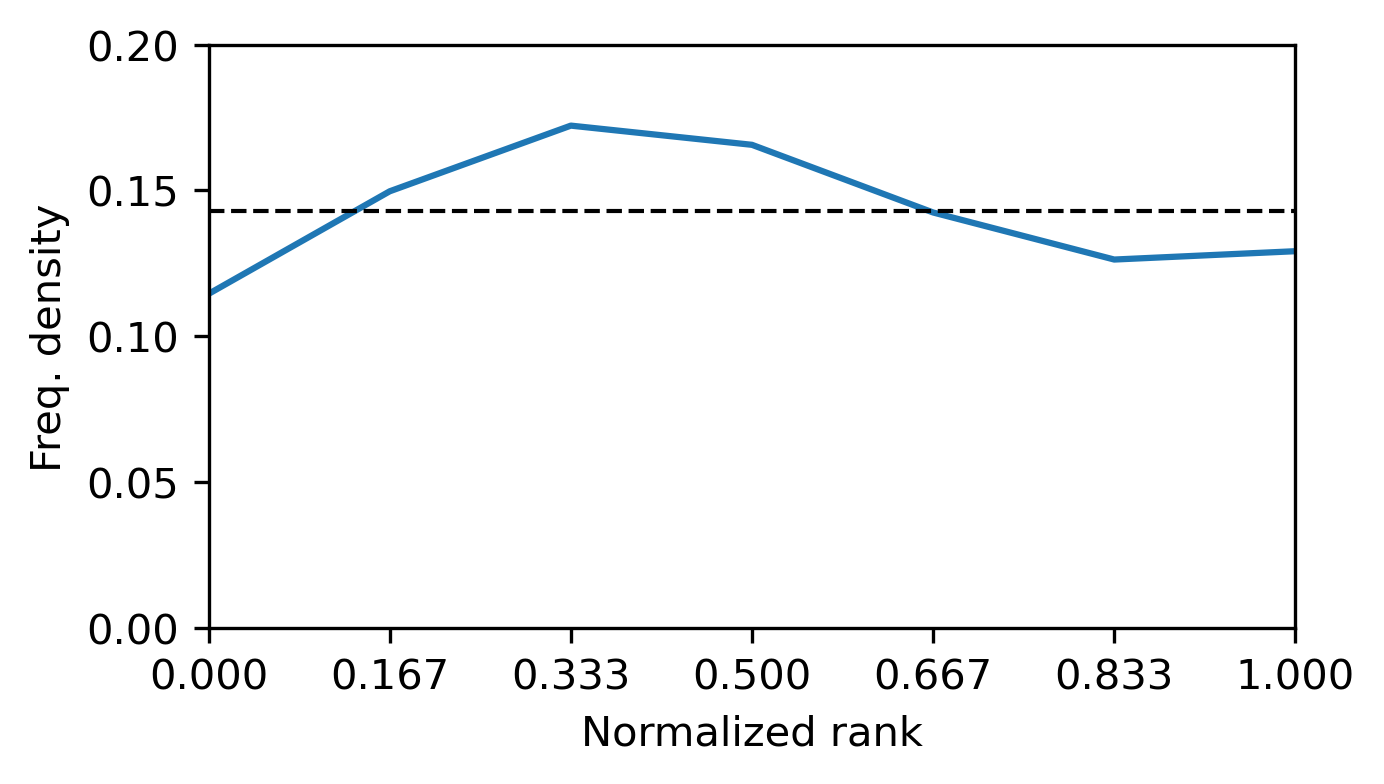}
    \caption{\textbf{Rank histogram of CPMGEM\_cCPM}. The frequency for each day and grid box of the proportion of samples below the CPM precipitation at each grid box for CPMGEM with coarsened CPM inputs. Dashed line shows the proportion of 1/7 expected for a perfectly calibrated model generating six predictions.}
    \label{fig:rank-histograms}
\end{figure}

To further assess the spread of predictions, Figure~\ref{fig:rank-histograms} shows a rank histogram to evaluate the reliability of CPMGEM\_cCPM predictions \cite{hamill2020rankhistograms}.
For each 8.8km grid box and for each example in the test dataset, we computed the normalized rank: the proportion of samples where the predicted precipitation intensity was less than the CPM intensity ($\frac{N_r}{N_s}$, where $N_r$ is the number of samples below the CPM value and $N_s=6$ is the number CPMGEM samples). For a perfectly calibrated model, these ranks would be uniformly distributed over the seven possible outcomes. This is approximately the case for CPMGEM\_cCPM, showing further evidence of CPMGEM producing a well-calibrated spread of predictions.

\begin{figure}[htbp]
    \centering
    \includegraphics[scale=1.0]{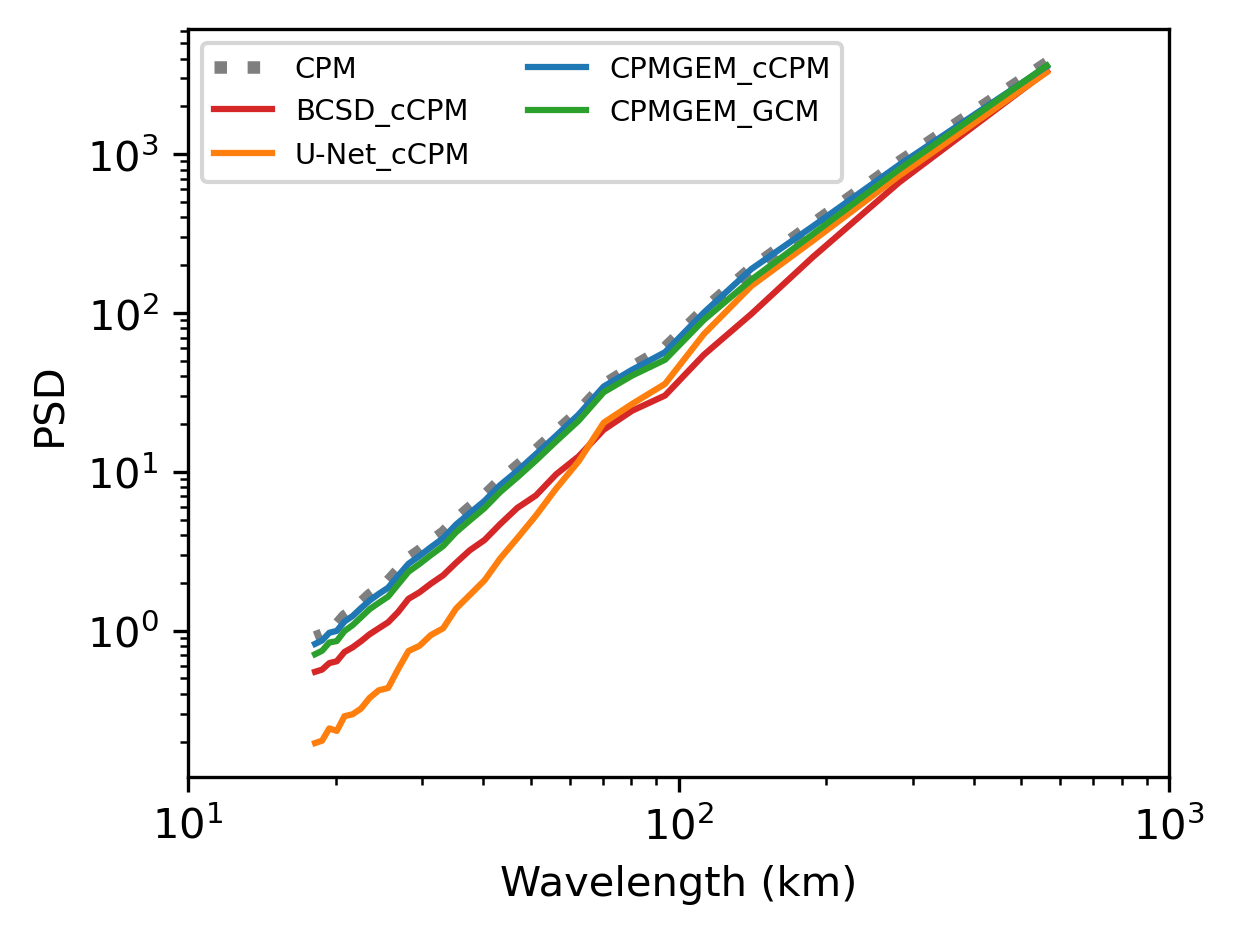}
    \caption{\textbf{Radially averaged spatial power spectral density (RAPSD)}. Shows the target CPM precipitation (grey dashes), the emulator samples, CPMGEM\_cCPM (blue line) and CPMGEM\_GCM (green line), U-Net\_cCPM (orange) and cCPM Bilinear (dark grey).}
    \label{fig:CPM-GCM-structure}
\end{figure}

To quantitatively evaluate the realism of the spatial structures in the emulator output, Figure~\ref{fig:CPM-GCM-structure} shows the radially averaged power spectral density (RAPSD; see Section~\ref{sec:methods:rapsd} for more detail). This quantifies the amount of variance across the range of spatial scales in the data. The RAPSD of the CPM and CPMGEM\_cCPM match closely for the full range of spatial scales, showing that CPMGEM\_cCPM produces structures which have similar variability at different spatial scales to the CPM simulation. The close agreement at small spatial scales reflects how the CPMGEM\_cCPM samples include a realistic degree of fine-scale structure, as seen in the samples in Figure~\ref{fig:CPM-samples}. The curve for U-Net generally displays too low variability, particularly at the smallest scales, corresponding to these predictions having unrealistically smooth small-scale structure
Unlike U-Net\_cCPM, the RAPSD of the BCSD\_cCPM samples does not have such steep a drop-off in variability at the finest scales, though its error is larger than for CPMGEM\_cCPM. At the shortest wavelength, RAPSD is underestimated by 39\% compared to CPM for BCSD\_cCPM and by 9\% for CPMGEM\_CPM.
At a wavelength of 62.6km, approximately the size of the coarse GCM grid boxes, the underestimation of the CPM's RASPD by BCSD\_cCPM is 47\%, while CPMGEM has a modest underestimation of 4\% and 12\% with cCPM and GCM inputs respectively. 
BCSD samples also have a consistent under-estimation of the variability compared to the CPM at scales up to several hundred kilometres.
The underestimation at small length scales may reflect that BCSD is unable to produce some structures like fine-scale convective showers, as seen in the samples shown in Figure~\ref{fig:CPM-samples}.
Comparing the results for the diffusion model and U-Net indicates that including a stochastic component in the emulator samples has helped to realistically represent small-scale variability.

\subsubsection{Regions, seasons and different precipitation types}
\label{sec:results:exp1:regions}

\begin{figure}[htbp]
    \centering
    \includegraphics[scale=1.0]{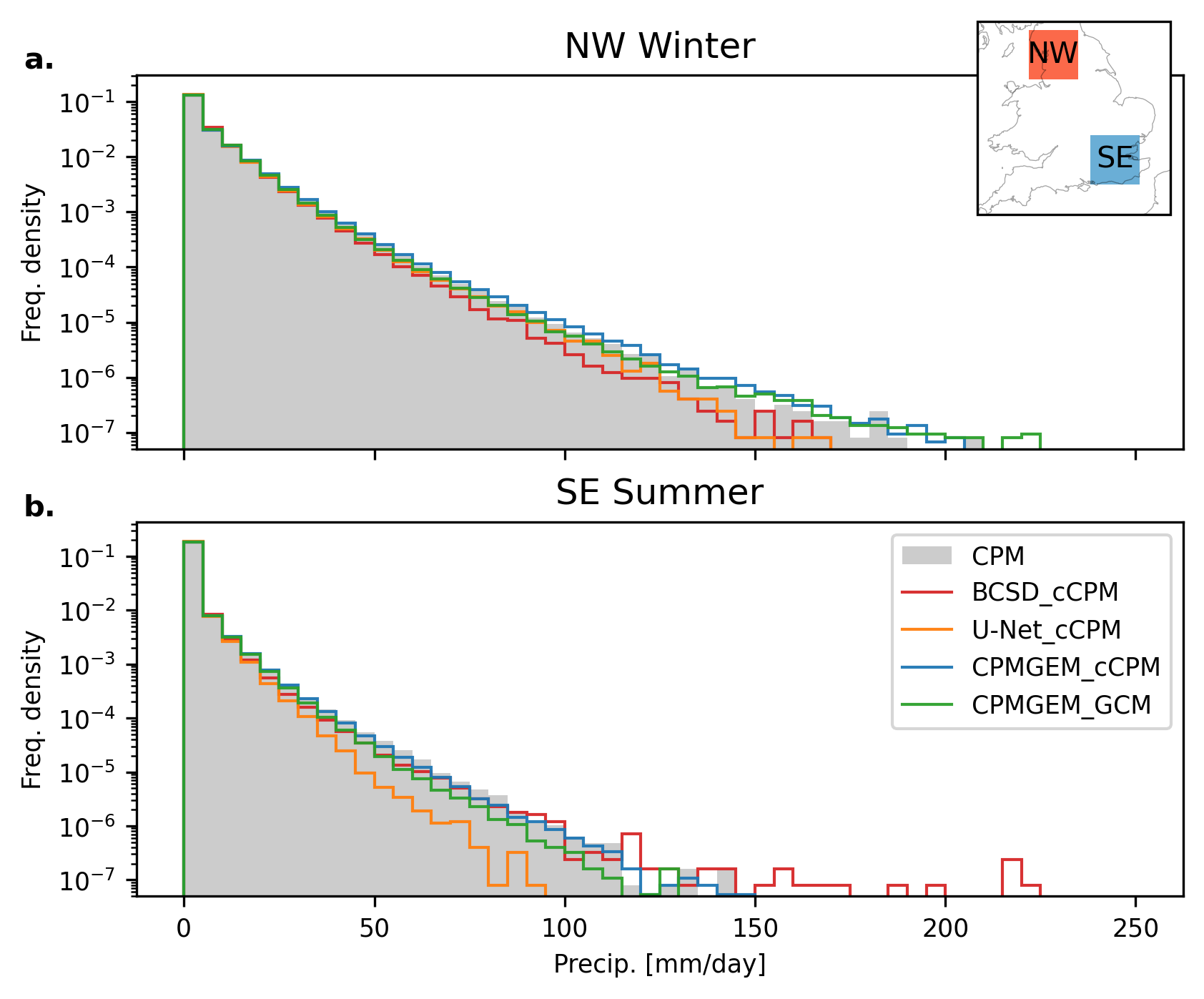}
    \caption{\textbf{Frequency distributions in different regions and seasons.} (a) Winter in northwest England, where precipitation is predominantly frontal and orographic; (b) summer in southeast England, which includes a substantial convective precipitation component. Plotted as in Figure~\ref{fig:CPM-GCM-dist}(a). The inset at the top right shows the regions in the context of the full spatial domain.}
    \label{fig:CPM-GCM-subregions}
\end{figure}

Different processes drive UK precipitation to different extents depending on the region and time of year. Therefore we evaluate whether the emulator reproduces these differing regional and seasonal characteristics of precipitation. Figure~\ref{fig:CPM-GCM-subregions} shows frequency distributions for precipitation in northwest England in winter and in the southeast in summer across individual points on the high-resolution grid. In the former case, precipitation is predominantly frontal and orographic, while the latter case has a larger convective component. Both regions are \(16\times16\) 8.8km grid boxes in size.

The frequency distribution of CPMGEM\_cCPM samples closely follows that of CPM precipitation for both cases.
The same is not true of samples from U-Net\_cCPM, which matches the northwest winter distribution well until intensities of about 125 mm/day, but substantially under-predicts the frequency of more extreme intensities, as for the full frequency distribution shown in Figure~\ref{fig:CPM-GCM-dist}(a). The agreement between the target CPM and U-Net\_cCPM is closer than for the full distribution, suggesting that the deterministic component of precipitation in this region and season accounts for a larger share of variability, perhaps due to the strong orographic component and importance of large-scale frontal systems, with the stochastic component being more apparent in the most intense extremes. 
In the southeast in summer, U-Net\_cCPM similarly underestimates the frequency of heavy precipitation. This indicates an important role for the stochastic component of precipitation here, which may be expected given the larger influence of small-scale convective rainfall systems than in the northwest winter case, as in the example in the third row in Figure~\ref{fig:CPM-samples}.
BCSD\_cCPM shows evidence of underestimating the frequency of precipitation above about 25mm/day for the northwest in winter, whilst producing excessively large intensities in the southeast in summer. This suggests it underestimates the intensities from large-scale frontal-systems and gives further information about its errors in reproducing summertime precipitation.

To further explore reproduction of extreme events, we estimated precipitation intensities at return levels up to 108 years at an individual grid box in each subregion. The grid boxes corresponded to the previously selected city locations (London for SE and Lancaster for NW), where extreme rainfall may have relatively large impacts. We found that the estimated CPM return periods are generally consistent with those of CPMGEM\_cCPM (see Supporting Information section S1, Figure~S2 and Table~S1).
The CPM intensities are within the spread of those estimated from each of the six CPMGEM\_cCPM samples for nearly all return periods, with there just being a slight underestimation for return periods close to 1 year in Lancaster in DJF.
In London, differences in percentage terms for CPMGEM\_cCPM from the CPM at the 1- and 9.8-year return times lie primarily within a range of -10\% to 5\% across all six sample sets and the same is true for Lancaster.
For 108-year return times, the range of the differences is larger at both locations (between about -27\% and 23\% for both locations and seasons, with two outliers for London in each season).
For the comparison approaches, BCSD\_cCPM shows a fit of comparable quality to typical CPMGEM samples, in line with the previous result that it reproduces the overall frequency distribution of precipitation well. U-Net\_cCPM shows systematic underestimation of the extreme return levels, again in line with previous result that it systematically underestimates the frequency of the highest rainfall events.

\subsection{Transfer to using GCM inputs}
\label{sec:gcm-xfer}

\begin{figure}[htbp]
    \centering
    \includegraphics[scale=1.0]{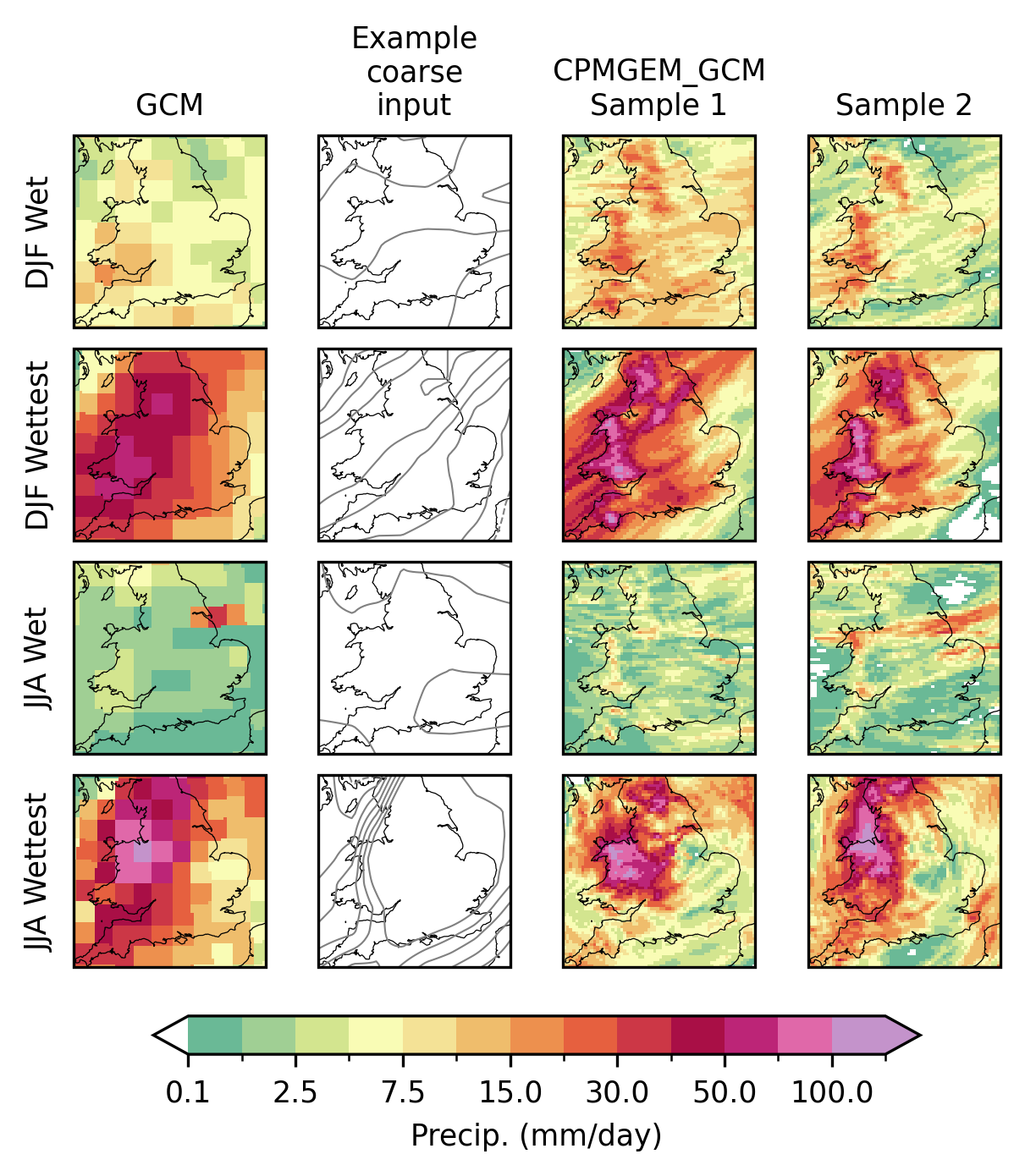}
    \caption{\textbf{Samples based on GCM input variables.} Each row shows results for one example day, chosen in a similar way to those in Figure~\ref{fig:CPM-samples}, based on domain-mean GCM precipitation. The first column shows the GCM precipitation, the second an example coarse input field (vorticity at 850hPa, contours in grey every \(2\times10^{-5}\textrm{s}^{-1}\) between \(-10^{-4}\textrm{s}^{-1}\) and \(+10^{-4}\textrm{s}^{-1}\), with dashed lines for negative values) and columns 3 and 4 are two samples from the emulator using GCM inputs (CPMGEM\_GCM). Note the emulator output is not constrained to match the coarse-resolution structures of the GCM precipitation, and may predict more realistic features at large as well as small scales.}
    \label{fig:GCM-samples}
\end{figure}

Next we examine how well our emulator transfers to the problem of downscaling GCM data, rather than the coarsened CPM data like that used for training. Having shown the greater realism of CPMGEM outputs compared to U-Net when using coarsened CPM variable inputs above, we focus here just on the performance of the CPMGEM\_GCM emulator.

Figure~\ref{fig:GCM-samples} shows example predictions from the CPMGEM\_GCM model. The GCM-sourced predictors have been adjusted to match the overall mean and variance of cCPM predictors at each location, as described in Section~\ref{sec:methods:gcm-adj}. The rows each correspond to a different day and are chosen in a similar way as in Figure~\ref{fig:CPM-samples}, but based on GCM precipitation, with ``DJF Wet'' and ``JJA Wet'' days corresponding to the 80th percentile of the domain-mean, and the other two rows to the wettest winter or summer day in the test subset respectively. The first two columns show data from the GCM for context: the low resolution precipitation and one input to the emulator, the coarsened vorticity at 850hPa. The third and fourth columns show two samples from CPMGEM\_GCM for the given predictor variables. It can be seen that they contain finer spatial detail. Also they do not in general have exactly the same coarse-scale spatial structure as the GCM simulation, as they have learnt a separate model of precipitation based on the CPM target, and so do more than just add spatial detail to the precipitation. For example, the CPMGEM\_GCM samples exhibit heavier precipitation in the western UK in the ``DJF Wet'' case, which may be associated with the GCM not resolving orography well there. And they show lighter precipitation in southwest England in the ``JJA Wettest'' case.
Note that it is not useful to compare with CPM output on these days as the position of precipitation features can be different between the CPM and GCM (see Section~\ref{sec:methods:coarse}).

Figure~\ref{fig:CPM-GCM-dist}(a) shows that when using GCM-derived inputs the emulator (CPMGEM\_GCM) continues to produce samples whose distribution of intensities is still very similar to the CPM precipitation target on the 8.8km grid. It also predicts a similar proportion of wet days (greater than 0.1 mm/day) to the CPM across all grid boxes: 54.7\% annually from the emulator compared to 53.0\% from the CPM. Seasonal differences in this proportion are also captured by the emulator: 68.1\% compared to 67.4\% in winter and 38.4\% compared to 36.9\% in summer. The biases in the mean, standard deviation and 99.9th percentile of precipitation of CPMGEM\_GCM are small relative to the CPM target throughout the spatial domain (Figure~\ref{fig:CPM-GCM-dist}(b), (c) and (d)) and the spatial power spectral density for CPMGEM\_GCM is very similar to CPMGEM\_cCPM and the CPM as well (Figure~\ref{fig:CPM-GCM-structure}). 
With GCM inputs, CPMGEM underestimates RASPD of CPM by 22\% at the smallest wavelength, suggesting possibly some slight smoothing compared with cCPM inputs but overall better recreation of fine scale details than BCSD\_cCPM and U-Net.
For the northwest winter and southeast summer (Figure~\ref{fig:CPM-GCM-subregions}), the frequency distributions of precipitation are also very similar, albeit with a slight underestimation of the frequency of days with more than 50 mm/day in the southeast in summer.
This indicates that the emulator produces precipitation samples with realistic structures and intensities when given bias-corrected GCM variables as input. Without this adjustment, the emulator had a modest dry bias across the whole domain---results demonstrating this are discussed by \citeA{kendon2025MLemulatorsperspective}.

The estimated extreme return periods for CPMGEM\_GCM remain quite consistent overall with those of the CPM (see Supporting Information section S1, Figure~S2 and Table~S1), though the spread of differences across estimates from each of the six CPMGEM\_GCM samples is generally slightly larger than for CPMGEM\_cCPM (around -15\% to 15\% for 1- and 9.8-year return times).
Exceptions are for London in DJF for return periods above about 36 years (for example, for the 108 year return period, the difference ranges from about -20\% to -6\%), for London in JJA at a return period of 9.8 years (when the CPM value lies just 0.1mm outside the spread, 0.2\% in relative terms), and for Lancaster in DJF and JJA for return periods of $\sim$1--10 years (6\% and 1\% beyond the spread at 1 and 9.8 year return period respectively). In these cases, the discrepancy with the CPM values is quite small.

\begin{figure}[htbp]
    \centering
    \includegraphics[scale=1.0]{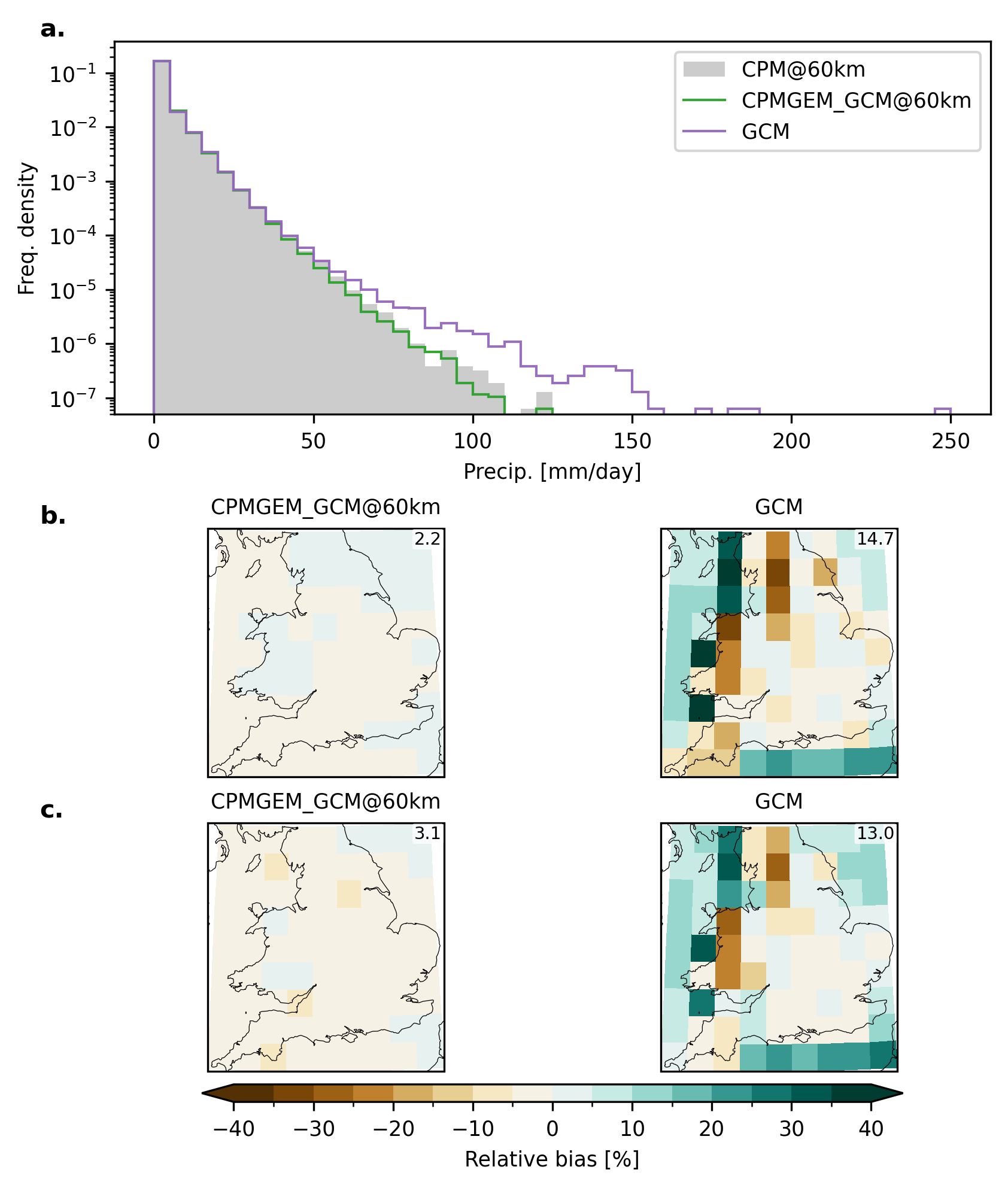}
    \caption{\textbf{Statistical comparison with GCM output.} (a) Histograms of daily precipitation values on the 60km grid, for the coarsened target CPM precipitation (grey filled area), the coarsened output of the CPMGEM\_GCM emulator (green), and GCM precipitation (purple). Note the vertical axis is logarithmic. (b) Relative bias in mean precipitation compared to the CPM for CPMGEM\_GCM (left) and GCM (right) on the coarse grid (in contrast to results shown for precipitation on the fine grid in Figure~\ref{fig:CPM-GCM-dist}(b)). (c) Same as (b) but for standard deviation bias.}
    \label{fig:CPM-GCM-60km-dist}
\end{figure}

The more realistic physics of the CPM compared to a GCM leads to changes in precipitation at coarse-scales as well as adding fine-scale detail.
To indicate the emulator's skill in matching this large-scale change by the CPM, Figure~\ref{fig:CPM-GCM-60km-dist} compares statistics of precipitation on the GCM's coarse grid between the GCM output and conservatively regridded CPMGEM\_GCM and CPM output.
The emulator produces a similar distribution of intensities to the CPM at this coarser scale right up to around the maximum coarsened CPM value from the test set, about 120 mm/day (panel (a)). 
The distribution of GCM precipitation, however, has a heavier tail for values larger than about 50 mm/day. 
The biases in the mean and standard deviation relative to the CPM are much smaller for the emulator than for the GCM, particularly in grid boxes covering upland areas, on parts of the western coast and in the English Channel (Figure~\ref{fig:CPM-GCM-60km-dist}(b) and~(c)). 

\subsection{Climate change comparison}
\label{sec:results:ccs}

\begin{figure}[htbp]
    \centering
    \includegraphics[scale=1.0]{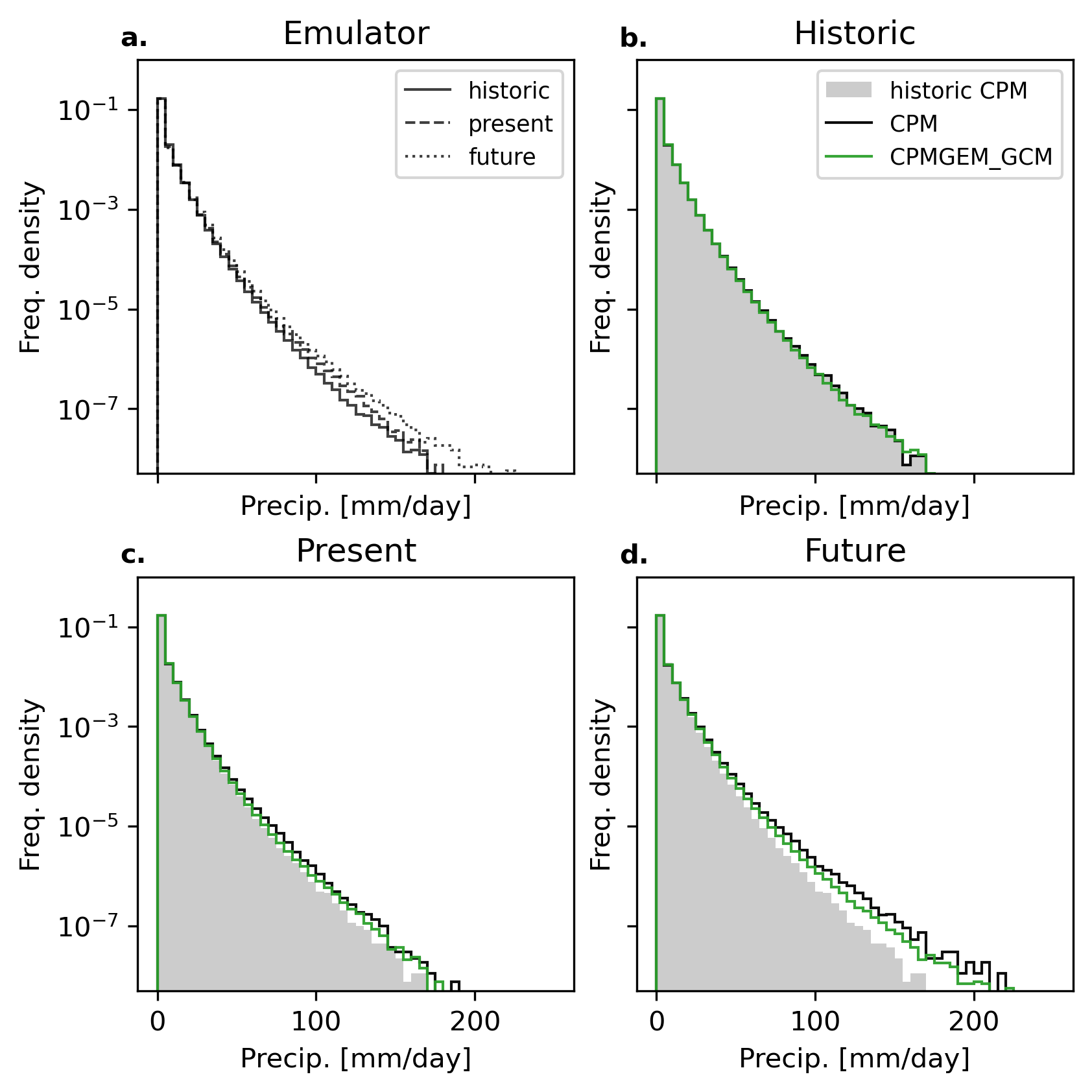}
    \caption{\textbf{Climate change effect on precipitation frequency distributions.} (a) shows the different CPMGEM\_GCM frequency densities for the three different time periods: Historic (solid black), Present (dashed black) and Future (dotted black). (b) shows a comparison of the frequency density histogram of CPMGEM\_GCM (green) with the CPM (black) for the Historic time period. (c) and (d) show the same as (b) but for Present and Future time periods respectively. In (b), (c) and (d) the filled grey histogram shows the CPM precipitation frequency distribution from the Historic time period to highlight the change in the precipitation distribution between the time periods.}
    \label{fig:GCM-CCS-hist}
\end{figure}

As mentioned in the introduction, a reason for training on high-resolution simulation data is the potential ability to learn a physically-based future climate change signal.
Figure~\ref{fig:GCM-CCS-hist} shows the emulator's ability to capture the effect of climate change on the precipitation frequency distribution, including extremes. Panel (a) shows the differences in the frequency distribution of precipitation from our emulator (CPMGEM\_GCM) for the three time periods (``Historic'' for 1981--2000, ``Present'' for 2021--2040 and ``Future'' for 2061--2080). There is an increase in intensities in the tail of the emulator's distribution as time increases. The other three panels compare distributions from CPMGEM\_GCM, the CPM target and the CPM in the Historic period. The emulator captures the shift away from the Historic distribution over time, as seen particularly clearly in the Future period. However, the increase in intensities is modestly underestimated.

\begin{figure}[htbp]
    \centering
    \includegraphics[scale=1.0]{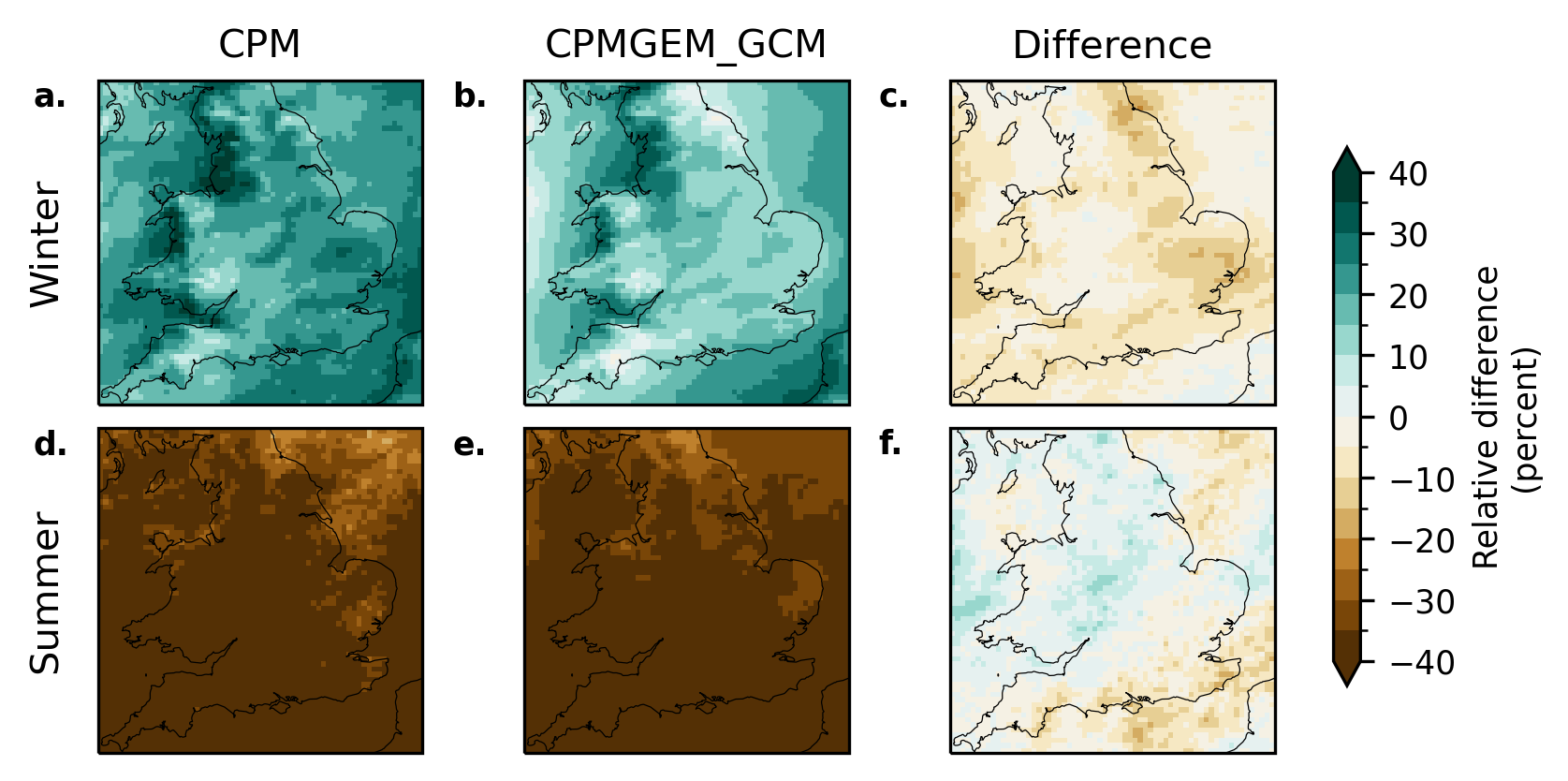}
    \caption{\textbf{Changes in seasonal mean precipitation from 1981--2000 to 2061--2080.} (a, b, c) show results for winter and (d, e, f) results for summer. The first column (a, d) is the difference between the future and historical means for the CPM and the second column (b, e) is the same for the emulator, CPMGEM\_GCM. The third column (c, f) is the difference between the CPMGEM\_GCM and CPM changes. In all cases the differences are shown as percentages of the Historic CPM seasonal mean. }
    \label{fig:GCM-CCS-mean}
\end{figure}

\begin{table}[]
    \centering
    \begin{tabulary}{0.9\textwidth}{c|R|R|R|R}
        Season &  CPM (\%) & Emulator (\%) & Difference (\% of CPM change) & Difference (\% of Historic CPM) \\
        \hline
        Winter &  23 &  17 &           -26 (\texttt{\phantom{1}-}41 to \texttt{-}10) & -5.9 (\texttt{-}9.5 to \texttt{-}2.3) \\
        Spring &   6 &   2 &           -61 (\texttt{-}134 to \texttt{+}13)           & -3.8 (\texttt{-}8.3 to \texttt{+}0.8) \\
        Summer & -40 & -39 &             2 (\texttt{\phantom{1}-}14 to \texttt{+}18)&  0.8 (\texttt{-}5.7 to \texttt{+}7.1) \\
        Autumn &  -4 &  -8 &          -114 (\texttt{-}240 to \texttt{+}13)                   & -4.5 (\texttt{-}9.4 to \texttt{+}0.5)
    \end{tabulary}
    \caption{\textbf{Climate change in the seasonal domain mean.} Changes from Historic to Future periods are shown for each season. The second and third columns contain the relative changes for the CPM and the CPMGEM\_GCM emulator respectively (relative to CPM Historic seasonal domain mean in both cases). The fourth column contains the difference between the change in the CPM and in CPMGEM\_GCM, relative to the change in CPM. The fifth column shows the same difference but relative to the Historic CPM seasonal domain mean. The values in brackets in the fourth and fifth columns are the bootstrapped 95\% confidence intervals of the difference.}
    \label{tab:GCM-CCS-domain-mean}
\end{table}

As well as an overall change in annual distribution, there are seasonal and spatial dependencies in the differences. It is projected that the UK will become drier in the summer and wetter in the winter \cite{kendon2021ukcpscienceupdate}. Figure~\ref{fig:GCM-CCS-mean} shows the relative change in the mean precipitation between Historic and Future time periods for winter (top row; a, b, c) and summer (bottom row; d, e, f) for both the CPM (left column; a, d) and CPMGEM\_GCM (middle column; b, e). The right column (c, f) shows the difference between the mean changes in the CPM and in the emulator.

The emulator reproduces the mean drying in summer well. Summer is a particularly important season for the emulator to represent well, as a large component of UK summer rainfall is convective, and the UK CPM predicts changes in rainfall intensities that are substantially larger than in climate models with lower resolutions \cite{kendon2017cpmprecipimprovements}. The emulator also captures most of the wettening in winter, though there is some underestimation.
Table~\ref{tab:GCM-CCS-domain-mean} shows the change in seasonal domain means. The change in summer is reproduced very accurately, but the winter change is underestimated by 26\% relative to the change in the CPM (with 95\% confidence interval 10--41\% from bootstrapping, described in Section~\ref{sec:methods:bootstrapping}). 
The results indicate that the changes in spring (March--May, MAM) and autumn (September--November, SON) are also reproduced with the correct sign, but there is some evidence of underestimation and overestimation of the magnitude respectively. However, the magnitude of the mean changes in these seasons is relatively small, giving high sampling uncertainty of the relative differences, and their 95\% confidence intervals overlap with zero. 
The values are similar for CPMGEM\_cCPM (see Table~\ref{tab:CPM-CCS-domain-mean} in \ref{sec:app:ccs-extra}).

\begin{table}[]
    \centering
    \begin{tabulary}{0.9\textwidth}{C|c|c|c}
        Season & Percentile & CPM & Emulator \\
        \hline
        Winter & 99   & 23.8 & 15.7 (12.4 to 17.8) \\
        Winter & 99.9 & 28.3 & 20.1 (16.8 to 24.1) \\
        \hline
        Summer & 99   & -7.0 & -11.9 (-14.2 to -9.2) \\
        Summer & 99.9 & 12.5 &   1.5 (-1.0 to 5.6) \\
    \end{tabulary}
    \caption{\textbf{Percentage changes in heavy precipitation intensities (in \%) between the Historic and Future time periods, for winter and summer.} Domain-mean of seasonal relative change in heavy daily-mean precipitation (99th and 99.9th percentile column 2) for CPM (column 3) and the CPMGEM\_GCM emulator (column 4). For the emulator, the value is the domain-mean of the percentage changes averaged over the 6 independent emulator sample runs (with the range across the samples given in brackets).}
    \label{tab:CCS-extreme-quantile-changes-GCM}
\end{table}

To quantify future changes in the tail of the distribution of seasonal intensities, Table~\ref{tab:CCS-extreme-quantile-changes-GCM} shows percentage changes in heavy precipitation (99th and 99.9 percentiles) for winter and summer.
Maps of the change in the 99th percentile can be found in \citeA{kendon2025MLemulatorsperspective}.
As with the change in mean precipitation, the emulator successfully reproduces seasonal differences in future changes in heavy precipitation. It shows substantial increases in both of these percentiles in winter and in summer it correctly shows a decrease in the 99th and an increase in the 99.9th percentile. The different signs of the changes in summer are associated with decreases in rainfall frequency reducing moderate percentiles, but increases in the intensity of rainfall events dominating the more extreme percentiles.
Results are similar for CPMGEM\_cCPM but with smaller error in the mean (Table~\ref{tab:CCS-extreme-quantile-changes-cCPM}).

However, in winter the emulator underestimates the future increases in these percentiles. In summer, the emulator's changes are less positive than in the CPM, with the decrease in the 99th percentile being overestimated and the increase in the 99.9th percentile underestimated.
The change in these percentiles when using cCPM inputs is in better agreement with the CPM results (see Table~\ref{tab:CCS-extreme-quantile-changes-cCPM}), suggesting that part of the reason for the discrepancy in CPMGEM\_GCM's predictions is differences between change signals in the GCM and coarsened CPM variables. However, the spread of predictions when using cCPM inputs still does not include the CPM changes in summer, indicating that the emulator did not fully learn the climate change signal in its training environment.

Changes in the frequency of wet and dry days are also critical to reproduce well. The emulator successfully reproduces the main features of seasonal differences in future changes in the frequency of wet days (precipitation greater than 0.1 mm). In winter, there is a small increase in wet days: a 2\% relative increase in frequency versus a 5\% increase in the CPM. In summer, a large decrease is predicted: a 38\% decrease according to the emulator versus 36\% in the CPM.

Overall, the emulator captures well the main qualitative aspects of the climate change signal of the UKCP Local dataset, reproducing seasonal differences in precipitation changes as predicted by the CPM.
However, there are quantitative errors in the emulator's predictions of future changes. The emulator produces less positive increases in the mean precipitation in winter and in high percentiles in both winter and summer than the CPM. Some contribution to this comes from predicted changes in the frequency of precipitation being slightly smaller than in the CPM. But it also indicates that the emulator underestimates increases in the intensity of precipitation events.

\subsection{Training with smaller datasets} \label{sec:results:small_data}

\begin{figure}[htbp]
    \centering
    \includegraphics[scale=0.95]{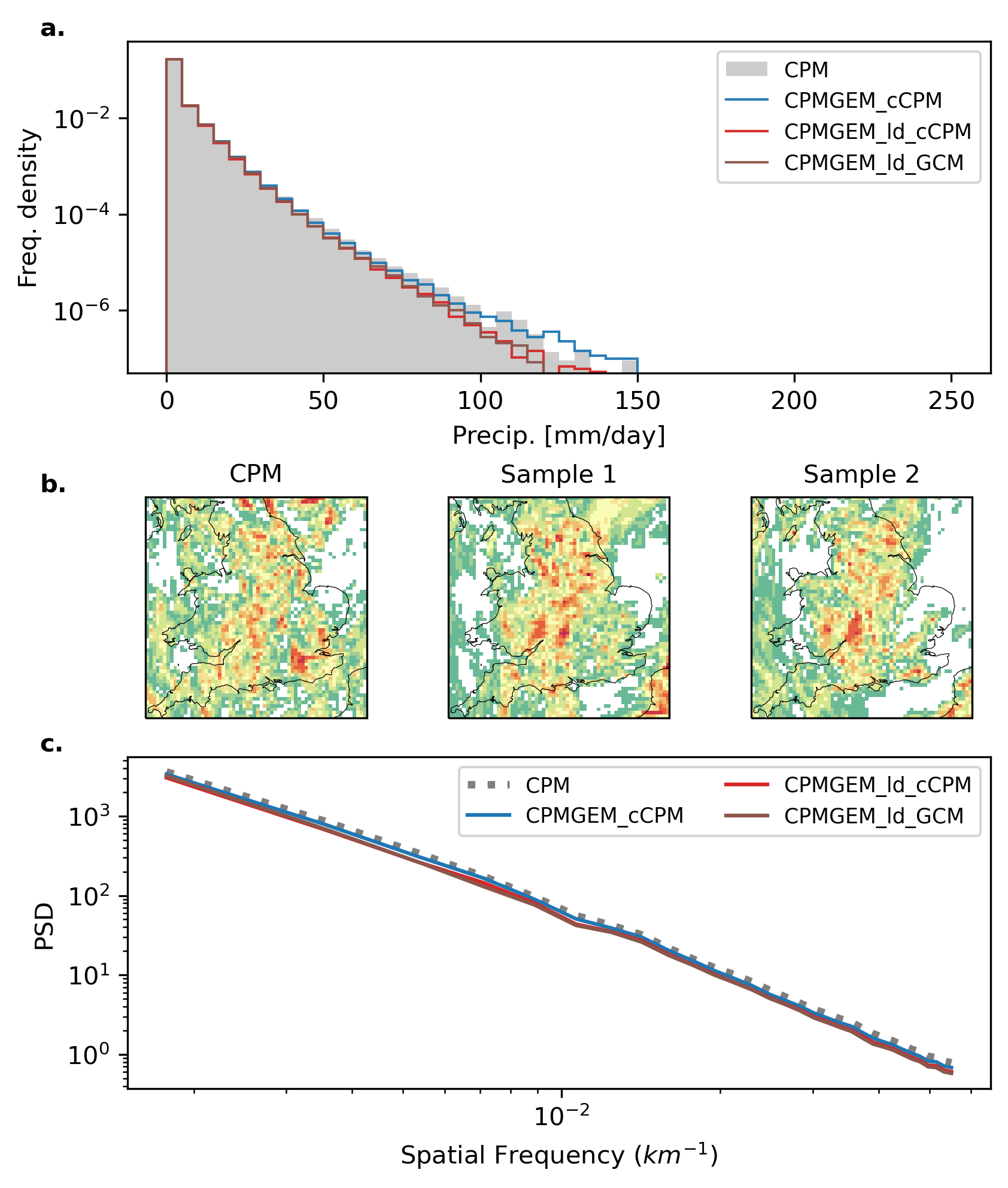}
    \caption{\textbf{Results for an emulator developed using a reduced (fourteen year) training dataset.} Both training and test data are from one ensemble member for the Historic time period only. (a) Histograms of precipitation values on the 8.8km grid: for the target CPM precipitation (grey filled area), the emulator developed using the full training data (CPMGEM\_cCPM, blue), and an emulator trained on the reduced data using coarsened CPM and GCM inputs: CPMGEM\_ld\_cCPM (purple) and CPMGEM\_ld\_GCM (dark green) respectively. Note the vertical axis is logarithmic. (b) Samples from the same ``JJA Wet'' day as in Figure~\ref{fig:CPM-samples}, from the CPM (left) and two samples from CPMGEM\_ld\_cCPM (middle and right). (c) Radially-averaged spatial power spectral density (PSD) for the target CPM precipitation (grey dashed) and each emulator (colours as (a)).}
    \label{fig:lo-data}
\end{figure}


We briefly discuss whether similar results could be obtained using CPM datasets of smaller size, $\sim$10--20 years, corresponding to the amount of data available from some projects \cite<e.g.,>{chan2020cpmeurope, kendon2019cpmafrica}, as opposed to the $\sim$500 years that we used to train the emulator discussed above. This is important for considering how widely the method could be applied. We train the low-data emulator CPMGEM\_ld on a much smaller dataset (14 years, one ensemble member in the Historic time period, $\sim$3\% of the full training dataset size). We test it on the portion of the test dataset from the Historic time period and the same ensemble member as used for training (3 years). We find that, using either cCPM or GCM inputs, this emulator is able to recreate the frequency distribution of the CPM precipitation well (Figure~\ref{fig:lo-data}(a)). Individual samples and the spatial power spectrum still appear realistic (Figure~\ref{fig:lo-data}(b,c)). 

Therefore the method does seem able to produce output matching the climatological CPM properties with this amount of data. The main challenge may be verification. With low amounts of high-resolution model data, it would be difficult to evaluate an emulator's performance on rare, high-impact weather events, although there are methods that may assist \cite{watson2022mlextremes}. It would also be difficult to verify the emulated climate change signal, since there would be high sampling variability with such small amounts of data.

\section{Discussion and conclusions}
\label{sec:conc}

We have demonstrated results from CPMGEM, an emulator of the Met Office UK CPM based on a diffusion model, a generative machine learning method. It is able to produce samples of high-resolution (8.8km), daily-mean precipitation with realistic properties, conditional on coarse-resolution (60km) variables from the Met Office GCM. It has a much lower computational cost than the CPM. The output is at high enough space and time resolution for informing applications such as flood inundation modelling \cite<e.g.>{bates2023ukfloodingrisk} and we have shown some evidence that the emulator produces predictions with realistic structure and climatological frequency for extreme events with return times up to $\sim$100 years. Further examination of the skill on extreme events would be a useful development in future work. 
The 21st century climate change signal is captured well in summer, the season when the CPM's representation of convective processes is most valuable, but there is evidence of some error in the mean change in other seasons.
We have used a $\sim$500 year training dataset for our main results, but also find that a good reproduction of climatological precipitation properties can also be obtained given only 14 years of training data.
This has potential applications such as downscaling GCM simulations where it is not practical or affordable to use the CPM, such as large-ensemble simulations to better sample meteorological variability.

The emulator is stochastic, and the stochastic component is reasonably well-calibrated, so that the emulator can create a range of plausible high-resolution samples for given large-scale conditions. From comparison to a deterministic U-Net model, the BCSD statistical method and other studies in the literature, the stochastic component is important for producing precipitation predictions with realistic small-scale structure (particularly in situations with many small-scale convective showers) and predicting the correct frequency of the most extreme intensities.

The main qualitative aspects of the 21st century climate change signal are captured well, including different patterns of change in summer compared to winter, and greater intensity increases for heavier events. The emulator captures well the mean change in summer, the season when the CPM's representation of convective processes is most valuable. However, there is some error in the predicted magnitude of the changes in winter and in increases in heavy precipitation intensities in summer. A possible reason is that the climate change signal does not account for a large fraction of the variance of precipitation between randomly chosen days, so that the standard ML approach of optimising a skill score evaluated on each prediction independently does not provide a strong enough signal to reduce errors in the mean change to near-zero. Therefore it may be valuable to develop modified training procedures that place more weight on capturing the climate change signal, such as using different loss functions. Including additional predictor variables may also give better performance. Development of methods to achieve a more faithful reproduction of the climate change signal whilst retaining the realistic spatial structure of samples would be highly valuable.

Further routes for development include increasing the output resolution and domain size in space and time yet further and increasing the number of output variables, to fully capture the value provided by the CPM.

Video-super-resolution techniques, such as ``video diffusion'' methods (e.g., \citeNP{blattmann2023videoldm, bartal2024videodiffusion, harvey2022videodiffusion}, could be applied to improve temporal coherence, which is likely to be important for producing high-quality sub-daily predictions. For example, \citeA{glawion2023spategan, bassetti2024diffesm} use a video-super-resolution approach to develop, respectively, a GAN and a diffusion model with temporal downscaling ability.

Generating multi-variate output is another key target for climate impacts modelling and would go further towards capturing the full value of CPM simulations. 
Work such as that by \citeA{lopezgomez2025rcmdiffusiondownscaling, mardani2023residualdiffkmscaledownscalingpublished} show that diffusion models are capable of skilfully producing multivariate output.

There is more to explore regarding the method's ability to produce an emulator that can generalize to other time periods or simulations.
Possible further tests include applying a model trained on the historic period only to data from future time periods to test the ability to extrapolate to unseen climates (as \citeA{rampal2024cganfutureextrapolotaion} do with a GAN). Another interesting test would be to train on data from one subset of ensemble members and evaluate on data from a different subset to investigate the relationship between dataset size and ability to generalize.
It would also be valuable to test using coarse predictors from other GCMs.
\citeA{kendon2025MLemulatorsperspective} did a preliminary exploration applying CPMGEM to other GCMs, comparing with dynamical downscaling using the target UK CPM.
Their results show variation in emulation errors depending on the GCM. How best to develop an emulator that can transfer skilfully between GCMs is an open question \cite{aich2024diffdownbc,banomedina2024emulatorxferxai,rampal2024rcmemulatorsreview,sun2024detdldownscalingreview}.
Another evaluation approach would be to assess how realistic are the results if the emulator's outputs are used to model impacts, such as flood inundation.

We have not attempted to explore the diffusion model method itself in detail. 
There is potential that computational costs may be reduced by using a neural network with many fewer parameters or sampling using fewer diffusion steps without an effect on emulator performance.
Preliminary results indicate that there is scope for reducing the size of the neural network, which would reduce memory requirements, in particular during training, or using fewer diffusion steps, which would reduce sampling time, without substantially degrading performance (not shown).
Approaches such as latent diffusion models \cite{rombach2022ldm} and patch diffusion \cite{wang2023patchdiffusion} may allow for increased resolution and domain size without GPU memory requirements outside the budget of many researchers by working on a compressed latent space or smaller patches of the domain, rather than the full size and resolution of the data as our diffusion model does.
Future work could explore performance improvements and computational cost reduction through tuning hyperparameters (e.g. number of diffusion steps) or use of more recent diffusion model developments \cite{karras2022edm, rombach2022ldm, mardani2023residualdiffkmscaledownscalingpublished, hess2025scaleadaptiveconsistencymodels, tomasi2025CPMdiff}.
A more detailed comparison to other advanced machine learning-based models that have been shown to be effective at simulating high-resolution precipitation in certain situations, such as GANs \cite{rampal2025cganadvlosssensitivity} or deterministic approaches with modified loss functions \cite{doury2024precipemulator,annau2023CPMdetGANwind}, could inform decisions about the method to choose in future studies. Relevant considerations are ease-of-use and computational resource requirements, on top of the quality of output. Whilst in this case we did not find that extensive tuning of hyperparameters of the diffusion model was required, in general this may be needed and could affect the ease of using this or other methods.

%
%

\section*{Open Research Section}

The data for training and evaluating the models is available on Zenodo \cite{addison2025cpmgemdataset}. 
Trained model weights and samples from ML models are also available on Zenodo \cite{addison2025cpmgemmodelsandsamples}.
The code for this work is available on GitHub in 3 repositories and snapshots have been deposited on Zenodo: for processing the data \cite{addison2025cpmgemdatacode}; for training and sampling from the ML models \cite{addison2025cpmgemcode}; and for evaluating the models \cite{addison2025cpmgemevalcode}.
In particular, the implementation of emulators (including full configuration) is developed on Github: \url{https://github.com/henryaddison/mlde}, which is a fork of \url{https://github.com/yang-song/score_sde_pytorch} \cite{song2021sbgmsde}. 
The version used for this manuscript is maintained in the v0.2-paper branch, which includes the v0.2.2 tag, the latest commit used by the manuscript.
The configuration of our final CPMGEM emulator can be found in \texttt{ukcp\_local\_pr\_12em\_cncsnpp\_continuous.py}.

\acknowledgments
HA was supported by the UKRI Centre for Doctoral Training in Interactive Artificial Intelligence under Grant EP/S022937/1.
EJK was supported by the Met Office Hadley Centre Climate Programme funded by DSIT under Grant GA01101.
PAGW was supported by a Natural Environment Research Council Independent Research Fellowship (grant no. NE/S014713/1).
We would like to thank Dr Stewart for compute resources used in the project, along with Bristol's ACRC for maintaining the cluster.
We also thank the editor and reviewers, whose suggestions helped to improve the quality of this work.

\textbf{Author contributions:}
\begin{itemize}
    \item Conceptualization: HA, EJK, LA, PAGW
    \item Data curation: HA
    \item Formal Analysis: HA, LA, PAGW
	\item Methodology: HA, EJK, SR, LA, PAGW
	\item Investigation: HA, LA, PAGW
	\item Visualization: HA, LA, PAGW
    \item Resources: LA, PAGW
    \item Software: HA
	\item Supervision: EJK, SR, LA, PAGW
	\item Writing—original draft: HA, PAGW
	\item Writing—review \& editing: HA, EJK, SR, LA, PAGW
\end{itemize}

\textbf{Competing interests:} The authors declare that they have no competing interests.


\bibliography{references}

@techreport{kendon2021ukcpscienceupdate,
   author = {Kendon, E. J. and Short, C. and Pope, J. and Chan, S. and Wilkinson, J. and Tucker, S. and Bett, P. and Harris, G.},
   title = {Update to UKCP Local (2.2km) projections},
   url = {https://www.metoffice.gov.uk/pub/data/weather/uk/ukcp18/science-reports/ukcp18_local_update_report_2021.pdf},
   year = {2021},
   type = {Journal Article}
}

@techreport{murphy2018ukcp18land,
   author = {Murphy, JM and Harris, GR and Sexton, DMH and Kendon, EJ and Bett, PE and Clark, RT and Eagle, KE and Fosser, G and Fung, F and Lowe, JA},
   title = {UKCP18 land projections: science report},
   url = {https://www.metoffice.gov.uk/pub/data/weather/uk/ukcp18/science-reports/UKCP18-Land-report.pdf},
   year = {2018},
   type = {Journal Article}
}

@article{gutierrez2019sdcomparison,
   author = {Gutiérrez, J. M. and Maraun, D. and Widmann, M. and Huth, R. and Hertig, E. and Benestad, R. and Roessler, O. and Wibig, J. and Wilcke, R. and Kotlarski, S. and San Martín, D. and Herrera, S. and Bedia, J. and Casanueva, A. and Manzanas, R. and Iturbide, M. and Vrac, M. and Dubrovsky, M. and Ribalaygua, J. and Pórtoles, J. and Räty, O. and Räisänen, J. and Hingray, B. and Raynaud, D. and Casado, M. J. and Ramos, P. and Zerenner, T. and Turco, M. and Bosshard, T. and Štěpánek, P. and Bartholy, J. and Pongracz, R. and Keller, D. E. and Fischer, A. M. and Cardoso, R. M. and Soares, P. M. M. and Czernecki, B. and Pagé, C.},
   title = {An intercomparison of a large ensemble of statistical downscaling methods over Europe: Results from the VALUE perfect predictor cross-validation experiment},
   journal = {International Journal of Climatology},
   volume = {39},
   number = {9},
   pages = {3750-3785},
   DOI = {10.1002/joc.5462},
   year = {2019},
   type = {Journal Article}
}

@article{maraun2019sddownscaling,
   author = {Maraun, D. and Widmann, M. and Gutierrez, J. M.},
   title = {Statistical downscaling skill under present climate conditions: A synthesis of the VALUE perfect predictor experiment},
   journal = {International Journal of Climatology},
   volume = {39},
   number = {9},
   pages = {3692-3703},
   DOI = {10.1002/joc.5877},
   year = {2019},
   type = {Journal Article}
}

@article{vandal2018mldownscaling,
   author = {Vandal, Thomas and Kodra, Evan and Ganguly, Auroop R.},
   title = {Intercomparison of machine learning methods for statistical downscaling: the case of daily and extreme precipitation},
   journal = {Theoretical and Applied Climatology},
   volume = {137},
   number = {1-2},
   pages = {557-570},
   DOI = {10.1007/s00704-018-2613-3},
   year = {2018},
   type = {Journal Article}
}

@article{walton2015hybriddownscaling,
   author = {Walton, D. B. and Sun, F. P. and Hall, A. and Capps, S.},
   title = {A Hybrid Dynamical-Statistical Downscaling Technique. Part I: Development and Validation of the Technique},
   journal = {Journal of Climate},
   volume = {28},
   number = {12},
   pages = {4597-4617},
   DOI = {10.1175/Jcli-D-14-00196.1},
   year = {2015},
   type = {Journal Article}
}

@article{chan2018precippredictors,
   author = {Chan, S. C. and Kendon, E. J. and Roberts, N. and Blenkinsop, S. and Fowler, H. J.},
   title = {Large-Scale Predictors for Extreme Hourly Precipitation Events in Convection-Permitting Climate Simulations},
   journal = {Journal of Climate},
   volume = {31},
   number = {6},
   pages = {2115-2131},
   ISSN = {0894-8755},
   DOI = {10.1175/Jcli-D-17-0404.1},
   year = {2018},
   type = {Journal Article}
}

@article{Leinonen2020GANsd,
   author = {Leinonen, Jussi and Nerini, Daniele and Berne, Alexis},
   title = {Stochastic Super-Resolution for Downscaling Time-Evolving Atmospheric Fields With a Generative Adversarial Network},
   journal = {IEEE Transactions on Geoscience and Remote Sensing},
   pages = {1–13},
   year = {2020},
   type = {Journal Article}
}

@article{grover2018flowgan,
   author = {Grover, A. and Dhar, M. and Ermon, S.},
   title = {Flow-GAN: Combining Maximum Likelihood and Adversarial Learning in Generative Models},
   journal = {Thirty-Second Aaai Conference on Artificial Intelligence / Thirtieth Innovative Applications of Artificial Intelligence Conference / Eighth Aaai Symposium on Educational Advances in Artificial Intelligence},
   volume = {32},
   number = {1},
   pages = {3069-3076},
   year = {2018},
   type = {Journal Article}
}

@article{ravuri2021deepgenprecip,
   author = {Ravuri, S. and Lenc, K. and Willson, M. and Kangin, D. and Lam, R. and Mirowski, P. and Fitzsimons, M. and Athanassiadou, M. and Kashem, S. and Madge, S. and Prudden, R. and Mandhane, A. and Clark, A. and Brock, A. and Simonyan, K. and Hadsell, R. and Robinson, N. and Clancy, E. and Arribas, A. and Mohamed, S.},
   title = {Skilful precipitation nowcasting using deep generative models of radar},
   journal = {Nature},
   volume = {597},
   number = {7878},
   pages = {672-677},
   DOI = {10.1038/s41586-021-03854-z},
   year = {2021},
   type = {Journal Article}
}

@article{Ronneberger2015unet,
   author = {Ronneberger, O. and Fischer, P. and Brox, T.},
   title = {U-Net: Convolutional Networks for Biomedical Image Segmentation},
   journal = {Medical Image Computing and Computer-Assisted Intervention, Pt Iii},
   volume = {9351},
   pages = {234-241},
   ISSN = {0302-9743},
   DOI = {10.1007/978-3-319-24574-4_28},
   year = {2015},
   type = {Journal Article}
}

@article{creswell2018ganreview,
   author = {Creswell, A. and White, T. and Dumoulin, V. and Arulkumaran, K. and Sengupta, B. and Bharath, A. A.},
   title = {Generative Adversarial Networks An overview},
   journal = {Ieee Signal Processing Magazine},
   volume = {35},
   number = {1},
   pages = {53-65},
   ISSN = {1053-5888},
   DOI = {10.1109/Msp.2017.2765202},
   year = {2018},
   type = {Journal Article}
}

@inproceedings{sohldickstein2015difforigin,
   author = {Sohl-Dickstein, Jascha and Weiss, Eric and Maheswaranathan, Niru and Ganguli, Surya},
   title = {Deep unsupervised learning using nonequilibrium thermodynamics},
   booktitle = {International conference on machine learning},
   publisher = {PMLR},
   pages = {2256-2265},
   year= {2015},
   type = {Conference Proceedings}
}

@inproceedings{song2021sbgmsde,
   author = {Song, Y. and Sohl-Dickstein, Jascha and Diederik and Kumar, Abhishek and Ermon, Stefano and Poole, Ben},
   title = {Score-Based Generative Modeling through Stochastic Differential Equations},
   booktitle = {ICLR},
   DOI = {arxiv:2011.13456},
   type = {Conference Proceedings},
   year = {2021}
}

@inproceedings{dharwial2021diffbeatsgans,
   author = {Dhariwal, Prafulla and Nichol, Alexander},
   title = {Diffusion models beat GANs on image synthesis},
   booktitle = {Advances in Neural Information Processing Systems},
   volume = {34},
   url = {https://proceedings.neurips.cc/paper/2021/file/49ad23d1ec9fa4bd8d77d02681df5cfa-Paper.pdf},
   year = {2021},
   type = {Conference Proceedings}
}

@inproceedings{ho2020ddpm,
   author = {Ho, Jonathan and Jain, Ajay and Abbeel, Pieter},
   title = {Denoising diffusion probabilistic models},
   booktitle = {Advances in Neural Information Processing Systems},
   volume = {33},
   pages = {6840-6851},
   url = {https://proceedings.neurips.cc/paper/2020/hash/4c5bcfec8584af0d967f1ab10179ca4b-Abstract.html},
   type = {Conference Proceedings},
   year = {2020}
}

@inproceedings{song2019smld,
   author = {Song, Yang and Ermon, Stefano},
   title = {Generative modeling by estimating gradients of the data distribution},
   booktitle = {Advances in Neural Information Processing Systems},
   volume = {32},
   url = {https://proceedings.neurips.cc/paper/2019/hash/3001ef257407d5a371a96dcd947c7d93-Abstract.html},
   type = {Conference Proceedings}
}

@article{doury2023RCMemulator,
   author = {Doury, Antoine and Somot, Samuel and Gadat, Sebastien and Ribes, Aurélien and Corre, Lola},
   title = {Regional climate model emulator based on deep learning: concept and first evaluation of a novel hybrid downscaling approach},
   journal = {Climate Dynamics},
   volume = {60},
   number = {5-6},
   pages = {1751-1779},
   ISSN = {0930-7575
1432-0894},
   DOI = {10.1007/s00382-022-06343-9},
   url = {https://doi.org/10.1007/s00382-022-06343-9},
   year = {2023},
   type = {Journal Article}
}

@article{doury2024precipemulator,
   author = {Doury, A. and Somot, S. and Gadat, S.},
   title = {On the suitability of a convolutional neural network based RCM-emulator for fine spatio-temporal precipitation},
   journal = {Climate Dynamics},
   volume = {62},
   number = {9},
   pages = {8587-8613},
   ISSN = {0930-7575},
   DOI = {10.1007/s00382-024-07350-8},
   url = {<Go to ISI>://WOS:001277003700002},
   year = {2024},
   type = {Journal Article}
}

@article{Widmann2019sptialvariability,
   author = {Widmann, M. and Bedia, J. and Gutierrez, J. M. and Bosshard, T. and Hertig, E. and Maraun, D. and Casado, M. J. and Ramos, P. and Cardoso, R. M. and Soares, P. M. M. and Ribalaygua, J. and Page, C. and Fischer, A. M. and Herrera, S. and Huth, R.},
   title = {Validation of spatial variability in downscaling results from the VALUE perfect predictor experiment},
   journal = {International Journal of Climatology},
   volume = {39},
   number = {9},
   pages = {3819-3845},
   DOI = {10.1002/joc.6024},
   year = {2019},
   type = {Journal Article}
}

@article{Harris2022cgandownscaling,
   author = {Harris, L. and McRae, A. T. T. and Chantry, M. and Dueben, P. D. and Palmer, T. N.},
   title = {A Generative Deep Learning Approach to Stochastic Downscaling of Precipitation Forecasts},
   journal = {J Adv Model Earth Syst},
   volume = {14},
   number = {10},
   pages = {e2022MS003120},
   DOI = {10.1029/2022MS003120},
   url = {https://www.ncbi.nlm.nih.gov/pubmed/36590321},
   year = {2022},
   type = {Journal Article}
}

@article{Vosper2023cgandownscalingcyclones,
   author = {Vosper, E. and Watson, P. and Harris, L. and McRae, A. and Santos-Rodriguez, R. and Aitchison, L. and Mitchell, D.},
   title = {Deep Learning for Downscaling Tropical Cyclone Rainfall to Hazard-Relevant Spatial Scales},
   journal = {Journal of Geophysical Research-Atmospheres},
   volume = {128},
   number = {10},
   pages = {e2022JD038163},
   DOI = {10.1029/2022JD038163},
   year = {2023},
   type = {Journal Article}
}

@article{scahller2020resolutionrole,
   author = {Schaller, N. and Sillmann, J. and Muller, M. and Haarsma, R. and Hazeleger, W. and Hegdahl, T. J. and Kelder, T. and van den Oord, G. and Weerts, A. and Whan, K.},
   title = {The role of spatial and temporal model resolution in a flood event storyline approach in western Norway},
   journal = {Weather and Climate Extremes},
   volume = {29},
   pages = {100259},
   DOI = {10.1016/j.wace.2020.100259},
   year = {2020},
   type = {Journal Article}
}

@article{leutbecher2008ensembleforecasting,
   author = {Leutbecher, M. and Palmer, T. N.},
   title = {Ensemble forecasting},
   journal = {Journal of Computational Physics},
   volume = {227},
   number = {7},
   pages = {3515-3539},
   DOI = {10.1016/j.jcp.2007.02.014},
   url = {https://www.sciencedirect.com/science/article/pii/S0021999107000812},
   year = {2008},
   type = {Journal Article}
}

@article{haynes2023uncertainty,
   author = {Haynes, Katherine and Lagerquist, Ryan and McGraw, Marie and Musgrave, Kate and Ebert-Uphoff, Imme},
   title = {Creating and Evaluating Uncertainty Estimates with Neural Networks for Environmental-Science Applications},
   journal = {Artificial Intelligence for the Earth Systems},
   volume = {2},
   number = {2},
   pages = {220061},
   ISSN = {2769-7525},
   DOI = {10.1175/aies-d-22-0061.1},
   url = {https://journals.ametsoc.org/view/journals/aies/2/2/AIES-D-22-0061.1.xml},
   year = {2023},
   type = {Journal Article}
}

@article{archer2024CPMflooding,
   author = {Archer, L. and Hatchard, S. and Devitt, L. and Neal, J. C. and Coxon, G. and Bates, P. D. and Kendon, E. J. and Savage, J.},
   title = {Future Change in Urban Flooding Using New Convection-Permitting Climate Projections},
   journal = {Water Resources Research},
   volume = {60},
   number = {1},
   pages = {e2023WR035533},
   DOI = {https://doi.org/10.1029/2023WR035533},
   url = {https://agupubs.onlinelibrary.wiley.com/doi/abs/10.1029/2023WR035533},
   year = {2024},
   type = {Journal Article}
}

@misc{sobolowski2023EUROCORDEXdesign,
  author       = {Sobolowski, Stefan and
                  Somot, Samuel and
                  Fernandez, Jesus and
                  Evin, Guillaume and
                  Maraun, Douglas and
                  Kotlarski, Sven and
                  Jury, Martin and
                  Benestad, Rasmus E. and
                  Teichmann, Claas and
                  Christensen, Ole B. and
                  Katharina, Bülow and
                  Buonomo, Erasmo and
                  Katragkou, Eleni and
                  Steger, Christian and
                  Sørland, Silje and
                  Nikulin, Grigory and
                  McSweeney, Carol and
                  Dobler, Andreas and
                  Palmer, Tamzin and
                  Wilke, Renate and
                  Boé, Julien and
                  Brunner, Lukas and
                  Ribes, Aurélien and
                  Qasmi, Said and
                  Nabat, Pierre and
                  Sevault, Florence and
                  Oudar, Thomas and
                  Brands, Swen},
  title        = {{EURO-CORDEX CMIP6 GCM Selection \& Ensemble Design: 
                   Best Practices and Recommendations}},
  month        = feb,
  year         = 2023,
  publisher    = {Zenodo},
  doi          = {10.5281/zenodo.7673400},
  url          = {https://doi.org/10.5281/zenodo.7673400}
}

@article{maher2021largeensembles,
   author = {Maher, Nicola and Milinski, Sebastian and Ludwig, Ralf},
   title = {Large ensemble climate model simulations: introduction, overview, and future prospects for utilising multiple types of large ensemble},
   journal = {Earth System Dynamics},
   volume = {12},
   number = {2},
   pages = {401-418},
   ISSN = {2190-4987},
   year = {2021},
   type = {Journal Article}
}

@article{leach2022extremesamplegeneration,
   author = {Leach, Nicholas J. and Watson, Peter A. G. and Sparrow, Sarah N. and Wallom, David C. H. and Sexton, David M. H.},
   title = {Generating samples of extreme winters to support climate adaptation},
   journal = {Weather and Climate Extremes},
   volume = {36},
   pages = {100419},
   ISSN = {22120947},
   DOI = {10.1016/j.wace.2022.100419},
   url = {https://www.sciencedirect.com/science/article/pii/S2212094722000111},
   year = {2022},
   type = {Journal Article}
}

@article{eyring2016CMIP6,
   author = {Eyring, Veronika and Bony, Sandrine and Meehl, Gerald A. and Senior, Catherine A. and Stevens, Bjorn and Stouffer, Ronald J. and Taylor, Karl E.},
   title = {Overview of the Coupled Model Intercomparison Project Phase 6 (CMIP6) experimental design and organization},
   journal = {Geoscientific Model Development},
   volume = {9},
   number = {5},
   pages = {1937-1958},
   note = {GMD},
   ISSN = {1991-9603},
   DOI = {10.5194/gmd-9-1937-2016},
   url = {https://gmd.copernicus.org/articles/9/1937/2016/},
   year = {2016},
   type = {Journal Article}
}

@article{bates2023ukfloodingrisk,
   author = {Bates, Paul D. and Savage, James and Wing, Oliver and Quinn, Niall and Sampson, Christopher and Neal, Jeffrey and Smith, Andrew},
   title = {A climate-conditioned catastrophe risk model for UK flooding},
   journal = {Natural Hazards and Earth System Sciences},
   volume = {23},
   number = {2},
   pages = {891-908},
   note = {NHESS},
   ISSN = {1684-9981},
   DOI = {10.5194/nhess-23-891-2023},
   url = {https://nhess.copernicus.org/articles/23/891/2023/},
   year = {2023},
   type = {Journal Article}
}

@article{watson2022mlextremes,
   author = {Watson, Peter A. G.},
   title = {Machine learning applications for weather and climate need greater focus on extremes},
   journal = {Environmental Research Letters},
   volume = {17},
   number = {11},
   pages = {111004},
   ISSN = {1748-9326},
   DOI = {10.1088/1748-9326/ac9d4e},
   url = {https://dx.doi.org/10.1088/1748-9326/ac9d4e},
   year = {2022},
   type = {Journal Article}
}

@book{maraun2018downscalingbook,
   author = {Maraun, Douglas and Widmann, Martin},
   title = {Statistical downscaling and bias correction for climate research},
   publisher = {Cambridge University Press},
   ISBN = {1108340644},
   year = {2018},
   type = {Book}
}

@article{schoof2013downscalingclimatology,
   author = {Schoof, Justin T.},
   title = {Statistical Downscaling in Climatology},
   journal = {Geography Compass},
   volume = {7},
   number = {4},
   pages = {249-265},
   ISSN = {1749-8198
1749-8198},
   DOI = {10.1111/gec3.12036},
   url = {https://compass.onlinelibrary.wiley.com/doi/abs/10.1111/gec3.12036},
   year = {2013},
   type = {Journal Article}
}

@article{maraun2017climchangesimbiascorrection,
   author = {Maraun, Douglas and Shepherd, Theodore G. and Widmann, Martin and Zappa, Giuseppe and Walton, Daniel and Gutiérrez, José M. and Hagemann, Stefan and Richter, Ingo and Soares, Pedro M. M. and Hall, Alex and Mearns, Linda O.},
   title = {Towards process-informed bias correction of climate change simulations},
   journal = {Nature Climate Change},
   volume = {7},
   number = {11},
   pages = {764-773},
   ISSN = {1758-678X},
   DOI = {10.1038/nclimate3418},
   year = {2017},
   type = {Journal Article}
}

@article{boe2022hybriddownscalingRCMemulator,
   author = {Boé, Julien and Mass, Alexandre and Deman, Juliette},
   title = {A simple hybrid statistical–dynamical downscaling method for emulating regional climate models over Western Europe. Evaluation, application, and role of added value?},
   journal = {Climate Dynamics},
   volume = {61},
   number = {1-2},
   pages = {271-294},
   ISSN = {0930-7575
1432-0894},
   DOI = {10.1007/s00382-022-06552-2},
   url = {https://doi.org/10.1007/s00382-022-06552-2},
   year = {2022},
   type = {Journal Article}
}

@article{wang2023dlforbiascorrectanddownscaling,
   author = {Wang, Fang and Tian, Di and Carroll, Mark},
   title = {Customized deep learning for precipitation bias correction and downscaling},
   journal = {Geoscientific Model Development},
   volume = {16},
   number = {2},
   pages = {535-556},
   note = {GMD},
   ISSN = {1991-9603},
   DOI = {10.5194/gmd-16-535-2023},
   url = {https://gmd.copernicus.org/articles/16/535/2023/},
   year = {2023},
   type = {Journal Article}
}

@article{mardani2023residualdiffkmscaledownscalingpublished,
   author = {Mardani, Morteza and Brenowitz, Noah and Cohen, Yair and Pathak, Jaideep and Chen, Chieh-Yu and Liu, Cheng-Chin and Vahdat, Arash and Nabian, Mohammad Amin and Ge, Tao and Subramaniam, Akshay and Kashinath, Karthik and Kautz, Jan and Pritchard, Mike},
   title = {Residual corrective diffusion modeling for km-scale atmospheric downscaling},
   journal = {Communications Earth \& Environment},
   volume = {6},
   number = {1},
   pages = {124},
   ISSN = {2662-4435},
   DOI = {10.1038/s43247-025-02042-5},
   url = {https://doi.org/10.1038/s43247-025-02042-5},
   year = {2025},
   type = {Journal Article}
}

@inproceedings{addison2022neuripsccai,
  title={Machine learning emulation of a local-scale UK climate model},
  author={Addison, Henry and Kendon, E. J. and Ravuri, Suman and Aitchison, Laurence and Watson, Peter},
  booktitle={NeurIPS 2022 Workshop on Tackling Climate Change with Machine Learning},
  url={https://www.climatechange.ai/papers/neurips2022/21},
  year={2022}
}

@article{kendon2014heaviersummerrain,
   author = {Kendon, E. J. and Roberts, Nigel M. and Fowler, Hayley J. and Roberts, Malcolm J. and Chan, Steven C. and Senior, Catherine A.},
   title = {Heavier summer downpours with climate change revealed by weather forecast resolution model},
   journal = {Nature Climate Change},
   volume = {4},
   number = {7},
   pages = {570-576},
   ISSN = {1758-678X
1758-6798},
   DOI = {10.1038/nclimate2258},
   url = {https://doi.org/10.1038/nclimate2258},
   year = {2014},
   type = {Journal Article}
}

@article{schulz2021dlvsnumweather,
   author = {Schultz, M. G. and Betancourt, C. and Gong, B. and Kleinert, F. and Langguth, M. and Leufen, L. H. and Mozaffari, A. and Stadtler, S.},
   title = {Can deep learning beat numerical weather prediction?},
   journal = {Philosophical Transactions of the Royal Society A: Mathematical, Physical and Engineering Sciences},
   volume = {379},
   number = {2194},
   pages = {20200097},
   ISSN = {1364-503X},
   DOI = {10.1098/rsta.2020.0097},
   year = {2021},
   type = {Journal Article}
}

@article{kendon2023uklocaltrends,
   author = {Kendon, E. J. and Fischer, E. M. and Short, C. J.},
   title = {Variability conceals emerging trend in 100yr projections of UK local hourly rainfall extremes},
   journal = {Nat Commun},
   volume = {14},
   number = {1},
   pages = {1133},
   ISSN = {2041-1723 (Electronic)
2041-1723 (Linking)},
   DOI = {10.1038/s41467-023-36499-9},
   url = {https://www.ncbi.nlm.nih.gov/pubmed/36882408},
   year = {2023},
   type = {Journal Article}
}

@article{kendon2019cpmafrica,
   author = {Kendon, E. J. and Stratton, R. A. and Tucker, S. and Marsham, J. H. and Berthou, S. and Rowell, D. P. and Senior, C. A.},
   title = {Enhanced future changes in wet and dry extremes over Africa at convection-permitting scale},
   journal = {Nat Commun},
   volume = {10},
   number = {1},
   pages = {1794},
   ISSN = {2041-1723 (Electronic)
2041-1723 (Linking)},
   DOI = {10.1038/s41467-019-09776-9},
   url = {https://www.ncbi.nlm.nih.gov/pubmed/31015416},
   year = {2019},
   type = {Journal Article}
}

@article{chan2020cpmeurope,
   author = {Chan, S. C. and Kendon, E. J. and Berthou, S. and Fosser, G. and Lewis, E. and Fowler, H. J.},
   title = {Europe-wide precipitation projections at convection permitting scale with the Unified Model},
   journal = {Clim Dyn},
   volume = {55},
   number = {3},
   pages = {409-428},
   ISSN = {0930-7575 (Print)
1432-0894 (Electronic)
0930-7575 (Linking)},
   DOI = {10.1007/s00382-020-05192-8},
   url = {https://www.ncbi.nlm.nih.gov/pubmed/32713994},
   year = {2020},
   type = {Journal Article}
}

@article{kendon2012vhrrcmprecip,
   author = {Kendon, E. J. and Roberts, N. M. and Senior, C. A. and Roberts, M. J.},
   title = {Realism of Rainfall in a Very High-Resolution Regional Climate Model},
   journal = {Journal of Climate},
   volume = {25},
   number = {17},
   pages = {5791-5806},
   ISSN = {0894-8755},
   DOI = {10.1175/Jcli-D-11-00562.1},
   url = {<Go to ISI>://WOS:000308633500010},
   year = {2012},
   type = {Journal Article}
}

@article{kendon2020futureCPMwinterprecip,
   author = {Kendon, Elizabeth J. and Roberts, Nigel M. and Fosser, Giorgia and Martin, Gill M. and Lock, Adrian P. and Murphy, James M. and Senior, Catherine A. and Tucker, Simon O.},
   title = {Greater Future U.K. Winter Precipitation Increase in New Convection-Permitting Scenarios},
   journal = {Journal of Climate},
   volume = {33},
   number = {17},
   pages = {7303-7318},
   ISSN = {0894-8755
1520-0442},
   DOI = {10.1175/jcli-d-20-0089.1},
   url = {https://journals.ametsoc.org/view/journals/clim/33/17/jcliD200089.xml},
   year = {2020},
   type = {Journal Article}
}

@inproceedings{blattmann2023videoldm,
   author = {Blattmann, Andreas and Rombach, Robin and Ling, Huan and Dockhorn, Tim and Kim, Seung Wook and Fidler, Sanja and Kreis, Karsten},
   title = {Align your latents: High-resolution video synthesis with latent diffusion models},
   booktitle = {Proceedings of the IEEE/CVF Conference on Computer Vision and Pattern Recognition},
   pages = {22563-22575},
   year = {2023},
   type = {Conference Proceedings}
}

@article{bartal2024videodiffusion,
   author = {Bar-Tal, Omer and Chefer, Hila and Tov, Omer and Herrmann, Charles and Paiss, Roni and Zada, Shiran and Ephrat, Ariel and Hur, Junhwa and Li, Yuanzhen and Michaeli, Tomer},
   title = {Lumiere: A space-time diffusion model for video generation},
   journal = {arXiv preprint arXiv:2401.12945},
   year = {2024},
   type = {Journal Article}
}

@inproceedings{harvey2022videodiffusion,
   author = {Harvey, William and Naderiparizi, Saeid and Masrani, Vaden and Weilbach, Christian and Wood, Frank},
   title = {Flexible diffusion modeling of long videos},
   journal = {Advances in Neural Information Processing Systems},
   volume = {35},
   pages = {27953-27965},
   year = {2022},
   type = {Conference Proceedings}
}

@article{klaver2020effectiveresolution,
   author = {Klaver, Remko and Haarsma, Rein and Vidale, Pier Luigi and Hazeleger, Wilco},
   title = {Effective resolution in high resolution global atmospheric models for climate studies},
   journal = {Atmospheric Science Letters},
   volume = {21},
   number = {4},
   pages = {e952},
   ISSN = {1530-261X},
   DOI = {10.1002/asl.952},
   year = {2020},
   type = {Journal Article}
}

@article{eurocordexdomain,
   author = {Jacob, D. and Petersen, J. and Eggert, B. and Alias, A. and Christensen, O. B. and Bouwer, L. M. and Braun, A. and Colette, A. and Déqué, M. and Georgievski, G. and Georgopoulou, E. and Gobiet, A. and Menut, L. and Nikulin, G. and Haensler, A. and Hempelmann, N. and Jones, C. and Keuler, K. and Kovats, S. and Kröner, N. and Kotlarski, S. and Kriegsmann, A. and Martin, E. and van Meijgaard, E. and Moseley, C. and Pfeifer, S. and Preuschmann, S. and Radermacher, C. and Radtke, K. and Rechid, D. and Rounsevell, M. and Samuelsson, P. and Somot, S. and Soussana, J. F. and Teichmann, C. and Valentini, R. and Vautard, R. and Weber, B. and Yiou, P.},
   title = {EURO-CORDEX: new high-resolution climate change projections for European impact research},
   journal = {Regional Environmental Change},
   volume = {14},
   number = {2},
   pages = {563-578},
   ISSN = {1436-3798},
   DOI = {10.1007/s10113-013-0499-2},
   year = {2014},
   type = {Journal Article}
}

@article{kendon2025MLemulatorsperspective,
   author = {Kendon, Elizabeth J. and Addison, Henry and Doury, Antoine and Somot, Samuel and Watson, Peter A. G. and Booth, Ben B. B. and Coppola, Erika and Gutiérrez, José Manuel and Murphy, James and Scullion, Calum},
   title = {Potential for Machine Learning Emulators to Augment Regional Climate Simulations in Provision of Local Climate Change Information},
   journal = {Bulletin of the American Meteorological Society},
   volume = {106},
   number = {6},
   pages = {E1175-E1203},
   ISSN = {0003-0007
1520-0477},
   DOI = {10.1175/bams-d-24-0114.1},
   url = {https://journals.ametsoc.org/view/journals/bams/106/6/BAMS-D-24-0114.1.xml},
   year = {2025},
   type = {Journal Article}
}

@article{kendon2017cpmprecipimprovements,
   author = {Kendon, Elizabeth J. and Ban, Nikolina and Roberts, Nigel M. and Fowler, Hayley J. and Roberts, Malcolm J. and Chan, Steven C. and Evans, Jason P. and Fosser, Giorgia and Wilkinson, Jonathan M.},
   title = {Do Convection-Permitting Regional Climate Models Improve Projections of Future Precipitation Change?},
   journal = {Bulletin of the American Meteorological Society},
   volume = {98},
   number = {1},
   pages = {79-93},
   DOI = {10.1175/bams-d-15-0004.1},
   year = {2017},
   type = {Journal Article}
}

@article{hess2022phiganesmprecip,
   author = {Hess, P. and Druke, M. and Petri, S. and Strnad, F. M. and Boers, N.},
   title = {Physically constrained generative adversarial networks for improving precipitation fields from Earth system models},
   journal = {Nature Machine Intelligence},
   volume = {4},
   number = {10},
   pages = {828-839},
   DOI = {10.1038/s42256-022-00540-1},
   year = {2022},
   type = {Journal Article}
}

@article{sha2020unetdsprecip,
   author = {Sha, Y. K. and Gagne, D. J. and West, G. and Stull, R.},
   title = {Deep-Learning-Based Gridded Downscaling of Surface Meteorological Variables in Complex Terrain. Part II: Daily Precipitation},
   journal = {Journal of Applied Meteorology and Climatology},
   volume = {59},
   number = {12},
   pages = {2075-2092},
   DOI = {10.1175/Jamc-D-20-0058.1},
   year = {2020},
   type = {Journal Article}
}

@book{efron1982bootstrap,
   author = {Efron, Bradley},
   title = {The jackknife, the bootstrap and other resampling plans},
   publisher = {SIAM},
   ISBN = {0898711797},
   year = {1982},
   type = {Book}
}

@article{harris2001rapsd,
   author = {Harris, D. and Foufoula-Georgiou, E. and Droegemeier, K. K. and Levit, J. J.},
   title = {Multiscale statistical properties of a high-resolution precipitation forecast},
   journal = {Journal of Hydrometeorology},
   volume = {2},
   number = {4},
   pages = {406-418},
   DOI = {10.1175/1525-7541(2001)002<0406:Mspoah>2.0.Co;2},
   year = {2001},
   type = {Journal Article}
}

@article{sinclair2005rapsd,
   author = {Sinclair, S. and Pegram, G. G. S.},
   title = {Empirical Mode Decomposition in 2-D space and time: a tool for space-time rainfall analysis and nowcasting},
   journal = {Hydrology and Earth System Sciences},
   volume = {9},
   number = {3},
   pages = {127-137},
   DOI = {DOI 10.5194/hess-9-127-2005},
   year = {2005},
   type = {Journal Article}
}

@article{vandermeer2023unetrcmemulator,
   author = {van der Meer, Marijn and de Roda Husman, Sophie and Lhermitte, Stef},
   title = {Deep Learning Regional Climate Model Emulators: A Comparison of Two Downscaling Training Frameworks},
   journal = {Journal of Advances in Modeling Earth Systems},
   volume = {15},
   number = {6},
   pages = {e2022MS003593},
   DOI = {10.1029/2022ms003593},
   year = {2023},
   type = {Journal Article}
}

@article{dee2011erainterim,
   author = {Dee, D. P. and Uppala, S. M. and Simmons, A. J. and Berrisford, P. and Poli, P. and Kobayashi, S. and Andrae, U. and Balmaseda, M. A. and Balsamo, G. and Bauer, P. and Bechtold, P. and Beljaars, A. C. M. and van de Berg, L. and Bidlot, J. and Bormann, N. and Delsol, C. and Dragani, R. and Fuentes, M. and Geer, A. J. and Haimberger, L. and Healy, S. B. and Hersbach, H. and Hólm, E. V. and Isaksen, L. and Kållberg, P. and Köhler, M. and Matricardi, M. and McNally, A. P. and Monge-Sanz, B. M. and Morcrette, J. J. and Park, B. K. and Peubey, C. and de Rosnay, P. and Tavolato, C. and Thépaut, J. N. and Vitart, F.},
   title = {The ERA-Interim reanalysis: configuration and performance of the data assimilation system},
   journal = {Quarterly Journal of the Royal Meteorological Society},
   volume = {137},
   number = {656},
   pages = {553-597},
   DOI = {10.1002/qj.828},
   year = {2011},
   type = {Journal Article}
}

@article{weaver2017climaterisk,
    doi = {10.1088/1748-9326/aa7494},
    url = {https://dx.doi.org/10.1088/1748-9326/aa7494},
    year = {2017},
    month = {jul},
    publisher = {IOP Publishing},
    volume = {12},
    number = {8},
    pages = {080201},
    author = {Weaver, C P and Moss, R H and Ebi, K L and Gleick, P H and Stern, P C and Tebaldi, C and Wilson, R S and Arvai, J L},
    title = {Reframing climate change assessments around risk: recommendations for the US National Climate Assessment},
    journal = {Environmental Research Letters},
}

@misc{addison2025cpmgemdataset,
  author       = {Addison, Henry and
                  Kendon, Elizabeth and
                  Ravuri, Suman and
                  Aitchison, Laurence and
                  Watson, Peter AG},
  title        = {Met Office UKCP Local CPM precipitation ML
                   emulator dataset
                  },
  month        = oct,
  year         = 2025,
  publisher    = {Zenodo},
  doi          = {10.5281/zenodo.17459721},
  url          = {https://doi.org/10.5281/zenodo.17459721},
  type         = {dataset},
}

@misc{addison2025cpmgemmodelsandsamples,
  author       = {Addison, Henry and
                  Kendon, Elizabeth and
                  Ravuri, Suman and
                  Aitchison, Laurence and
                  Watson, Peter},
  title        = {Models and samples for ML emulators of Met Office
                   UK CPM
                  },
  month        = nov,
  year         = 2025,
  publisher    = {Zenodo},
  doi          = {10.5281/zenodo.17466547},
  url          = {https://doi.org/10.5281/zenodo.17466547},
  type         = {dataset},
}

@misc{addison2025cpmgemcode,
  author       = {Henry Addison and
                  Yang Song},
  title        = {henryaddison/mlde: v0.2.2},
  month        = oct,
  year         = 2025,
  publisher    = {Zenodo},
  version      = {v0.2.2},
  doi          = {10.5281/zenodo.17466814},
  url          = {https://doi.org/10.5281/zenodo.17466814},
  swhid        = {swh:1:dir:63328679b276d2400b6768c275c0b75e1b790305
                   ;origin=https://doi.org/10.5281/zenodo.15096140;vi
                   sit=swh:1:snp:de979401019aeda2c4cc2db94861d7d22041
                   0245;anchor=swh:1:rel:7f31e9a28afdf4204a2abe66b917
                   f057336cadae;path=henryaddison-mlde-11fecda
                  },
  type         = {software},
}

@misc{addison2025cpmgemevalcode,
  author       = {Henry Addison},
  title        = {henryaddison/cpmgem-paper-evaluation: v0.2.1},
  month        = oct,
  year         = 2025,
  publisher    = {Zenodo},
  version      = {v0.2.1},
  doi          = {10.5281/zenodo.17467036},
  url          = {https://doi.org/10.5281/zenodo.17467036},
  type         = {ComputationalNotebook},
}

@misc{addison2025cpmgemdatacode,
  author       = {Henry Addison},
  title        = {henryaddison/mlde-data: v0.3.0},
  month        = mar,
  year         = 2025,
  publisher    = {Zenodo},
  version      = {v0.3.0},
  doi          = {10.5281/zenodo.15102004},
  url          = {https://doi.org/10.5281/zenodo.15102004},
  swhid        = {swh:1:dir:e38e98a953ade5d4fd6a6b9166fd7fe11a07344f
                   ;origin=https://doi.org/10.5281/zenodo.15096356;vi
                   sit=swh:1:snp:c25c0957217364e5c5c0bd57bd462bbed97a
                   bbe3;anchor=swh:1:rel:980f5bca7e209ddff0db130fa1e1
                   a12c55ef8bbd;path=henryaddison-mlde-data-277219e
                  },
  type         = {software},
}

@article{rampal2024rcmemulatorsreview,
   author = {Rampal, Neelesh and Hobeichi, Sanaa and Gibson, Peter B. and Baño-Medina, Jorge and Abramowitz, Gab and Beucler, Tom and González-Abad, Jose and Chapman, William and Harder, Paula and Gutiérrez, José Manuel},
   title = {Enhancing Regional Climate Downscaling through Advances in Machine Learning},
   journal = {Artificial Intelligence for the Earth Systems},
   volume = {3},
   number = {2},
   pages = {230066},
   ISSN = {2769-7525},
   DOI = {10.1175/aies-d-23-0066.1},
   url = {https://journals.ametsoc.org/view/journals/aies/3/2/AIES-D-23-0066.1.xml},
   year = {2024},
   type = {Journal Article}
}

@article{lopezgomez2025rcmdiffusiondownscaling,
   author = {Lopez-Gomez, I. and Wan, Z. Y. and Zepeda-Nunez, L. and Schneider, T. and Anderson, J. and Sha, F.},
   title = {Dynamical-generative downscaling of climate model ensembles},
   journal = {Proc Natl Acad Sci U S A},
   volume = {122},
   number = {17},
   pages = {e2420288122},
   ISSN = {1091-6490 (Electronic)
0027-8424 (Print)
0027-8424 (Linking)},
   DOI = {10.1073/pnas.2420288122},
   url = {https://www.ncbi.nlm.nih.gov/pubmed/40279391},
   year = {2025},
   type = {Journal Article}
}

@inproceedings{karras2022edm,
 author = {Karras, Tero and Aittala, Miika and Aila, Timo and Laine, Samuli},
 booktitle = {Advances in Neural Information Processing Systems},
 editor = {S. Koyejo and S. Mohamed and A. Agarwal and D. Belgrave and K. Cho and A. Oh},
 pages = {26565--26577},
 publisher = {Curran Associates, Inc.},
 title = {Elucidating the Design Space of Diffusion-Based Generative Models},
 url = {https://proceedings.neurips.cc/paper_files/paper/2022/file/a98846e9d9cc01cfb87eb694d946ce6b-Paper-Conference.pdf},
 volume = {35},
 year = {2022}
}

@inproceedings{rombach2022ldm,
   author = {Rombach, Robin and Blattmann, Andreas and Lorenz, Dominik and Esser, Patrick and Ommer, Björn},
   title = {High-resolution image synthesis with latent diffusion models},
   booktitle = {Proceedings of the IEEE/CVF conference on computer vision and pattern recognition},
   pages = {10684-10695},
   year = {2022},
   type = {Conference Proceedings}
}

@article{moralesbrotons2024ema,
    title={Exponential Moving Average of Weights in Deep Learning: Dynamics and Benefits},
    author={Daniel Morales-Brotons and Thijs Vogels and Hadrien Hendrikx},
    journal={Transactions on Machine Learning Research},
    issn={2835-8856},
    year={2024},
    url={https://openreview.net/forum?id=2M9CUnYnBA},
}

@article{saharia2023sr3,
   author = {Saharia, C. and Ho, J. and Chan, W. and Salimans, T. and Fleet, D. J. and Norouzi, M.},
   title = {Image Super-Resolution via Iterative Refinement},
   journal = {IEEE Trans Pattern Anal Mach Intell},
   volume = {45},
   number = {4},
   pages = {4713-4726},
   ISSN = {1939-3539 (Electronic)
0098-5589 (Linking)},
   DOI = {10.1109/TPAMI.2022.3204461},
   url = {https://www.ncbi.nlm.nih.gov/pubmed/36094974},
   year = {2023},
   type = {Journal Article}
}

@article{li2022srdiff,
   author = {Li, Haoying and Yang, Yifan and Chang, Meng and Chen, Shiqi and Feng, Huajun and Xu, Zhihai and Li, Qi and Chen, Yueting},
   title = {SRDiff: Single image super-resolution with diffusion probabilistic models},
   journal = {Neurocomputing},
   volume = {479},
   pages = {47-59},
   ISSN = {0925-2312},
   DOI = {https://doi.org/10.1016/j.neucom.2022.01.029},
   url = {https://www.sciencedirect.com/science/article/pii/S0925231222000522},
   year = {2022},
   type = {Journal Article}
}

@inproceedings{wang2023patchdiffusion,
 author = {Wang, Zhendong and Jiang, Yifan and Zheng, Huangjie and Wang, Peihao and He, Pengcheng and Wang, Zhangyang "Atlas" and Chen, Weizhu and Zhou, Mingyuan},
 booktitle = {Advances in Neural Information Processing Systems},
 editor = {A. Oh and T. Naumann and A. Globerson and K. Saenko and M. Hardt and S. Levine},
 pages = {72137--72154},
 publisher = {Curran Associates, Inc.},
 title = {Patch Diffusion: Faster and More Data-Efficient Training of Diffusion Models},
 url = {https://proceedings.neurips.cc/paper_files/paper/2023/file/e4667dd0a5a54b74019b72b677ed8ec1-Paper-Conference.pdf},
 volume = {36},
 year = {2023}
}

@article{hamill2020rankhistograms,
   author = {Hamill, T. M.},
   title = {Interpretation of rank histograms for verifying ensemble forecasts},
   journal = {Monthly Weather Review},
   volume = {129},
   number = {3},
   pages = {550-560},
   ISSN = {0027-0644},
   DOI = {Doi 10.1175/1520-0493(2001)129<0550:Iorhfv>2.0.Co;2},
   url = {<Go to ISI>://WOS:000167329000013},
   year = {2001},
   type = {Journal Article}
}

@article{bayram2018sdemethods,
   author = {Bayram, M. and Partal, T. and Buyukoz, G. O.},
   title = {Numerical methods for simulation of stochastic differential equations},
   journal = {Advances in Difference Equations},
   volume = {2018},
   number = {1},
   pages = {17},
   ISSN = {1687-1847},
   DOI = {ARTN 17
10.1186/s13662-018-1466-5},
   url = {<Go to ISI>://WOS:000422722900003},
   year = {2018},
   type = {Journal Article}
}

@article{bassetti2024diffesm,
   author = {Bassetti, S. and Hutchinson, B. and Tebaldi, C. and Kravitz, B.},
   title = {DiffESM: Conditional Emulation of Temperature and Precipitation in Earth System Models With 3D Diffusion Models},
   journal = {Journal of Advances in Modeling Earth Systems},
   volume = {16},
   number = {10},
   pages = {e2023MS004194},
   DOI = {ARTN e2023MS004194
10.1029/2023MS004194},
   url = {<Go to ISI>://WOS:001334285500001},
   year = {2024},
   type = {Journal Article}
}

@article{saoulis2025fathomdiffusion,
   author = {Saoulis, Alexandros Angelos and Lucas, Chris and Lord, Natalie S and Addor, Nans and Moraga, Jorge Sebastián},
   title = {Diffusion Models for Climate Data Surpass alternative Statistical Downscaling Techniques},
   journal = {Authorea Preprints},
   year = {2025},
   type = {Journal Article}
}

@article{hess2025scaleadaptiveconsistencymodels,
   author = {Hess, P. and Aich, M. and Pan, B. X. and Boers, N.},
   title = {Fast, scale-adaptive and uncertainty-aware downscaling of Earth system model fields with generative machine learning},
   journal = {Nature Machine Intelligence},
   volume = {7},
   number = {3},
   pages = {363-373},
   ISSN = {2522-5839},
   DOI = {10.1038/s42256-025-00980-5},
   url = {<Go to ISI>://WOS:001445160800001},
   year = {2025},
   type = {Journal Article}
}

@article{ling2024eastasiadiffusion,
   author = {Ling, F. H. and Lu, Z. Y. and Luo, J. J. and Bai, L. and Behera, S. K. and Jin, D. C. and Pan, B. X. and Jiang, H. D. and Yamagata, T.},
   title = {Diffusion model-based probabilistic downscaling for 180-year East Asian climate reconstruction},
   journal = {Npj Climate and Atmospheric Science},
   volume = {7},
   number = {1},
   pages = {131},
   ISSN = {2397-3722},
   DOI = {ARTN 131
10.1038/s41612-024-00679-1},
   url = {<Go to ISI>://WOS:001248079800002},
   year = {2024},
   type = {Journal Article}
}

@article{rampal2025cganadvlosssensitivity,
   author = {Rampal, N. and Gibson, P. B. and Sherwood, S. and Abramowitz, G. and Hobeichi, S.},
   title = {A Reliable Generative Adversarial Network Approach for Climate Downscaling and Weather Generation},
   journal = {Journal of Advances in Modeling Earth Systems},
   volume = {17},
   number = {1},
   pages = {e2024MS004668},
   DOI = {ARTN e2024MS004668
10.1029/2024MS004668},
   url = {<Go to ISI>://WOS:001387216800001},
   year = {2025},
   type = {Journal Article}
}

@article{rampal2024cganfutureextrapolotaion,
   author = {Rampal, N. and Gibson, P. B. and Sherwood, S. and Abramowitz, G.},
   title = {On the Extrapolation of Generative Adversarial Networks for Downscaling Precipitation Extremes in Warmer Climates},
   journal = {Geophysical Research Letters},
   volume = {51},
   number = {23},
   pages = {e2024GL112492},
   ISSN = {0094-8276},
   DOI = {ARTN e2024GL112492
10.1029/2024GL112492},
   url = {<Go to ISI>://WOS:001370011600001},
   year = {2024},
   type = {Journal Article}
}

@article{miralles2022CPMGANwind,
   author = {Miralles, Ophélia and Steinfeld, Daniel and Martius, Olivia and Davison, Anthony C.},
   title = {Downscaling of Historical Wind Fields over Switzerland Using Generative Adversarial Networks},
   journal = {Artificial Intelligence for the Earth Systems},
   volume = {1},
   number = {4},
   pages = {e220018},
   ISSN = {2769-7525},
   DOI = {10.1175/aies-d-22-0018.1},
   url = {https://journals.ametsoc.org/view/journals/aies/1/4/AIES-D-22-0018.1.xml},
   year = {2022},
   type = {Journal Article}
}

@article{annau2023CPMdetGANwind,
   author = {Annau, Nicolaas J. and Cannon, Alex J. and Monahan, Adam H.},
   title = {Algorithmic Hallucinations of Near-Surface Winds: Statistical Downscaling with Generative Adversarial Networks to Convection-Permitting Scales},
   journal = {Artificial Intelligence for the Earth Systems},
   volume = {2},
   number = {4},
   pages = {e230015},
   ISSN = {2769-7525},
   DOI = {10.1175/aies-d-23-0015.1},
   url = {https://journals.ametsoc.org/view/journals/aies/2/4/AIES-D-23-0015.1.xml},
   year = {2023},
   type = {Journal Article}
}

@article{tomasi2025CPMdiff,
   author = {Tomasi, Elena and Franch, Gabriele and Cristoforetti, Marco},
   title = {Can AI be enabled to perform dynamical downscaling? A latent diffusion model to mimic kilometer-scale COSMO5.0\_CLM9 simulations},
   journal = {Geoscientific Model Development},
   volume = {18},
   number = {6},
   pages = {2051-2078},
   ISSN = {1991-9603},
   DOI = {10.5194/gmd-18-2051-2025},
   url = {https://gmd.copernicus.org/articles/18/2051/2025/},
   year = {2025},
   type = {Journal Article}
}

@article{glawion2023spategan,
   author = {Glawion, Luca and Polz, Julius and Kunstmann, Harald and Fersch, Benjamin and Chwala, Christian},
   title = {spateGAN: Spatio-Temporal Downscaling of Rainfall Fields Using a cGAN Approach},
   journal = {Earth and Space Science},
   volume = {10},
   number = {10},
   pages = {e2023EA002906},
   DOI = {https://doi.org/10.1029/2023EA002906},
   url = {https://agupubs.onlinelibrary.wiley.com/doi/abs/10.1029/2023EA002906},
   year = {2023},
   type = {Journal Article}
}

@article{bettolli2025stationcnnberngamma,
   author = {Bettolli, M. L. and da Rocha, R. P. and Milovac, J. and Fernandez, J. and Balmaceda-Huarte, R. and Baño-Medina, J. and Blázquez, J. and Rodrigues, D. C. and Chou, S. C. and Coppola, E. and da Silva, M. L. and Doyle, M. and Llorente, J. M. G. and Olmo, M. and Prein, A. F. and Raffaele, F. and Solman, S. and Cuadra, S. V.},
   title = {High-Resolution Deep-Learning and Dynamical Climate Downscaling for Impact Modeling in Southeast South America},
   journal = {Earth Systems and Environment},
   ISSN = {2509-9426},
   DOI = {10.1007/s41748-025-00661-8},
   url = {<Go to ISI>://WOS:001499156400001},
   year = {2025},
   type = {Journal Article}
}

@article{aich2024diffdownbc,
   author = {Aich, Michael and Hess, Philipp and Pan, Baoxiang and Bathiany, Sebastian and Huang, Yu and Boers, Niklas},
   title = {Conditional diffusion models for downscaling \& bias correction of Earth system model precipitation},
   journal = {arXiv preprint arXiv:2404.14416},
   year = {2024},
   type = {Journal Article}
}

@article{banomedina2024emulatorxferxai,
   author = {Baño-Medina, Jorge and Iturbide, Maialen and Fernández, Jesús and Gutiérrez, José Manuel},
   title = {Transferability and Explainability of Deep Learning Emulators for Regional Climate Model Projections: Perspectives for Future Applications},
   journal = {Artificial Intelligence for the Earth Systems},
   volume = {3},
   number = {4},
   pages = {e230099},
   ISSN = {2769-7525},
   DOI = {10.1175/aies-d-23-0099.1},
   url = {https://journals.ametsoc.org/view/journals/aies/3/4/AIES-D-23-0099.1.xml},
   year = {2024},
   type = {Journal Article}
}

@article{sun2024detdldownscalingreview,
   author = {Sun, Y. J. and Deng, K. F. and Ren, K. J. and Liu, J. and Deng, C. J. and Jin, Y. J.},
   title = {Deep learning in statistical downscaling for deriving high spatial resolution gridded meteorological data: A systematic review},
   journal = {Isprs Journal of Photogrammetry and Remote Sensing},
   volume = {208},
   pages = {14–38},
   ISSN = {0924-2716},
   DOI = {10.1016/j.isprsjprs.2023.12.011},
   url = {<Go to ISI>://WOS:001156403800001},
   year = {2024},
   type = {Journal Article}
}

@article{maurerUtilityDailyLargescale2010,
  title = {The Utility of Daily Large-Scale Climate Data in the Assessment of Climate Change Impacts on Daily Streamflow in {{California}}},
  author = {Maurer, E. P. and Hidalgo, H. G. and Das, T. and Dettinger, M. D. and Cayan, D. R.},
  year = 2010,
  month = jun,
  journal = {Hydrol. Earth Syst. Sci.},
  volume = {14},
  number = {6},
  pages = {1125--1138},
  publisher = {Copernicus Publications},
  issn = {1607-7938},
  doi = {10.5194/hess-14-1125-2010}
}

@article{woodHydrologicImplicationsDynamical2004,
  title = {Hydrologic {{Implications}} of {{Dynamical}} and {{Statistical Approaches}} to {{Downscaling Climate Model Outputs}}},
  author = {Wood, A. W. and Leung, L. R. and Sridhar, V. and Lettenmaier, D. P.},
  year = 2004,
  month = jan,
  journal = {Climatic Change},
  volume = {62},
  number = {1},
  pages = {189--216},
  issn = {1573-1480},
  doi = {10.1023/B:CLIM.0000013685.99609.9e},
  urldate = {2025-10-22},
  langid = {english},
}

\newpage

\appendix

\section{Diffusion models}
\label{sec:app:diff}

Probabilistic models assume that observed data, such as high-resolution precipitation over the UK, is drawn from an unknown distribution \(p^*(\mathbf{x})\). A conditional model such as our high-resolution precipitation emulator conditioned on coarse inputs \(p^*(\mathbf{x}|\mathbf{y})\) can also be considered but for simplicity we will stick with the unconditional version.

\citeA{song2021sbgmsde} combine earlier approaches \cite{song2019smld, ho2020ddpm} into a single framework called Score-Based Generative Models with Stochastic Differential Equations (SDE). The idea is to imagine a diffusion process \(\{\mathbf{x}(t)_{t=0}^{T}\}\) modelled by an SDE:

\begin{equation}
  \mathrm{d}\mathbf{x} = \mathbf{f}(\mathbf{x}, t)\mathrm{d}t + g(t)d\mathbf{w}
\end{equation}

When run forward, a sample, \(\mathbf{x}(0)\), from the data distribution, \(p_0\), is gradually perturbed over time into a sample from a final noise distribution, \(p_T\). The final distribution is chosen as something tractable for sampling, usually a Gaussian.

More interesting for us is running the reverse diffusion process:

\begin{equation} \label{eqn:mlbg:reversesde}
  \mathrm{d}\mathbf{x} = [\mathbf{f}(\mathbf{x}, t) - g(t)^2 \nabla_{\mathbf{x}}\log{p_{t}(x)}]{d}t + g(t)d\bar{\mathbf{w}}
\end{equation}

By solving this, samples from \(p_T\) (which are easy to produce by design) can be converted into samples from the original data distribution. This requires two steps: calculating the score, \(\nabla_{\mathbf{x}}\log{p_{t}(x)}\), and then applying numerical approaches to solve Equation \ref{eqn:mlbg:reversesde}.

The score is estimated as a neural net \(s_\theta(\mathbf{x}, t)\) where \(\theta\) are determined by minimizing:

\begin{equation}
  \mathbb{E}_t \{
    \lambda(t) \mathbb{E}_{\mathbf{x}(0)} \mathbb{E}_{\mathbf{x}(0) | \mathbf{x}(t)}
      \left[
        || s_\theta(\mathbf{x}(t), t) -  \nabla_{\mathbf{x}(t)}\log{p_{0t}(\mathbf{x}(t) | \mathbf{x}(0))} ||_2^2
      \right
    \}
\end{equation}
where \(\lambda\) is a positive weighting function that is chosen along with f and g.

\citeA{song2021sbgmsde} summarize three approaches for solving the reverse SDE. General-purpose numerical methods can be used to find approximate solutions to the SDE. Predictor-Corrector sampling takes this a step further by using making use of estimated score at each timestep to apply a correction to the sample estimated at that timestep by the general purpose solver. Alternatively the problem can be reformulated as a deterministic process without affecting the trajectory probabilities and in turn solved using an ODE solver.

\section{Further climate change results}
\label{sec:app:ccs-extra}

Table~\ref{tab:CPM-CCS-domain-mean} and Table~\ref{tab:CCS-extreme-quantile-changes-cCPM} show climate change related results based on cCPM inputs rather than GCM inputs as used in the main text.

\begin{table}[h]
    \centering
    \begin{tabulary}{0.9\textwidth}{c|R|R|R|R}
    Season &  CPM (\%) & Emulator (\%) & Difference (\% of CPM change) & Difference (\% of Historic CPM) \\
        \hline
        DJF &  23 &  18 & -21 (\texttt{-}33 to \texttt{\phantom{1}\phantom{1}-}9) & -6 (\texttt{-}7.7 to \texttt{-}2.0) \\
        MAM &   6 &   4 & -31 (\texttt{-}78 to \texttt{\phantom{1}+}16)           & -2 (\texttt{-}4.8 to \texttt{+}1.0) \\
        JJA & -40 & -39 &   2 (\texttt{-}7 to \texttt{\phantom{1}\phantom{1}+}10) &  1 (\texttt{-}2.9 to \texttt{+}3.9) \\
        SON &  -4 &  -6 & -59 (\texttt{\phantom{1}-}129 to \texttt{+}9)           & -2 (\texttt{-}5.1 to \texttt{+}0.4)
    \end{tabulary}
    \caption{\textbf{Change in seasonal domain mean for CPMGEM\_cCPM.} As Table~\ref{tab:GCM-CCS-domain-mean} for CPMGEM\_cCPM. Shows changes in the domain mean from Historic to Future periods for each season. The second and third columns contain the relative changes for the CPM and the CPMGEM\_cCPM emulator respectively (relative to CPM Historic seasonal domain mean in both cases). The fourth column contains the difference between the change in the CPM and CPMGEM\_cCPM, relative to the change in CPM. The fifth column shows the same difference but relative to the Historic CPM seasonal domain mean. The values in brackets in these columns are the bootstrapped 95\% confidence intervals of the difference.}
    \label{tab:CPM-CCS-domain-mean}
\end{table}

\begin{table}[]
    \centering
    \begin{tabulary}{0.9\textwidth}{C|c|c|c|c|c}
        Season & Percentile & CPM & Emulator \\
        \hline
        Winter & 99   & 23.8 & 18.7 (15.3 to 24.6) \\
        Winter & 99.9 & 28.3 & 25.0 (20.6 to 31.0) \\
        \hline
        Summer & 99   & -7.0 & -10.9 (-12.3 to -9.5) \\
        Summer & 99.9 & 12.5 &   9.9 (9.3 to 11.3) \\
    \end{tabulary}
    \caption{\textbf{Percentage changes in heavy precipitation intensities (in \%) between the Historic and Future time periods, for winter and summer using CPMGEM\_cCPM predictions.} As in Table~\ref{tab:CCS-extreme-quantile-changes-GCM} for CPMGEM\_cCPM. Domain-mean of seasonal relative change in heavy daily-mean precipitation (99th and 99.9th percentile column 2) for CPM (column 3) and the CPMGEM\_cCPM emulator (column 4). For the emulator, the value is the domain-mean of the percentage changes averaged over the 6 independent emulator sample runs (with the range across the samples given in brackets).}
    \label{tab:CCS-extreme-quantile-changes-cCPM}
\end{table}

\end{document}


%
%


\title{Supporting Information for "Machine learning emulation of precipitation from km-scale UK regional climate simulations using a diffusion model"}

%
%



\authors{Henry Addison\affil{1}, Elizabeth J. Kendon\affil{2,1}, Suman Ravuri\affil{3}, Laurence Aitchison\affil{1}, Peter A.G. Watson\affil{1}}

\affiliation{1}{University of Bristol, Bristol, UK}
\affiliation{2}{Met Office Hadley Centre, Exeter, UK}
\affiliation{3}{NVIDIA Corporation}

%
%

%

\begin{article}

%
%

\noindent\textbf{Contents of this file}
\begin{enumerate}
\item Text S1
\item Figures S1 to S8
\item Table S1 
\end{enumerate}

\noindent\textbf{S1. Return Level Estimates}

Winter and summer return levels were estimated for an individual grid box in each of the SE and NW domains  (see Section 3.1.2 in manuscript). The chosen grid boxes correspond to the locations of cities London and Lancaster, respectively, chosen as locations where heavy rainfall can have a large impact due to large populations.  Return periods for precipitation equalling or exceeding values in the data are estimated by sorting the time series of precipitation at the given grid box for a given season by intensity and taking the nth entry in this sorted list as the estimate of the n-in-108 year return level (e.g. the maximum is 1-in-108 year return level, the second largest is 1-in-56 year return level and so on). 

\begin{figure}[htbp]
    \centering
    \includegraphics[width=\textwidth]{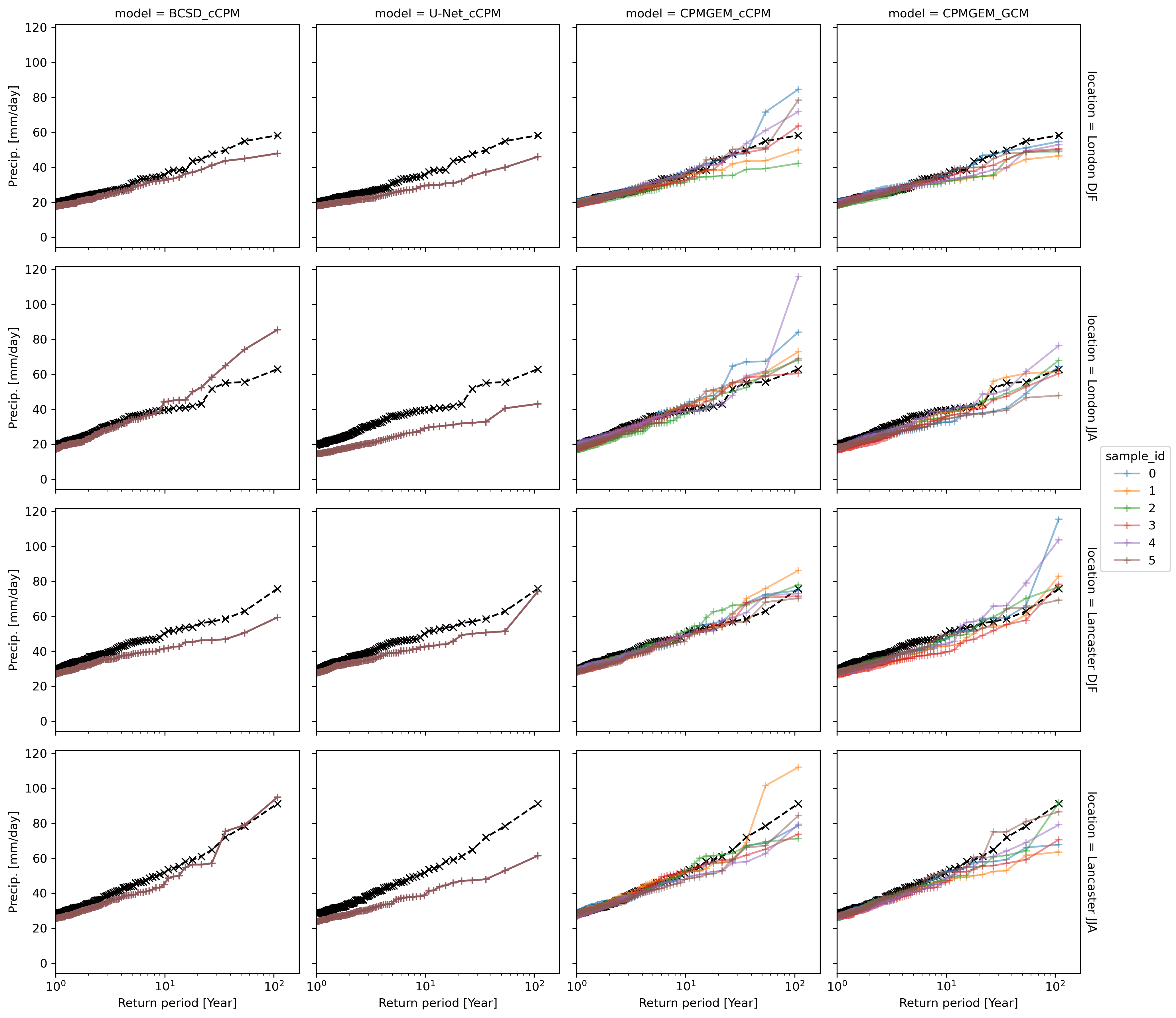}
    \caption{\textbf{Return level estimates for individual grid boxes}, with intensity plotted on the vertical axis and return period plotted horizontally, for grid points at the locations of London (top two rows) and Lancaster (bottom two rows) for winter (rows 1 and 3) and summer (rows 2 and 4). CPM return levels are shown in black. For CPMGEM, each colour shows a separate return period estimate from using one of the 6 generated samples for each day, indicating sampling variability. See S1 above for more.}
    \label{fig:return-levels}
\end{figure}

Figure~\ref{fig:return-levels} shows these return level estimates from 1-in-1 year to 1-in-108 year return periods for the London grid box in summer and winter (rows one and two) and the Lancaster grid box (rows three and four). Crosses mark data points from the CPM simulations and CPMGEM samples, with lines a simple piecewise interpolation joining them. Results are shown for BCSD\_cCPM (column one), CPMGEM driven by GCM and cCPM inputs (columns two and three; coloured lines each represent one of the six samples sets generated from our stochastic emulator) and U-Net\_cCPM (column four). The black dashed line shows the corresponding return levels for the CPM output. The differences in performance between the methods were similar when computing return levels based on the domain maximum for the whole SE and NW domains (not shown).

The error (and spread for CPMGEM) for the four cases is summarized in Table~\ref{tab:return-levels} for return periods of 1, 9.8 and 108 years.

\end{article}
\clearpage

\begin{figure}[htbp]
    \centering
    \includegraphics[width=\textwidth]{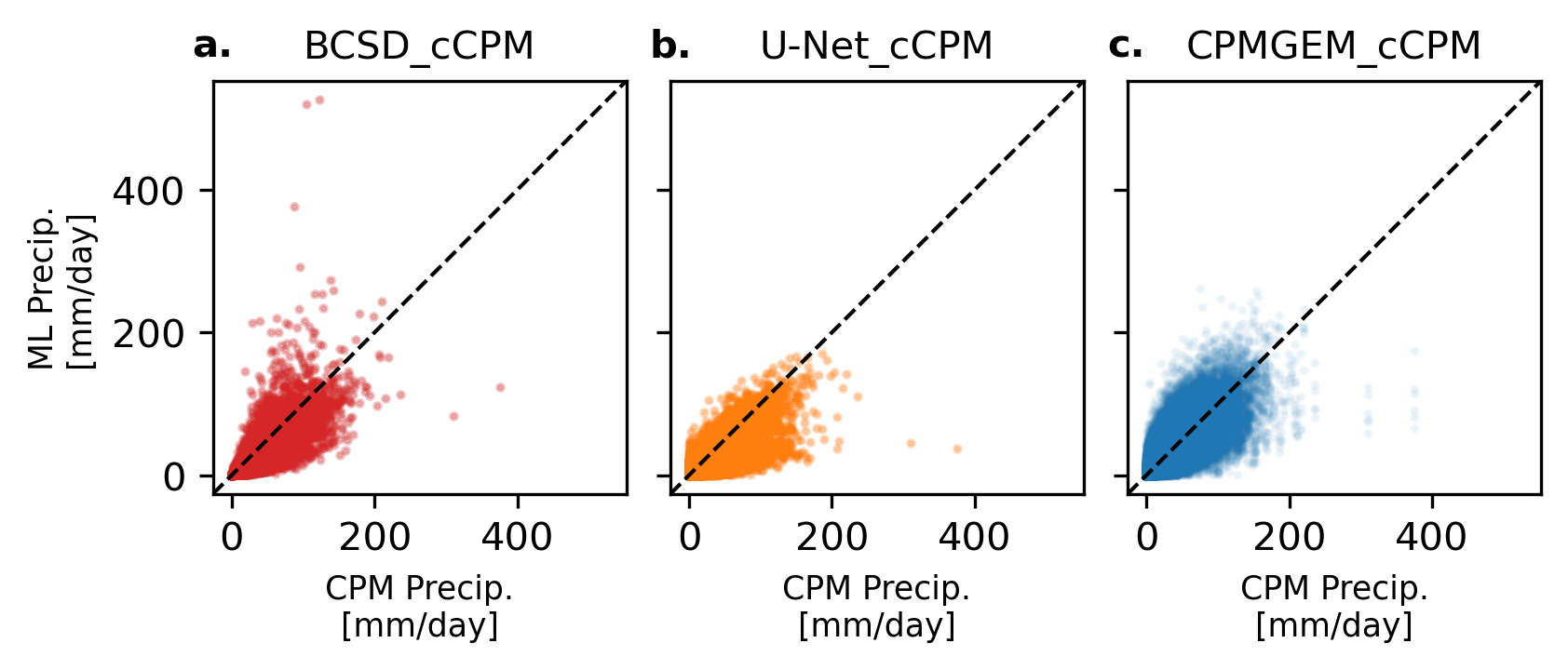}
    \caption{\textbf{Daily emulated vs CPM domain-maximum precipitation.} Scatter plots of domain maximum precipitation for emulator samples versus target CPM value for U-Net\_cCPM (left) and CPMGEM\_cCPM (right, separate points for each of the six sample runs for CPMGEM\_cCPM). Plots combine data from all three time periods.}
    \label{fig:domain-max}
\end{figure}

\begin{figure}[htbp]
    \centering
    \includegraphics[width=\textwidth]{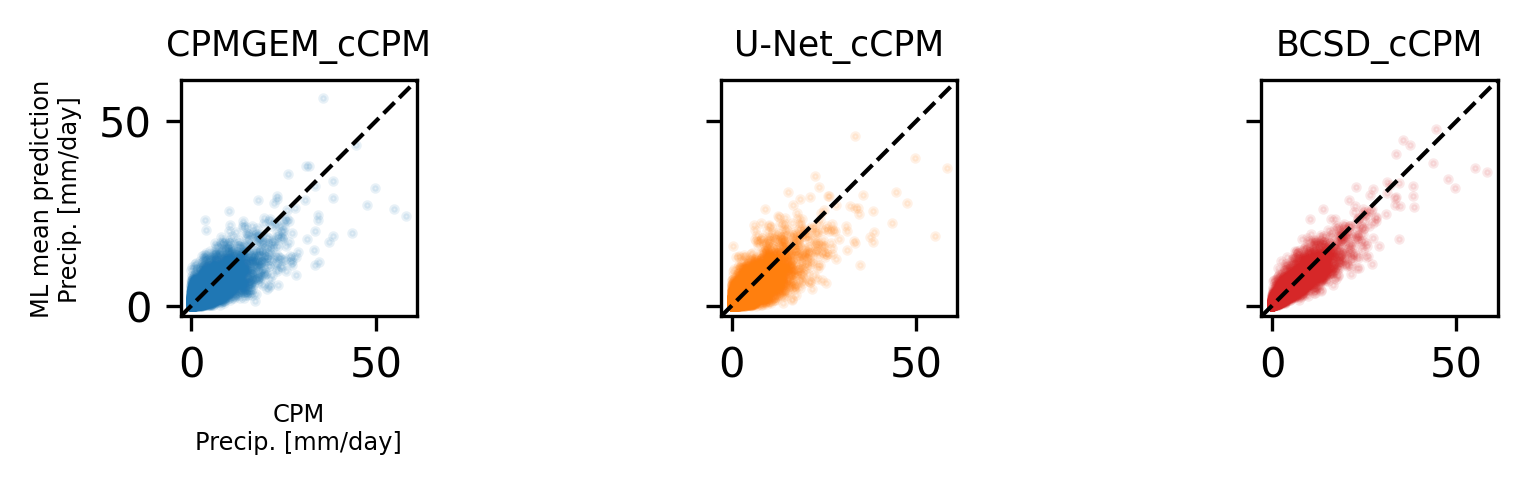}
    \caption{\textbf{Daily emulated vs CPM precipitation for a single grid box in London for winter}. Scatter plots of daily precipitation at a single grid box in London for predictions versus target CPM values.}
    \label{fig:scatter:london-djf}
\end{figure}

\begin{figure}[htbp]
    \centering
    \includegraphics[width=\textwidth]{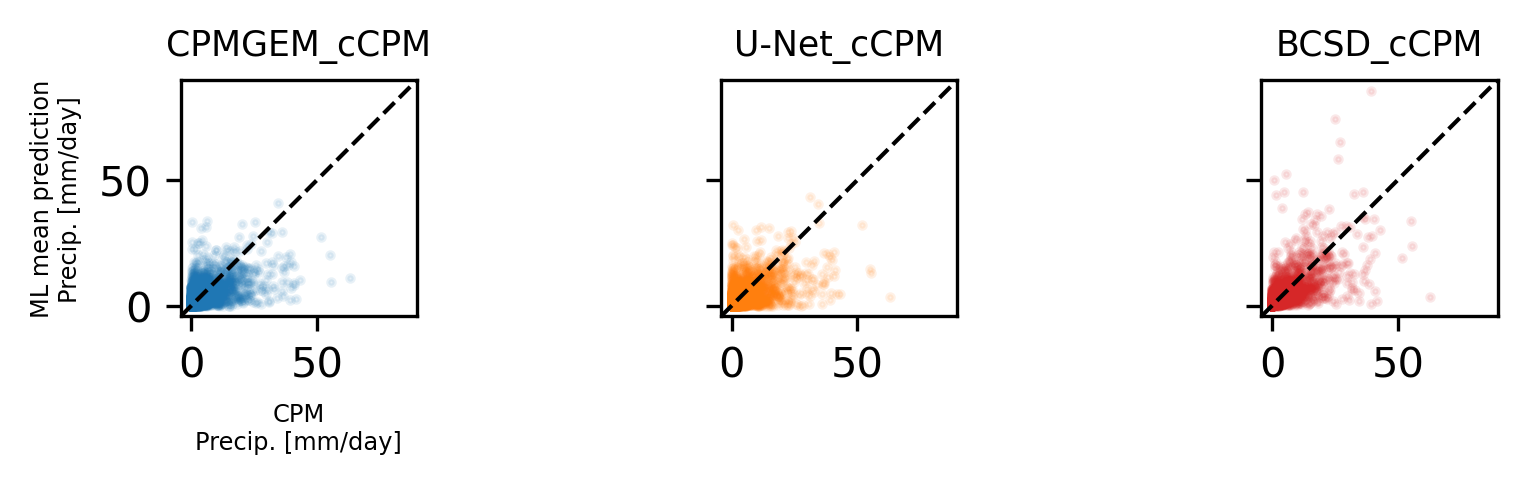}
    \caption{\textbf{Daily emulated vs CPM precipitation for a single grid box in London for summer}. As Figure~\ref{fig:scatter:london-djf} for summer.}
    \label{fig:scatter:london-jja}
\end{figure}

\begin{figure}[htbp]
    \centering
    \includegraphics[width=\textwidth]{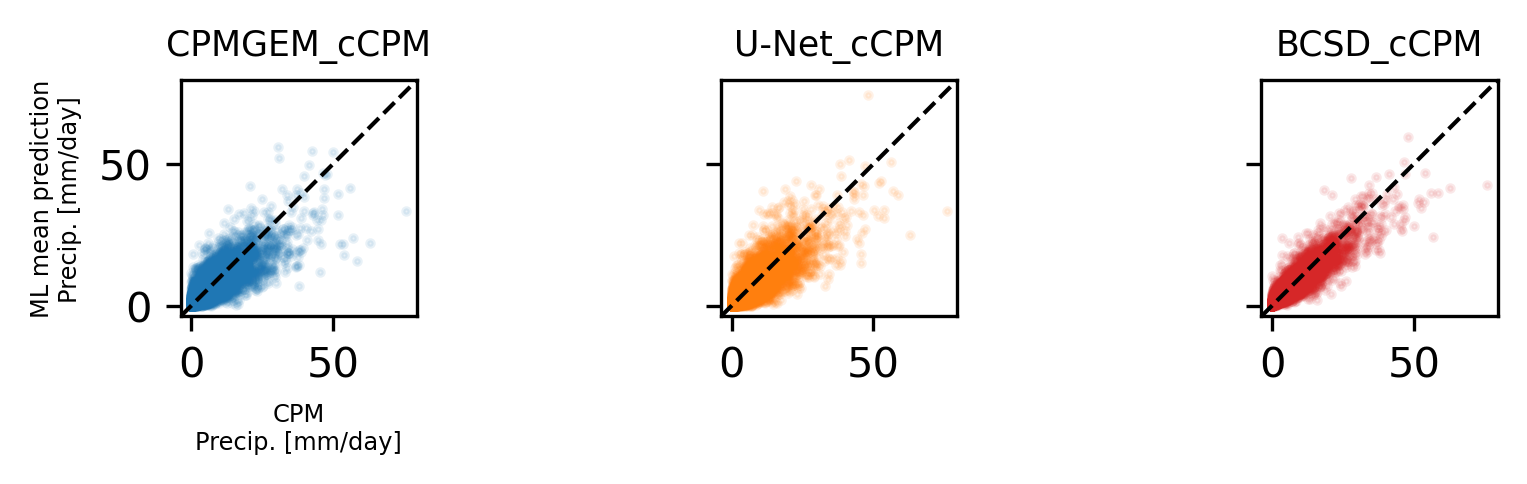}
    \caption{\textbf{Daily emulated vs CPM precipitation for a single grid box in Lancaster for winter}. As Figure~\ref{fig:scatter:london-djf} for summer.}
    \label{fig:scatter:lanc-djf}
\end{figure}

\begin{figure}[htbp]
    \centering
    \includegraphics[width=\textwidth]{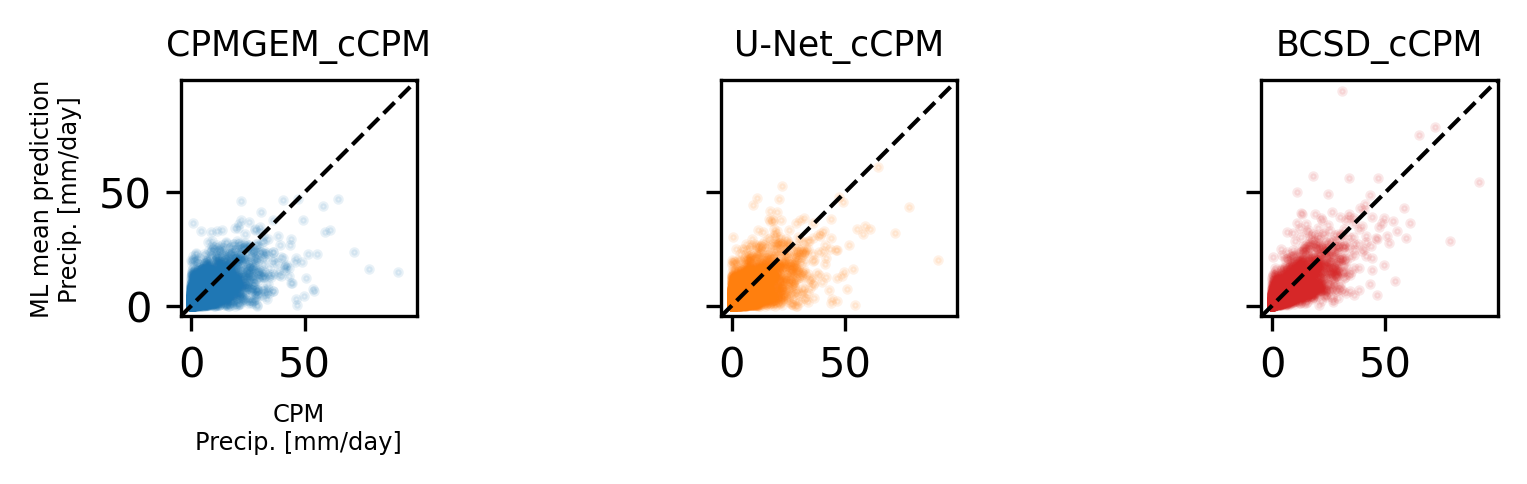}
    \caption{\textbf{Daily emulated vs CPM precipitation for a single grid box in Lancaster for summer}. As Figure~\ref{fig:scatter:london-djf} for summer.}
    \label{fig:scatter:lanc-jja}
\end{figure}

\begin{figure}[htbp]
    \centering
    \includegraphics[width=\textwidth]{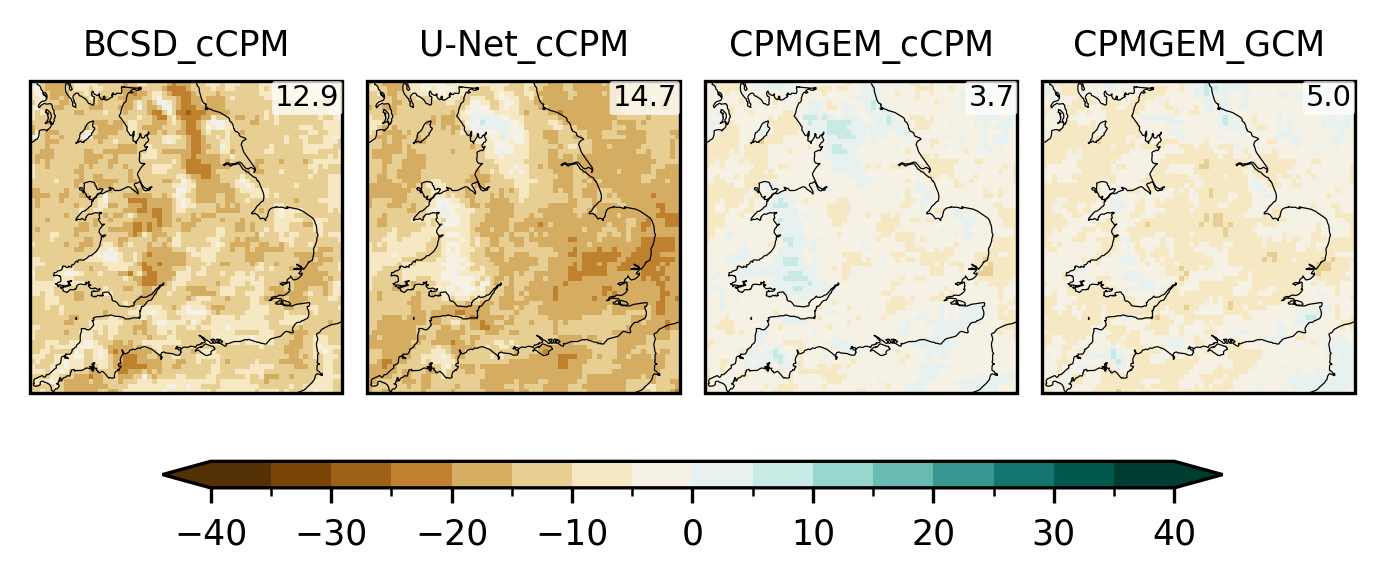}
    \caption{\textbf{Percentage bias in mean RX1Day relative to the CPM}. For CPMGEM, RX1Day is calculated for each of the six generated samples and the mean over these is used to calculate the bias. Root mean squared bias (\%) over the domain is shown in top right of each spatial plot.}
    \label{fig:rx1day-bias}
\end{figure}

\begin{figure}[htbp]
    \centering
    \includegraphics[width=\textwidth]{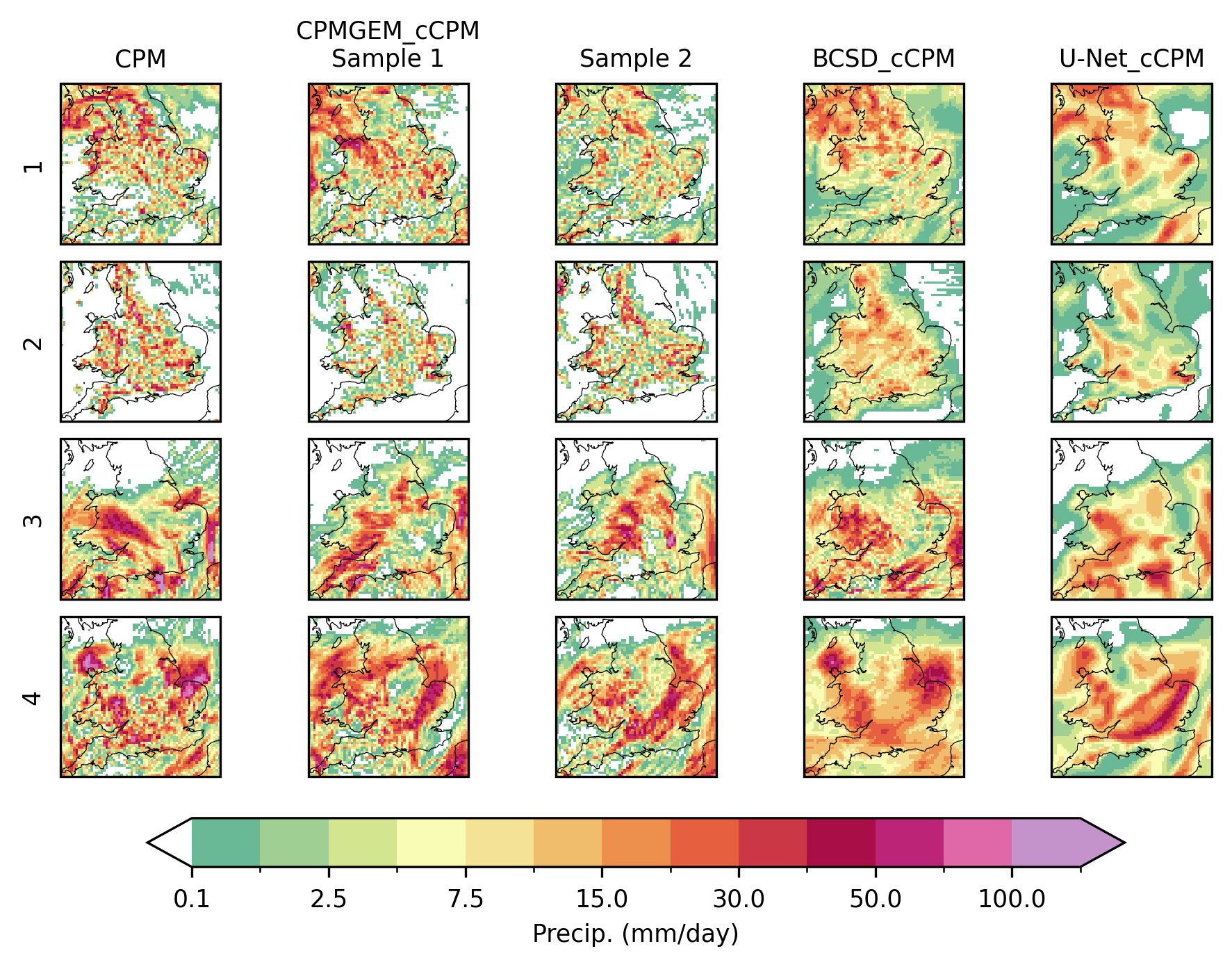}
    \caption{\textbf{Predictions from examples of days with high variance at small spatial scales}. The CPM examples were chosen by selecting the four examples with the highest RAPSD at the smallest spatial scale (17.6km). Results are presented as in Figure~2 (without the example input column).}
    \label{fig:samples:hi-small-scale-variance}
\end{figure}

\clearpage

\begin{sidewaystable}
    \centering
    \caption{\textbf{Error in return level estimates (mm/day).} Error in winter and summer return level estimates between predictions and CPM for individual grid boxes (London, top six rows; Lancaster bottom six rows) for select return periods. For CPMGEM (columns five and six), error is given as a range across the six sets of samples generated and values are shown in bold when this range include 0 error.}
    \label{tab:return-levels}
    \begin{tabular}{c|p{3cm}|c|c|c|c}
\textbf{Location} & \centering \textbf{Return period \newline (Years)} & \textbf{BCSD\_cCPM} & \textbf{U-Net\_cCPM} & \textbf{CPMGEM\_cCPM} & \textbf{CPMGEM\_GCM} \\
\hline
\textbf{London DJF} & \centering 1 & -2.1 & -1.7 & \textbf{-1.1 to 0.5} & \textbf{-1.4 to 0.3} \\
 & \centering 9.8 & -3.0 & -6.2 & \textbf{-4.3 to 1.4} & \textbf{-6.9 to 0.9} \\
 & \centering 108 & -10.4 & -12.3 & \textbf{-16 to 26.3} & -11.8 to -3.6 \\
 \hline
\textbf{London JJA}  & \centering 1 & -2.5 & -5.1 & \textbf{-2.8 to 0.0} & -3.0 to -1.8 \\
 & \centering 9.8 & 4.6 & -10.4 & -6.9 to -0.1 & \textbf{-2.5 to 3.0} \\
 & \centering 108 & 22.5 & -19.9 & \textbf{-2.4 to 52.9} & \textbf{-15.1 to 13.4} \\
 \hline
\textbf{Lancaster DJF} & \centering 1 & -2.6 & -2.0 & -3.3 to -1.7 & -1.3 to -0.5 \\
 & \centering 9.8 & -8.6 & -7.6 & \textbf{-4.5 to 1.1} & \textbf{-10.4 to 0.6} \\
 & \centering 108 & -16.4 & -1.6 & \textbf{-6.5 to 39.7} & \textbf{-5.5 to 10.4} \\
 \hline
\textbf{Lancaster JJA} & \centering 1 & -2.7 & -4.7 & \textbf{-1.4 to 0.6} & -2.7 to -1.5 \\
 & \centering 9.8 & -6.8  & -12.8 & \textbf{-3.8 to 1.7} & \textbf{-5.5 to 0.6} \\
 & \centering 108 & 3.7  & -29.9 & \textbf{-17.5 to 20.8} & \textbf{-27.7 to 0.4}
    \end{tabular}
    
\end{sidewaystable}